\documentclass[11pt,Bold,letterpaper]{mcgilletdclassmine}
\usepackage[dvips,final]{graphicx}
\usepackage{latexsym,amssymb, amsmath,amsfonts,epsfig, epic, eepic,subfigure}
\usepackage[dvips,letterpaper,left=3.8cm,top=3.2cm,bottom=2.5cm,right=2.5cm,includefoot]{geometry}
\usepackage{slashed}
\usepackage{rotating}
\usepackage{subfigure}

\newcommand{\roughly}[1]{\mathrel{\raise.3ex\hbox{$#1$\kern-0.85em
\lower1ex\hbox{$\sim$}}}}

\def\gsim{\roughly>}

\SetTitle{\Huge\bfseries{String theory in the early universe\\\vspace{1cm}}}%
\SetAuthor{\vspace{1.5cm}\Large\bfseries{Rhiannon Gwyn}}%
\SetDepartment{\Large{Physics Department}}
\SetUniversity{\Large{McGill University}}%
\SetUniversityAddr{\Large{Montreal, Quebec}}%
\SetThesisDate{\Large{June 2009}}%
\SetRequirements{\vspace{2cm}\Large{A thesis submitted in partial fulfilment \\ of 
the requirements for the degree of }}%
\SetDegreeType{\vspace{0.3cm}\Large{Doctor of Philosophy}}%
\SetCopyright{\hspace{20mm}}%

\newcommand{\newchapter}[1]
{\chapter{#1} \markboth{Chapter \thechapter. {#1}\hfill \today\hfill}{Chapter \thechapter. {#1}\hfill \today 
\hfill}}

\begin{document}

\maketitle%


\begin{romanPagenumber}{1}%
\SetAbstractEnName{Abstract}%
\SetAbstractEnText{
String theory is a rich and elegant framework which many believe furnishes a UV-complete unified theory of the fundamental interactions, including gravity. However, if true, it holds at energy scales out of the reach of any terrestrial particle accelerator. While we cannot observe the string regime directly, we live in a universe which has been evolving from the string scale since shortly after the Big Bang. It is possible that string theory underlies cosmological processes like inflation, and that cosmology could confirm or constrain stringy physics in the early universe. This makes the intersection of string theory with the early universe a potential window into otherwise inaccessible physics. \\
\hspace*{0.5cm}
The results of three papers at this intersection are presented in this thesis. First, we address a longstanding problem: the apparent incompatibility of the experimentally constrained axion decay constant with most string theoretic realisations of the axion. Using warped compactifications in heterotic string theory, we show that the axion decay constant can be lowered to acceptable values by the warp factor.\\
\hspace*{0.5cm}
Next, we move to the subject of cosmic strings: linelike topological defects formed during phase transitions in the early universe. It was realised recently that cosmic superstrings are produced in many models of brane inflation, and that cosmic superstrings are stable and can have tensions within the observational bounds. Although they are now known not to be the primary generators of primordial density perturbations leading to structure formation, the evolution of cosmic string networks could have important consequences for astrophysics and cosmology. In particular, there are quantitative differences between cosmic superstring networks and GUT cosmic string networks. \\
\hspace*{0.5cm}
We investigate the properties of cosmic superstring networks in warped backgrounds, where they are expected to be produced at the end of brane inflation. We give the tension and properties of three-string junctions of different kinds in these networks. Finally, we examine the possibility that cosmic strings in heterotic string theory could be responsible for generating the galactic magnetic fields that seeded those observed today. We were able to construct suitable strings from wrapped M5-branes. Demanding that they support charged zero modes forces us into a more general heterotic M--theory picture, in which the moduli of a large moduli space of M-theory compactifications are time dependent and evolve cosmologically. Thus a string theory solution of this problem both implies constraints on the string theory construction and has cosmological implications which might be testable with future observations. 
The breadth of topics covered in this thesis is a reflection of the importance of the stringy regime in the early universe, the effects of which may be felt in many different contexts today. The intersection of string theory with cosmology is thus a complex and exciting field in the study of fundamental particle physics. }
\AbstractEn%
\SetAcknowledgeName{{Acknowledgements}}%
\SetAcknowledgeText{I would like to thank my supervisor, Keshav Dasgupta, for his seemingly boundless time and help. I am indebted to him for patiently teaching me string theory and guiding my work in all the projects undertaken during my Ph.D. and for his support and encouragement throughout.  \\
\hspace*{0.5cm} I would also like to thank the other faculty members in the high energy theory group at McGill, from whom I have learnt a great deal. I am indebted to Jim Cline, Alex Maloney and Guy Moore, and especially Robert Brandenberger. My Master's supervisor Robert de Mello Koch's support and encouragement have been indispensable. During my Ph.D. I received financial support from the Physics department at McGill University, my supervisor Keshav Dasgupta, a McGill Major's Chalk-Rowles fellowship and a Schulich fellowship.\\
\hspace*{0.5cm} Anke Knauf has been a source of support, an inspiration and a font of wisdom. Thanks also to Hassan Firouzjahi, Andrew Frey, Omid Saremi and Bret Underwood. I would also like to thank my collaborators Stephon Alexander, Josh Guffin and Sheldon Katz; my officemates and peers Neil Barnaby, Aaron Berndsen, Simon Carot-Huot, Racha Cheaib, Lynda Cockins, Rebecca Danos, Paul Franche, Johanna Karouby,  Nima Lashkari, Dana Lindemann, Subodh Patil, Natalia Shuhmaher, James Sully, Aaron Vincent, Alisha Wissanji, Hiroki Yamashita and especially Ra'ad Mia; the lecturers and organisers of the Jerusalem winter school, PiTP, Les Houches and TASI; and my fellow students there Michael Abbott, Murad Alim, Ines Aniceto, Chris Beem, Adam Brown, Alejandra Castro, Paul Cook, Sophia Domokos, Lisa Dyson, Damien George, Manuela Kuraxizi, Louis Leblond, Nelia Mann, Arvind Murugan, Jonathan Pritchard, Rakib Rahman, Sarah Shandera, Alex Sellerholm, Jihye Seo, Julian Sonner, David Starr, Linda Uruchurtu, Amanda Weltman Murugan, Ketan Vyas, and Navin Sivanandam in particular. \\
Thanks are also due to my friends outside of string theory - listed elsewhere - and to my family. I'd like to thank Rhys and Lludd, Fiona and Marianne Ackerman, Cathleen Mawdsley-Inggs, and especially my parents Gwyn Campbell and Judith Inggs, to whom I owe everything. This thesis is dedicated to Nannie and Grandad, Mamgu, and Iago and Iestyn Grwndi.\\
\hspace*{0.7cm} 
  }%
\Acknowledge
\TOCHeading{\huge\bf{Table of Contents}\vspace{1cm}}%
\LOFHeading{List of Figures}%
\LOTHeading{List of Tables}%
\tableofcontents %
\listoffigures %
\listoftables %
 
 \end{romanPagenumber}

\newchapter{Introduction: String theory and cosmology}
\label{introduction}
It is often said that the two great pillars of twentieth century physics were the theories of quantum mechanics, formulated by Heisenberg, Schrodinger, Born and others in the 1920s, and the theory of general relativity, developed by Einstein in 1916. These impressive theoretical works have been confirmed by every conceivable experiment and have resulted in technological advances which shaped the history of the last century, such as transistors and global satellite devices. They represent massive advancement of our knowledge of the world on either side of the human scale, pushing back the frontiers on the scales of the very small (atoms and their constituents) as well as the very large (galaxies and galactic clusters). And, counterintuitive as it may initially seem, attempting to push either one of those frontiers back still further - to gain either a complete understanding of the universe's evolution or the quantum world - leads the theoretical physicist to a regime where the two are intertwined.
\section{Overview}
In this thesis I present the work and findings of a series of projects at the intersection of string theory with `real-world' physics in cosmology and particle physics. These projects were undertaken during my Ph.D. and published in the papers \cite{Dasgupta:2007ds, Dasgupta:2008hb, Gwyn:2008fe}. Other work published by myself together with collaborators in this period has some relevance to the topics presented here and is cited where necessary, but I have chosen to focus on the projects dealing with string theory and cosmology here so as to limit the length and tighten the scope of the thesis. The other work undertaken during my Ph.D. \cite{Dasgupta:2006yd, Dasgupta:2006sg} focussed mainly on geometric transitions \cite{Gopakumar:1998ki, Atiyah:2000zz, Cachazo:2001jy, Vafa:2000wi} and is reviewed in the article by Gwyn and Knauf \cite{Gwyn:2007qf}.

I begin in this introduction by explaining the relevance of string theory to early universe physics. If string theory is the correct theory at shortest distances and highest energies, it should also be the correct theory at the earliest times, which means that cosmology and particle physics necessarily intersect. In Chapter \ref{throat} I review flux compactifications and the setup of the Klebanov-Strassler throat, which will be needed for Chapters \ref{chapter:axion} and \ref{lumps}. The original work in this thesis begins in Chapter \ref{chapter:axion} where we present the investigation of  a viable axion model in string theory. This investigation at the intersection of string theory with particle physics was published in \cite{Dasgupta:2008hb}. The second two projects concern string cosmology and specifically  cosmic strings arising from string theory. Chapter \ref{chapter:CS} is an introduction and literature review of cosmic strings and their networks. In Chapter \ref{lumps} we investigate the properties of cosmic strings in warped geometries, while superconducting heterotic cosmic strings and the possibility of generating primordial galactic magnetic fields from them are explained in Chapter \ref{magneto}. Chapters \ref{lumps} and \ref{magneto} are based on the papers \cite{Dasgupta:2007ds} and \cite{Gwyn:2008fe} respectively. We end with a Conclusion.
\section{String theory}
\subsection{Motivation}
\subsubsection{The Standard Model}
Combining special relativity and quantum mechanics led in the middle half of the last century to quantum field theory, the theoretical framework for our current model of particle physics (excluding gravity), known as the Standard Model (SM). Quantum mechanics and electromagnetism were unified by quantum electrodynamics (QED), a quantum field theory developed by Dirac and Dyson (among others) and finalised by Feynman, Schwinger and Tomonaga in the 1940s. QED was confirmed to many decimal places by experiments in the 1950s. In the 1960s it was discovered by Sheldon Glashow, Steven Weinberg and Abdus Salam that QED and the theory of the weak interaction (which governs left-handed leptons and flavour-changing processes like beta decay) are the disparate low-energy descriptions of a more symmetric unified electroweak theory, in which (at energies higher than the electroweak symmetry-breaking scale) photons and vector bosons are indistinguishable. The theory of the strong interaction, quantum chromodynamics or QCD, was finalised in the mid 1970s after experimental evidence that nucleons are made up of fractionally charged quarks. Together, these theories make up the SM, a description of all particle physics:  the three generations of quarks and leptons and the gauge bosons mediating the strong, weak and electromagnetic interactions. It is a gauge field theory with gauge group $SU(3) \times SU(2) \times U(1)$. The SM has been subjected to many tests. Together with general relativity, the SM is consistent with almost all known physics, down to the smallest scale we can probe with particle accelerators. It has been confirmed by repeated experimental verification of its predictions, for instance the existence and properties of the top quark, discovered at Fermilab in 1995; and the W and Z bosons, discovered at CERN in 1983. 
\subsubsection{Problems with the Standard Model}
Despite its successes, the Standard Model has a number of faults and weaknesses that have left theoretical particle physicists searching for a deeper and more fundamental theory of nature, and experimental particle physicists building bigger and more powerful particle accelerators. The most recent and ambitious of these is the Large Hadron Collider or LHC at CERN, from which results are expected in the next year or two.\footnote{It has been pointed out by several people at this stage that this statement, true at the time of writing, is a time-independent one. We remain optimistic.}It is hoped that the LHC's results for collisions with a centre of mass energy of 14 TeV will shed light on some of the mysteries that remain unexplained by the SM. These discrepancies can be roughly divided into two categories: experimental and theoretical.
\vskip 0.5 cm
\BUTitle{Experimental Discrepancies}
\vskip 0.5 cm
\begin{itemize}
\item An essential part of the SM is electroweak symmetry breaking (EWSB), by which electroweak gauge symmetry (the $SU(2) \times U(1)$ part of the SM gauge group) is spontaneously broken to the $U(1)$ of electromagnetism. In the SM, this is believed to proceed via the Higgs mechanism \cite{Higgs:1964ia, Higgs:1964pj, Englert:1964et, Guralnik:1964eu}, which produces a neutral scalar known as the Higgs boson.
This is the only fundamental particle predicted by our current model of particle physics which has not yet been discovered; the exact dynamics responsible for electroweak symmetry breaking are thus still unknown. It is possible that the Higgs mechanism should be extended.\footnote{In some models, like the MSSM (Minimal Supersymmetric Standard Model), there are two complex Higgs doublets (instead of one), leading to 5 physical Higgs bosons after EWSB. The light neutral Higgs boson will be difficult to distinguish from the SM Higgs, but detection of the others would be a signature of the MSSM and therefore of physics beyond the Standard Model \cite{Aad:2009wy}.}The mass of the Higgs is not predicted by the SM, but it has an upper bound of around 1.4 TeV dictated by demanding unitarity in the Standard Model \cite{Lee:1977eg, Lee:1977yc}. If one assumes the Standard Model, a global fit to all existing EW data leads to the limit $m_H < 144 $ GeV (with 95 \% confidence level); taking into account the lower bound of 114.4GeV found at LEP \cite{Barate:2003sz} with the same confidence gives a higher bound of 182 GeV \cite{Alcaraz:2007ri}.\footnote{See for instance \cite{Aad:2009wy} for limits given different assumptions, and the corresponding references.}Given all this, it is widely expected that the Higgs will be discovered at the LHC. There is a high discovery potential for Higgs bosons in both the SM and the MSSM over the full parameter range \cite{Aad:2009wy}. The discovery may lead to modification of the Standard Model. 
 \item There is by now established evidence that neutrino masses are non-zero \cite{GonzalezGarcia:2008ru}. This is the only way to explain flavour-changing in neutrinos, and implies a lepton mass mixing matrix. However, the SM predicts neutrino masses to be exactly zero. The correct modification of the Standard Model that can account for non-zero neutrino masses is not yet known. The matter is further complicated by the fact that neutrino masses are at least 6 orders of magnitude smaller than the electron mass, with an unexplained gap between them (unlike in the spectrum of charged fermions) and that the lepton mixing matrix is qualitatively unlike the quark mixing matrix. A better understanding of the physics responsible for electroweak symmetry breaking is crucial, so that results from the LHC may point us in the right direction. A thorough summary of the theoretical implications is given in \cite{Mohapatra:2005wg}. See \cite{deGouvea:2009gx} for an accessible recent review of the evidence, its implications for theory, and a summary of other relevant experimental searches.
\end{itemize}
\vskip 0.5 cm
\BUTitle{Theoretical Discrepancies}
\vskip 0.5 cm
\begin{itemize}
\item The 20 or so ``free" parameters of the Standard Model (masses and couplings that are experimental inputs to the theory)  make it vulnerable to accusations of arbitrariness, especially when initially compared to the dynamically determined masses and couplings arising from string theory. However, one should note that many parameters need to be tuned to give a particular solution of string theory, of which there are a very large number and no way of uniquely selecting one that corresponds to our universe, as is discussed in Section \ref{landscape} below. Still, the aesthetic desire to reduce the apparent arbitrariness of the Standard Model was historically part of the motivation to seek a more fundamental and dynamically determined theory (see for instance \cite{Polchinski:1998rq}), so we include this argument here for completeness. 
\item As well as being ``free" in the sense that they are not predicted by the theory, some of the parameters in the Standard Model are unnaturally small. For instance, it is not known why the Higgs boson mass should be so much smaller than the Planck scale (or, in other words, why the weak force is so much stronger than gravity). This is known as the \em hierarchy problem.\em 
\end{itemize}

There exist many possible attempts to modify the Standard Model in such a way as to explain neutrino oscillations and the hierarchy problem; these include extra dimensions and supersymmetry (SUSY). However, the possibility that the Standard Model is only applicable below a certain energy scale and that a more fundamental UV-complete theory might be found that would tie up these loose ends is a very attractive one, which deserves investigation. This impulse is fuelled by the many unifications in theoretical physics achieved in the last hundred years. The weak and electromagnetic interactions were most recently unified in the electroweak theory: electroweak symmetry breaking can give rise to the the $SU(2) \times U(1)$ piece of the SM gauge group. Is there some larger symmetry group that includes all of the SM gauge group factors? Can gravity be included with the other three fundamental interactions in a unified framework?

The strongest signal that such an underlying theory is needed is the apparent incompatibility of quantum field theory with general relativity, the theory of the gravitational interaction. General relativity is, like electromagnetism, a classical field theory, but quantising this theory fails because the resulting theory is nonrenormalisable. This can easily be seen by noting that Newton's constant $G_N$ in the Einstein-Hilbert action
\begin{eqnarray}
S& = & \frac{1}{16 \pi G_N} \int d^4 x \sqrt{-g} R
\end{eqnarray}
has mass dimension $-2$. $G_N = \frac{1}{M_{Pl}^2}$, where $M_{Pl}$ is the Planck mass: $M_{Pl} = 10^{19}$ GeV. 

Any scattering amplitude (between two particles interacting gravitationally) will therefore have a factor of $E^2$ for each factor of the coupling constant, in order to make it dimensionless. The corrections at each order increase with energy, and at $E > M_{Pl}$, perturbation theory breaks down. For a given number of gravitons, the sum of corrections is divergent in the UV. Furthermore, this problem grows worse at each order in perturbation theory.  This signals the need for a different, UV complete, description of both theories, which reduces to general relativity and quantum mechanics at their respective limits, and resolves the existing difficulties when they coincide. The search for such a theory has become the defining question of theoretical particle physics, and many believe that string theory is the best, and possibly only, candidate.
\subsubsection{How does string theory help?}
\begin{figure}[htp]
\centering
\includegraphics[scale=0.4]{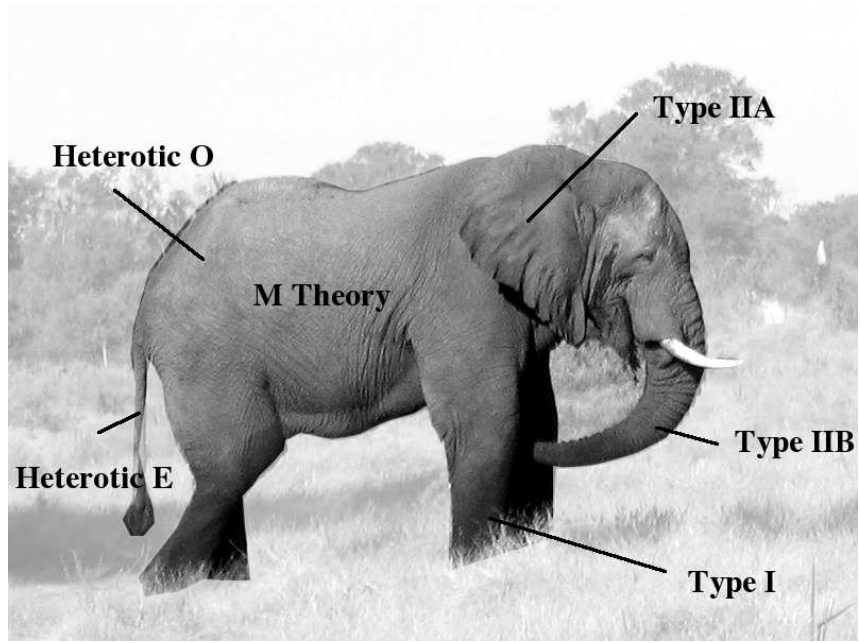}
\caption[A web of string dualities]{The parable of the elephant which appears different to different blind (or ant-sized) observers, seems well-suited to the web of dualities linking what appeared to be distinct string theories but are actually particular limits of M-theory.}
\label{elephant}
\end{figure}

String theory was first studied in the late 1960s as a model of quark confinement. The spectrum of excitations of a vibrating one dimensional string was matched to the spectrum of hadrons by Veneziano and others. However, problems with the model, and the confirmation of QCD as the correct theory of the strong interaction, relegated string theory to the sidelines for almost two decades. One of its drawbacks was the unavoidable prediction of a massless spin-2 excitation as one of the vibrational modes of the strings. In 1974 it was pointed out that these behave like gravitons \cite{Scherk:1974ca}, meaning that string theory naturally includes gravity.
It was not until the first superstring revolution in 1985 that an anomaly-free supersymmetric model of string theory in 10 dimensions was given by Green and Schwarz \cite{Green:1984sg}, and string theory became an active field of research.  Since then, five distinct string theories have been written down, and shown to be related to each other by a web of dualities. Furthermore they are all understood to be low-energy limits of the same 11-dimensional M-theory - see Figure \ref{elephant}.\footnote{The parable illustrated in this figure was introduced to the English-speaking world by John Godfrey Saxe in his poem \em The Blindmen and the Elephant\em. Brian Greene connects the parable to the web of string dualities in his popular book \em The Elegant Universe \em  \cite{Greene:1999kj}.} The field content and basic properties of these theories are described briefly in Section \ref{basics}.

\begin{figure}[htp]
\centering
\subfigure[Two graviton exchange between propagating particles]{
\includegraphics[scale=0.7]{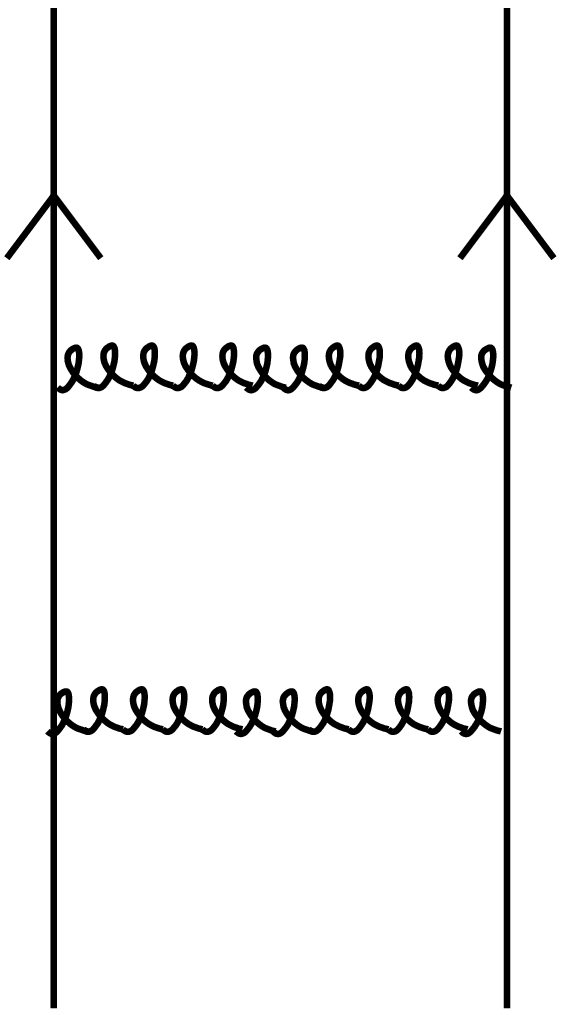}
\label{fig:grav}}
\subfigure[The same process in string theory]{
\includegraphics[scale=0.5]{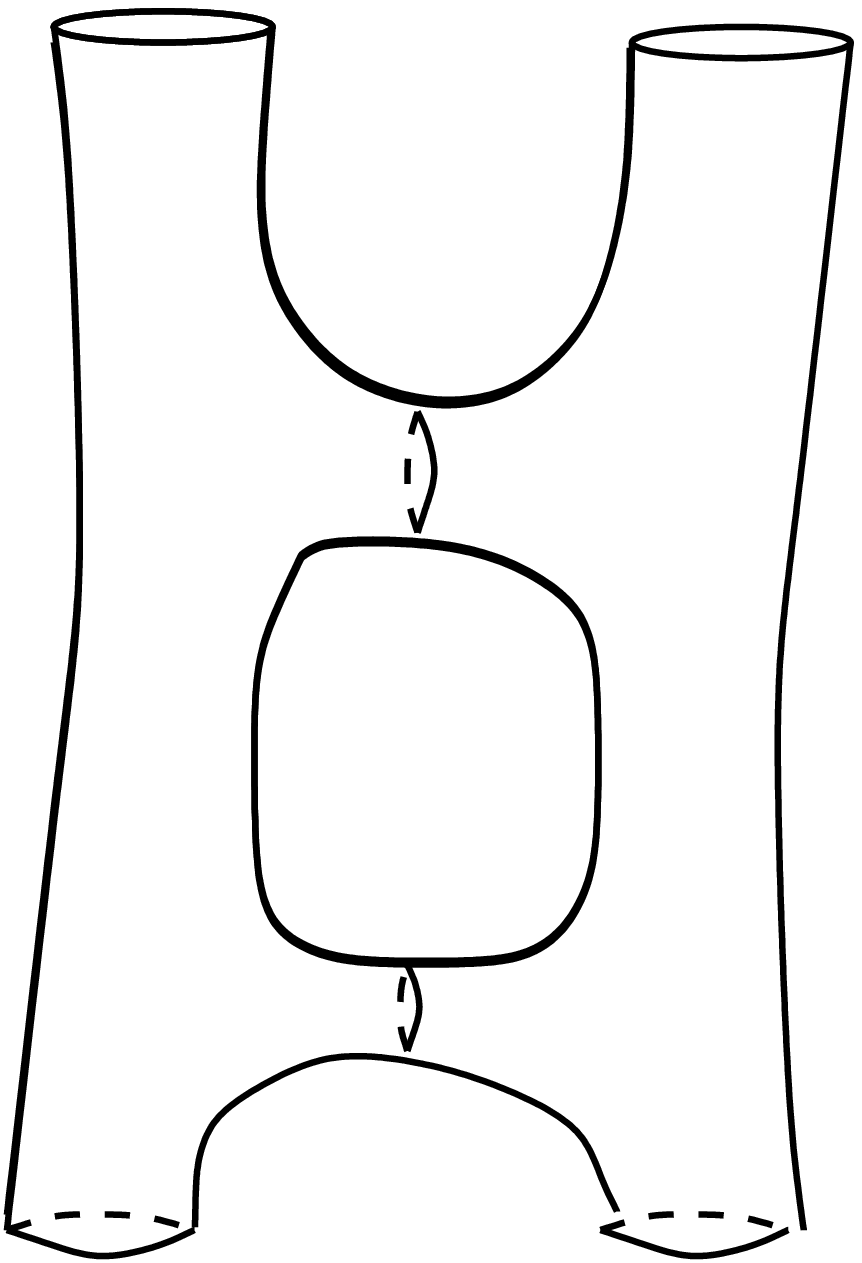}\label{fig:string}}
\caption[String interactions]{Smearing out of interactions by strings. Adapted from \cite{Polchinski:1998rq}.}
\label{smearing}
\end{figure}

String theory solves the UV-divergence problems of quantum gravity by avoiding the short distances at which divergences arise. Because strings are extended objects, they cannot interact at a point vertex like the pointlike particles in quantum field theory. Instead, the Feynman diagrams representing string interactions involve tubes or worldsheets joining and crossing seamlessly, as shown in Figure \ref{smearing}. All known particles and fields result from different modes of excitation of these fundamental strings, so that the loss of this pointlike vertex is seen to be equivalent to the realisation that our existing theories, renormalisable or otherwise, should not be extrapolated to arbitrarily high energies. In Figure \ref{fig:grav} two particles interact gravitationally via exchange of two gravitons in a correction which we have seen grows larger with increasing energy, while in Figure \ref{fig:string} the same process is calculated for strings. In this case there is a UV cutoff: the interaction point is smeared out for the case of the string worldsheets.

The consequences of taking an object extended in one dimension as the fundamental unit of matter are dramatic. Not only is gravity automatically included, it is included in a consistent quantum framework, as we have just seen. In one sense string theory also avoids the charge of arbitrariness that was previously levelled at the Standard Model, since all the parameters of a given solution of string theory are dynamically determined (rather than measured experimentally and then substituted into a Lagrangian). However, a new problem of uniqueness soon emerges, since there is no known selection mechanism guaranteeing or even favouring a vacuum (a ground state corresponding to a certain set of parameters) that gives rise to physics resembling that of our world. In some sense experimental input is still required, if only to choose the solution we live in. This is discussed further in Section \ref{vacuum} from the point of view of using cosmology as evidence in the vacuum selection process.
Remarkably, string theory also contains several of the elements of previously suggested modifications to the Standard Model intended to iron out the discrepancies mentioned earlier.  Supersymmetry, previously conjectured as a way to set the cosmological constant to zero, is demanded by consistency of the theory; bosonic string theory is unstable to tachyon decay.\footnote{Simply put, supersymmetry is a symmetry whose transformations mix fermions and bosons, and which can be understood as an extension of the  Poincar\'{e} group to include spinor generators.\cite{Sohnius:1985qm, Wess:1992cp} and \cite{Bailin:1994qt} are useful references. It should be noted that although superstring theory, which has superseded bosonic string theory and is usually referred to simply as string theory, is formulated in a supersymmetric formalism, stable solutions which break supersymmetry can be constructed within it. This is of course desirable, since SUSY is broken in our world.}Further, to avoid ghosts, strings must live in ten dimensions. From the point of view of any four-dimensional effective theory, there are extra dimensions - hinted at in the context of unification by Kaluza Klein theory and invoked more recently to solve the hierarchy problem \cite{Randall:1999ee}. Furthermore, realistic physics can and has been obtained in string theoretic constructions \cite{Braun:2005nv, Blumenhagen:2006ux, Lebedev:2006kn, Lebedev:2007hv}. 
\subsection{String theory basics}
In the original formulation of string theory,\footnote{The canonical textbooks are Green, Schwarz and Witten \cite{Green:1987sp, Green:1987mn}, Polchinski \cite{Polchinski:1998rq, Polchinski:1998rr} and more recently Becker, Becker and Schwarz \cite{Becker:2007zj}.}five distinct consistent theories were known, called Type I, Type IIA, Type IIB, Heterotic SO(32) and Heterotic $E_8 \times E_8$. Each requires spacetime superymmetry in $9+1$ dimensions to be consistent, and has a specific spectrum of massless bosonic and fermionic fields. 

Type IIA and Type IIB have in common the massless NS-NS spectrum consisting of the (symmetric) metric tensor $g_{\mu \nu}$, the dilaton $\phi$ (a scalar) and $B_2^{NS-NS}$, an antisymmetric tensor. The corresponding field strength is usually denoted $H_3 = dB_2$.\footnote{In Type I and heterotic string theory, anomaly terms are added on the right-hand side.}The fermionic sector of these theories is made up of two spin $\frac{3}{2}$ and two spin $\frac{1}{2}$ states. The RR-sector of IIA contains $p$-forms with $p$ odd, denoted $A_\mu$ and $C_{\mu \nu \rho}$, while the RR sector of IIB contains $p$-forms with $p$ even, i.e. $C_0$ the axion, $C_{\mu \nu}$ and $C _{\mu \nu \rho \sigma}$. This means Dp-branes with $p$ even are stable in IIA and Dp-branes with $p$ odd are allowed in IIB; a $C_{p+1}$ form couples to the ($(p+1)$-dimensional) worldvolume of a Dp-brane in the same way that a photon couples to the worldline of an electrically charged point particle:
\begin{eqnarray*}
& \int_{\Sigma^{p+1}} C_{p+1}.&
\end{eqnarray*}
In general the field strength of $C_p$ is denoted $F_{p+1} = d C_p$. 

During the so-called second superstring revolution in the mid-nineties, it was realised that the five theories listed above are all connected by string dualities. Type IIA and Type IIB are T-dual to one another, while Type I can be obtained by orientifolding IIB. This projects out states that are odd under reversal of worldsheet orientation to give rise to a theory of unoriented strings. The spectrum of Type I consists of $g_{\mu \nu}, \phi, A_\mu$ and $C_{\mu \nu}$. It is related to the heterotic spectrum by S-duality, which interchanges $B_{\mu \nu}$ and $C_{\mu \nu}$. Spacetime-filling D9-branes are possible in Type I and NS-branes in the heterotic theory. Because of the web of dualities connecting the theories, a problem in one theory might look very different to the dual problem in another theory and yet be equivalent. The theories are different descriptions of the same thing, and just as the elephant in Figure \ref{elephant} is not just the ear or the tail that the blind man feels, these theories are actually understood to be different (low-energy) limits of a higher dimensional theory called M-theory, which is given by taking the strong coupling limit of Type IIA. 

In this thesis we work in different regimes of this web of dualities depending on the problem. In Chapter \ref{chapter:axion} we study heterotic compactifications, while in Chapter \ref{lumps} our focus is on Type IIB. In Chapter \ref{magneto} our construction is in the heterotic M-theory setup, described there, in which one can descend directly to heterotic string theory from M-theory. 

\label{basics}
\section{Intersection of string theory with cosmology}
We know that the universe today is expanding and cooling. Extrapolating the FRW metric backwards, we find an initial singularity, dubbed the Big Bang. The universe immediately after the Big Bang was extremely hot and dense and, if it indeed holds at the highest energy scales, the physics governing it must be string theory. What this implies for the universe's evolution since then is not yet fully known, but this early stringy regime may prove extremely important in our understanding of cosmology and, better yet, may furnish opportunities for indirect observation of stringy physics in the form of distinctive signatures of this regime. The universe and its evolution thus represent the biggest once-off particle physics experiment possible - one whose energies we can never harness on earth. Their study, cosmology, is therefore rich with data and implications for particle physics.

 If string theory is the correct theory in the earliest universe, then we should expect all the known results from cosmology to be embedded in a consistent string theory description of our universe, called a vacuum or string theory solution. Since string theory admits a huge number of solutions and there is no way to uniquely select the one corresponding to our world, this scenario suggests some very interesting questions: 
\begin{enumerate}
\item What can we learn about string theory from observation (cosmology)? Is the choice of string theory vacuum constrained by the value or evolution of the cosmological constant, or by the type of inflation undergone in the early universe?
\item Can cosmology give evidence of string theory?  Does string theory, if we assume it to be the correct description of nature at the earliest times, give rise to specifically stringy signatures that might still be observed with future astronomical observations?
\item Can a stringy description of the early universe provide a complete and elegant explanation of cosmological phenomena?
\end{enumerate}
Considering these, we see that investigating the intersection of string theory with cosmology could potentially teach us about both. Not only should string theory be necessary for a complete understanding of the very early universe, but the universe's evolution might be our best shot for obtaining direct or indirect evidence for string theory, reaching scales that cannot be probed by our terrestrial particle accelerators. Both new astrophysical and new particle physics data are expected very soon, and may allow us to make inroads on these questions.
 In Section \ref{vacuum} we discuss the first, and in Section \ref{infl} we discuss inflation in the context of the second two questions. Possible signatures from cosmic strings are discussed in Chapter \ref{chapter:CS}. The three research projects presented in the rest of the thesis touch on all three questions with varying emphasis.

\subsection{Choosing a string theory vacuum: cosmological inputs to string theory}
\label{vacuum}
\subsubsection{String compactifications}
Thus far, we have only been able to detect 4 dimensions - 3 spatial and 1 temporal. Mathematically consistent superstring theory is a 10-dimensional string theory, which means that the strings and other objects in it have a ten-dimensional spacetime in which to live and interact. In order for string theory to give rise to the physics describing our world, its extra 6 dimensions must be curled up somehow, forming what is called an internal or compactification manifold. Which kind of manifold is allowed is tightly constrained, so we proceed carefully. The physics of any given $ 3+ 1$ dimensional theory will be dependent on the internal six-dimensional manifold. Because of this, one can restrict the allowed internal manifold by making demands on the nature of the physics experienced by a 4-dimensional observer. 

Historically, string theorists began by considering compactifications of heterotic string theory on Calabi-Yau manifolds (see Chapter \ref{throat} for more details). Calabi-Yaus arise when one demands unbroken ${\cal N} = 1$ SUSY (i.e. minimal supersymmetry) remain in the four-dimensional theory after compactification on a background without fluxes and with constant dilaton.
These manifolds are characterised by a finite number of parameters which determine the Kahler form and complex structure of the Calabi-Yau. 
The parameter space in which these can take values is called the moduli space for the compactification. If the parameters are not stabilised by some potential, they appear as massless scalar fields in 4 dimensions - massless scalar fields we do not observe. There are thousands of possible 6-dimensional Calabi-Yau manifolds, each of which has size and shape moduli which make up an infinite moduli space. Thus the apparently determined nature of string theory and its answer to the arbitrariness of the Standard Model's 20-odd free parameters is replaced by a much larger number of free parameters. 

 Many of the moduli can be fixed by turning on fluxes, as discussed in Chapter \ref{throat}, but we are left with the problem of selecting which fluxes are turned on. Other moduli arise when branes are included, corresponding to their positions and orientations. Each set of tuned moduli corresponds to a particular background and compactification manifold and gives rise to different 4-dimensional physics in the compactified theory, determining the type and number of fields and their interactions. The set of these vacua or solutions is called the string landscape; selection mechanisms for finding our vacuum in it are discussed below.  
\subsubsection{The landscape}
\label{landscape}
Although string theory appears to be a fundamental and unique theory in the sense that no better candidate for a theory of quantum gravity is known, string theory is better thought of as a framework, like QFT, of which many solutions are possible. The solutions do not only correspond to the number of Calabi-Yaus or classes of internal manifolds which are possible, but are also parametrised by the fluxes of the background and the many scalar fields that typically arise upon dimensional reduction of the ten-dimensional theory. These give rise to a multi-dimensional moduli space whose peaks and valleys are given by the combination of potentials of the different scalars. The altitude of the moduli space corresponds to the potential which in a valley gives the cosmological constant of that vacuum. This moduli space parametrises a huge number of possible vacua or low-energy solutions of string theory. No top-down method to select our vacuum exists, and estimates of the number of possible habitable vacua are huge ($\sim 10^{500}$ \cite{Douglas:2003um}).

A couple of options for proceeding present themselves. 
\begin{enumerate}
\item We could set about exploring different solutions systematically, via classes of compactification manifolds or allowed fluxes. Without top-down guiding principles, this is akin to the proverbial search for the needle in a very large haystack.
\item A bottom-up approach is to try to engineer a string theory solution that reproduces the physics of our world. This has met with some success in the sense that it is certainly possible to arrive at the Standard Model or something close to it using specific brane configurations in string theory \cite{Braun:2005nv, Blumenhagen:2006ux, Lebedev:2006kn, Lebedev:2007hv}. However, without a selection principle, there is no guarantee of uniqueness, and no physically compelling reason to choose one such configuration over another. 
\end{enumerate}

Neither of these approaches is especially satisfying, nor are they ever likely to answer definitively which vacuum we are in or why. A conceptually different approach to navigating the landscape of possible vacua\footnote{The landscape was so dubbed by Susskind \cite{Susskind:2003kw} after an example in biology. It was strictly defined as the set of all string theory vacua with nonzero vacuum energy (in accordance with observation) rather than the flat plain of  ``supermoduli space" where the cosmological constant $\Lambda$ is zero, although this distinction is arguably unnecessary in most usage.}is essentially statistical. The anthropic principle, first put forward by Weinberg \cite{Weinberg:1987dv, Weinberg:1996xe}, asserts that an upper bound for the cosmological constant can be found using anthropic arguments. The argument is that in order for any sort of life to form in an initially homogeneous and isotropic universe, it is necessary for sufficiently large gravitationally bound systems to form. Although this sounds imprecise, the large number of vacua provided by string theory means one can take a statistical approach to finding a selection mechanism for picking out the vacuum corresponding to the physics of our world, using the cosmological constant and other observables, as suggested by Susskind \cite{Susskind:2003kw}. His argument rested on evidence that metastable $\Lambda \geq 0$ vacua are possible in string theory \cite{Kachru:2003aw} and that the number of such vacua is indeed large \cite{Bousso:2000xa}. Recent work on the landscape includes \cite{Denef:2004cf, Denef:2004ze}. This is an unexpected field of research in string theory, which overlaps with string cosmology questions, but is not discussed further in this thesis.
\subsection{Inflation}
\label{infl}
An important topic at the intersection of string theory and cosmology, which this thesis does not address in any detail, is inflation. Inflation \cite{Guth:1980zm, Linde:1981mu, Albrecht:1982wi} explains the large-scale homogeneity, isotropy and flatness of the universe as observed today. It also provides a model for structure formation (via fluctuations of the inflaton field) whose predictions \cite{Mukhanov:1981xt, Mukhanov:1985rz, Hawking:1982cz, Starobinsky:1982ee, Guth:1982ec, Bardeen:1983qw} for the nature of the inhomogeneities in the cosmic microwave background (CMB) are in impressive agreement with experiment \cite{Komatsu:2008hk}.\footnote{This data ruled out cosmic strings as the primary generators of primordial perturbations leading to structure formation - see Section \ref{struct}.} 

However, there is no real explanation for why inflation occurred. Why should the universe have undergone a period of exponential expansion some $10^{-36}$ seconds after the Big Bang? Most early models simply assumed the existence of a suitable low energy effective field theory (EFT) and examined different potentials for the inflaton. The resulting primordial perturbation spectra are extremely sensitive to the details of this potential.

Not surprisingly, inflation can also depend very sensitively on Planck-scale physics, and so should be studied in a UV-complete theory such as string theory. \cite{Baumann:2009ni} reviews the reasons for this and the current status of the most promising string models of inflation.  Inflation can be realised in string theory, but does not appear to be natural, either from the point of view of the EFT Lagrangian or from the point of view of initial conditions. The required flatness of the inflaton potential is a nontrivial condition because of quantum corrections to it, while large-field inflationary models (which could be distinguished observationally by their large gravitational wave signals) are especially sensitive to UV corrections, and are therefore difficult to construct.  Models are constrained theoretically, by consistency conditions, but also by the data \cite{Shandera:2006ax, Bean:2007hc, Bean:2007eh}. 

Still tighter observational constraints are expected from Planck and other experiments in the new future (see \cite{Baumann:2009ni} and references therein). Deviations from scale-invariance, gaussianity and adiabaticity of the CMB could rule out single-field slow roll inflation, while detection of B-mode polarisation would tightly constrain large-field inflation models. Thus string theoretical models of inflation could potentially constrain the string landscape of vacua, or even give rise to observational tests of string theory.

\newchapter{The Throat: Warped Compactifications}
\label{throat}
Realistic compactifications in which most of the moduli describing a compactification can be stabilised by fluxes, and which can give rise to large hierarchies, are called warped or flux compactifications. In the rest of the thesis, and specifically in Chapters \ref{chapter:axion} and \ref{lumps}, we will refer to a warped region or throat in the compactification geometry, which is described locally by a Klebanov-Strassler throat. Here we review the Klebanov-Strassler set-up and the theory of flux compactifications. Note that the Klebanov-Strassler solution is non-compact, but that a compact version of a warped compactification with fluxes is known to exist and was given by Giddings, Kachru and Polchinski \cite{Giddings:2001yu}. Future references to the throat will be locally valid statements, where it is assumed that the throat is glued onto some compact bulk where the global constraints discussed below are satisfied.

\section{Calabi-Yau compactifications}

As we have discussed, string theory is formulated in 10 spacetime dimensions, so that 4-dimensional physics can only be obtained by compactifiying the six extra dimensions as an internal manifold. The first class of internal manifolds to be studied was that of Calabi-Yaus. Calabi-Yaus arise when one demands unbroken ${\cal N} = 1$ SUSY (i.e. minimal supersymmetry) remain in the four-dimensional theory after compactification on a background without fluxes and with constant dilaton.
This requirement is convenient (a supersymmetric solution in 4 dimensions also satisfies the equations of motion \cite{Green:1987mn}) and phenomenologically promising (${\cal N} \geq 2$ does not allow chiral fermions). Demanding ${\cal N} = 1$ SUSY is equivalent to demanding that the SUSY transformations of all fermions vanish. This condition reduces to the requirement that a covariantly constant spinor be defined on the internal manifold. For a 6-dimensional Calabi-Yau 3-fold, this is true for the case of $SU(3)$ holonomy,\footnote{The holonomy group of a manifold is the group of all transformations undergone by a field upon being parallel transported along a closed curve.}giving a Kahler manifold\footnote{Kahler manifolds are complex manifolds with closed Kahler form $dJ = 0$, where $J = \imath G_{i \bar j} dz^i d\bar z^j$ is the Kahler form. A Calabi-Yau additionally has an exact Ricci form ${\cal R} = R_{i \bar j} dz^i d\bar z^j$. A good reference for complex manifolds and their properties is \cite{nak}.}with vanishing first Chern class, or a Calabi-Yau manifold. 
These manifolds are characterised by a finite number of parameters which determine the Kahler form and complex structure of the Calabi-Yau.  There are thousands of possible 6-dimensional Calabi-Yau manifolds, each of which has an infinite moduli space given by the K\"ahler and complex structure moduli. The problem of fixing these moduli is ameliorated by compactifications with fluxes switched on, as discussed in Section \ref{flux}. 

\section{Warped compactifications}
\subsection{Phenomenology of warped compactifications}
Warped compactifications 
\begin{eqnarray}
\label{warpedmetric}
ds^2 & = & e^{2 A(y)} \eta_{\mu \nu} dx^{\mu} dx^{\nu} + e^{-2 A(y)} \tilde g_{mn} dy^m dy^n,
\end{eqnarray}
where $\mu, \nu$ run over the 4 non-compact dimensions and $m,n$ label the internal manifold, are consistent with 4-dimensional Poincar\'{e} symmetry. $e^{2 A(y)}$ is called the warp factor; it gives the normalisation of the 4-dimensional metric and can vary in the transverse dimensions. As shown in the models of Randall-Sundrum \cite{Randall:1999vf, Randall:1999ee} and Verlinde \cite{Verlinde:1999fy}, warped compactifications naturally give rise to hierarchies in four dimensions. 

However, these hierarchies are functions of the unfixed moduli which parametrise the compactification. In string theory constructions, these will be tied to fluxes. The RS models are five dimensional, i.e. they have only one extra dimension. In a 10-dimensional string theory context, warping can arise in the presence of branes, as in the AdS/CFT correspondence \cite{Maldacena:1997re}. The original formulation relates a string theory construction consisting of a stack of D3-branes to a conformally invariant gauge theory with maximal (${\cal N} = 4$) supersymmetry. In our world, both supersymmetry and scale invariance are broken somehow. By placing the stack of D3-branes on a conifold, it is possible to break most of the SUSY in the dual field theory \cite{Klebanov:1998hh}. In the Klebanov-Strassler model \cite{Klebanov:2000hb}, conformal symmetry in the dual field theory is broken when fractional D3-branes are included in the construction. The gauge theory then exhibits confinement and chiral symmetry breaking in the far IR. The correct dual description of this far IR region is a deformed conifold with fluxes, given below. 

\subsection{The Klebanov-Strassler throat}
In the Klebanov-Strassler model, a stack of $N$ D3-branes is placed at the tip of a conifold, while $M$ D-branes wrap the $S^2$ of this conifold. The conifold\footnote{See the appendix of \cite{Gwyn:2007qf} for a detailed review.}is a Calabi-Yau threefold composed of a cone over a 5-dimensional base $T^{1,1}$, whose metric is given by \cite{Candelas:1989js}
\begin{eqnarray}
\label{T11}
d\Sigma_{T^{1,1}}^2 & = & \frac{1}{9} \left ( d \psi + \sum_{i=1}^2 \cos \theta_i d \phi_i \right ) ^2 + \frac{1}{6} \sum_{i = 1}^2 \left ( d \theta_i^2 + \sin^2 \theta_i d \phi_1^2 \right ). 
\end{eqnarray}
$T^{1,1}$ is a coset space $\frac{SU(2) \times SU(2)}{U(1)}$ with topology $S^2 \times S^3$. Its metric has isometry group $SU(2)\times SU(2) \times U(1)$. In (\ref{T11}), $\psi = \psi_1 + \psi_2$ where $(\psi_i, \phi_i, \theta_i)$ are the Euler angles of each $SU(2)$. The singular conifold then has the metric
\begin{eqnarray}
ds^2 & = & dr^2 + r^2 d \Sigma_{T^{1,1}}^2.
\end{eqnarray}
As in the original formulation of the AdS/CFT correspondence \cite{Maldacena:1997re}, there is a duality between the gauge theory living on the branes and the gravity theory, found by taking the near-horizon limit. The near-horizon geometry in this case is $AdS_5 \times T^{1,1}$ and the $SU(N) \times SU(M)$ gauge theory on the branes is non-conformal. In the IR limit the branes have cascaded away, leaving a deformed conifold with fluxes
 \begin{eqnarray}
\frac{1}{4 \pi^2 \alpha'} \int_{S^3} F_3 & = &  M;\\
\frac{1}{4 \pi^2 \alpha'} \int_{B} H_3 & = & - K,
\end{eqnarray}
where B is the 3-cycle dual to $S^3$. $H_3 = dB_2$ where $B_2$ is radially dependent because of the conformal symmetry breaking \cite{Klebanov:2000hb}, while $F_3$ is the magnetic flux due to the fractional D3-branes. The gauge group in the IR (before confinement) is $SU(M)$. In the case of a deformed conifold, the singularity is removed by blowing up the $S^3$ at the tip (when the singularity is removed by blowing up the $S^2$ instead a resolved conifold results). The radius of the finite $S^3$ at the tip is given by the deformation parameter $\epsilon$, which modifies the conifold equation:
\begin{eqnarray*}
\sum_{i = 1}^4 z_i^2 \, = \, 0 & \Rightarrow& \sum_{i = 1}^4 z_i^2 \, = \, \epsilon^2.
\end{eqnarray*}

The metric of the deformed conifold was studied in \cite{Candelas:1989js, Minasian:1999tt, Ohta:1999we, Herzog:2001xk}. The ten-dimensional metric is given by \cite{Herzog:2001xk}
\begin{eqnarray}
ds_{10}^2 & = & H^{-\frac{1}{2}} (\tau) \eta_{\mu \nu} dx^\mu dx^\nu + H^{\frac{1}{2}} (\tau) ds_6^2
\end{eqnarray}
where $\tau$ is the radial co-ordinate and $ds_6^2$ is the conifold metric. This is the usual form for warping due to a stack of D3-branes with harmonic function $H(\tau)$. In the case of the Klebanov-Strassler solution, as mentioned above, there are in addition $M$ wrapped D5-branes, which appear as fractional D3-branes. In the IR limit the D3-branes have cascaded away and the fractional branes must be replaced by fluxes
which are responsible for the deformation of the conifold. This can be seen in $ds_6$ and $H(\tau)$:
\begin{eqnarray}
ds_6^2 & = & \frac{1}{2} \epsilon^{\frac{4}{3}} K (\tau) \left [\frac{1}{3 K^3 (\tau) }( d \tau^2  + (g^5)^2) + \cosh^2 ( \frac{\tau}{2} )\left ( (g^3)^2 + (g_4)^2\right )\right . \\
&&\left. + \sinh^2 ( \frac{\tau}{2})((g^1)^2 + (g^2)^2) \right ]; \\
K(\tau)& = & \frac{(\sinh(2 \tau) - 2 \tau)^{\frac{1}{3}}}{2^{\frac{1}{3}} \sinh \tau};\\
H(\tau) & = & (g_s M \alpha')^2 2^{\frac{2}{3}} \epsilon^{- \frac{8}{3}} I ( \tau);
\end{eqnarray}
where $ds_6^2$ has been written in the basis \cite{Klebanov:2000hb}
\begin{eqnarray*}
g^1 & = & \frac{e^1 - e^3}{\sqrt{2}}\\
g^2 & = & \frac{e^2 - e^4}{\sqrt{2}}\\
g^3 & = & \frac{e^1 + e^3}{\sqrt{2}}\\
g^4 & = & \frac{e^2 + e^4}{\sqrt{2}}\\
g^5 & = & e^5.
\end{eqnarray*}
with vielbeins\footnote{Note that these vielbeins will not give a closed holomorphic 3-form on the deformed conifold, as pointed out in \cite{Gwyn:2007qf}. Thus if one uses the standard complex structure, these vielbeins do not display the CY property of the manifold.}
\begin{eqnarray*}
e^1 & = & - \sin \theta_1 d \phi_1\\
e^2 & = & d \theta_1\\
e^3 & = & \cos \psi \sin \theta_2 d \phi_2 - \sin \psi d \theta_2\\
e^4 & = & \sin \psi \sin \theta_2 d \phi_2 + \cos \psi d \theta_2\\
e^5 & = & d \psi + \cos \theta_1 d \phi_1 + \cos \theta_2 d \phi_2
\end{eqnarray*}
and
\begin{eqnarray}
I(\tau) & = & \int_{\tau}^\infty dx \frac{x \coth x - 1 }{\sinh^2 x} \left (\sinh (2x) - 2x \right) ^{\frac{1}{3}}.
\end{eqnarray}
$\theta_i, \phi_i$ and $\psi$ are the co-ordinates on the base of the singular conifold, $T^{1,1}$, which is topologically equivalent to $S^2 \times S^3$; the two $S^2$s are parametrised by $(\theta_i, \phi_i)$.
At the tip of the conifold, $ \tau \rightarrow 0$ and $ds_6^2$ degenerates to 
 \begin{eqnarray}
 ds_3^2 & = & \frac{1}{2} \epsilon^{\frac{4}{3}} \left ( \frac{2}{3} \right )^{\frac{1}{3}} \left [ \frac{1}{2} (g^5)^2 + (g^3)^2 + (g^4)^2 \right ],
 \end{eqnarray}
 which has the topology of a 3-sphere \cite{Minasian:1999tt}. $I(\tau) \rightarrow a_0 \approx 0.72$ when $\tau \rightarrow 0$.

 We can then write the metric in the tip as
 \begin{eqnarray}
ds^2 & = & H^{-\frac{1}{2}}(\tau \rightarrow 0)  \eta_{\mu \nu} dx^\mu dx^\nu + H^{\frac{1}{2} }( \tau \rightarrow 0) \frac{1}{2} \epsilon^{\frac{4}{3}} \left ( \frac{2}{3}\right )^{\frac{1}{2}} d \Omega_3^2, 
 \end{eqnarray}
 where $H(\tau \rightarrow 0) = 2^{\frac{2}{3}}(g_s M \alpha')^2 \epsilon^{- \frac{8}{3}} a_0$. As in \cite{Dasgupta:2007ds}, we can absorb numerical factors in the second term and write
 \begin{eqnarray}
 \label{metric0}
 ds^2 & = &  h^2\, \eta_{\mu \nu} dx^{\mu} dx^{\nu}+
  g_s M \alpha'(d\psi^2 +\sin^2 \psi\, d \Omega_2 ^2),
 \end{eqnarray}
 where \[h = H ( \tau \rightarrow 0)^{-\frac{1}{4}} = \epsilon^{\frac{2}{3}} 2^{-\frac{1}{6}} a_0^{- \frac{1}{4}}(g_s M \alpha')^{-\frac{1}{2}}\] and we have expanded the three-sphere metric in terms of new co-ordinates. $\psi$ is thus the usual polar co-ordinate in an $S^3$, and ranges from $0$ to $\pi$. This is the metric we will use for the throat in Chapter \ref{lumps}.

M gives the number of units of Ramond-Ramond fluxes $F_3$ turned on inside this $S^3$.
The two-form associated with $F_3$ is given by \cite{Firouzjahi:2006vp}:
\begin{eqnarray}
\label{C2}
C_2=  M \alpha' \,  \left(\psi-\frac{\sin (2\psi)}{2} \right) \sin \theta\, d \theta \, d\phi \, .
\end{eqnarray}
\subsection{Flux compactifications}
\label{flux}
Although the Klebanov-Strassler solution was constructed using branes and fractional branes on a conifold background, the final warped deformed conifold after the duality cascade on the field theory side can be understood directly as a flux compactification. Flux compactifications give a natural string embedding of the warped compactifications (and resulting hierarchies) of Randall-Sundrum. However, the KS solution is non-compact and therefore incomplete as a string compactification. A fully compact string compactification with fluxes was found by Giddings, Kachru and Polchinski \cite{Giddings:2001yu}, who showed that the presence of fluxes generates potentials for all (or all but one) of the moduli, stabilising the hierarchy. It should be noted that fixing the moduli corresponds to reducing the supersymmetry and breaking the conformal invariance in the dual gauge theory.

Compactifications in the presence of background fluxes had not been considered initially because of a no-go theorem \cite{deWit:1986xg, Maldacena:2000mw} which can be formulated in Type IIB supergravity \cite{Giddings:2001yu} (see also \cite{Frey:2003tf}): The type IIB metric with localised sources is given by \cite{Polchinski:1998rr}
\begin{eqnarray}
\nonumber S_{IIB} & = & \frac{1}{2 \kappa_{10}^2} \int d^{10}x \sqrt{-g_s} \left [e^{-2 \phi} \left ( R_s + 4 ( \nabla \phi )^2 \right ) - \frac{F_{(1)}^2}{2} - \frac{1}{2.3!} G_{(3)}\cdot \bar G_{(3)} - \frac{\tilde F_{(5)}^2}{4.5!}\right ] \\\label{twob}&&+\, \,  \frac{1}{8 \imath \kappa_{10}^2} \int e^\phi C_{(4)} \wedge G_{(3)} \wedge \bar G_{(3)} + S_{loc},
\end{eqnarray}
Here $g_s$ is the string frame metric, $\phi$ is the dilaton, and $G_{(3)}$ is the linear combination $G_{(3)} = F_{(3)} - \tau H_{(3)}$ where $\tau = C_{(0)} + \imath e^{- \phi}$ is the axion-dilaton. $\tilde F_{(5)}$ is defined as
\begin{eqnarray}
\tilde F_{(5)} & = & F_{(5)} - \frac{1}{2} C_{(2)} \wedge H_{(3)} + \frac{1}{2} B_{(2)} \wedge F_{(3)}.
\end{eqnarray}
As explained in the Introduction, $F_{p+1} = d C_p$ denotes the field strength of the RR $p$-forms, while $H_{(3)} = d B_{(2)}^{NS-NS}$. $\tilde F_{(5)}$ is subject to the self-duality condition $\tilde F_{(5)} = \star \tilde F_{(5)}$ which must be imposed by hand. The action (\ref{twob}) is given in the Einstein frame by
\begin{eqnarray}
\nonumber S_{IIB} & = & \frac{1}{2 \kappa_{10}^2} \int d^{10}x \sqrt{-g} \left [ R - \frac{\partial_M \tau \partial^M \bar \tau}{2 (\Im \, \tau)^2} - \frac{G_{(3)} \cdot \bar G_{(3)}}{12 \Im \, \tau} - \frac{\tilde F_{(5)}^2}{4 \cdot 5!}\right ] \\&&+ \frac{1}{8 \imath \kappa_{10}^2} \int \frac{C_{(4)} \wedge G_{(3)} \wedge \bar G _{(3)}}{\Im \, \tau} + S_{loc}.
\end{eqnarray}
The metric is taken to be of the form (\ref{warpedmetric}). $\tau = \tau(y)$ and $A = A(y)$ can both vary over the compact manifold. Only compact components of $G_{(3)}$ preserve 4-dimensional Poincar\'{e} invariance. In addition a five-form flux (see \cite{Giddings:2001yu})
\begin{eqnarray}
\tilde F_{(5)} & = & (1 + \star) \left [ d \alpha \wedge dx^0 \wedge dx^1 \wedge dx^2 \wedge dx^3\right ]
\end{eqnarray}
is allowed, where $\alpha$ is a function on the compact space. Allowing for some localised sources (such as D-branes), the Einstein equation results in the following constraint:
\begin{eqnarray}
\nonumber \tilde \nabla ^2 e^{4A} & = & e^{2A} \frac{G_{mnp}\, \bar G^{mnp}}{12 \, \Im \,  \tau} + e^{-6A} \left [ \partial_M \alpha \partial^M \alpha + \partial_m e^{4A} \partial^M e^{4A} \right ] \\ \label{nogoeqn} &&+ \frac{\kappa_{10}^2}{2} e^{2A} \left ( T_m^m - T_{\mu}^{\mu}\right )^{loc},
\end{eqnarray}
where a tilde denotes use of the metric on the internal space. Integrating over the compact internal manifold ${\cal M}_6$ on both sides gives zero on the left-hand side, and a positive definite quantity on the right-hand side, unless local sources with $\left ( T_m^m - T_{\mu}^{\mu}\right) $ negative are present. This is the reason that flux compactifications were ruled out in ordinary supergravity: no such objects exist and so this amounts to a no-go theorem for flux compactifications, setting fluxes to zero and the warp factor to be constant. 

However, objects for which   $\left ( T_m^m - T_{\mu}^{\mu}\right ) <0  $ exist in string theory. For a D$p$-brane wrapped on ($p-3$)-cycle $\Sigma$ of ${\cal M}_6$,
\begin{eqnarray}
\label{tension}
\left ( T_m^m - T_{\mu}^{\mu}\right )^{loc} & = & (7-p) T_p \, \delta(\Sigma),
\end{eqnarray}
where $T_p$ is the $p$-brane tension. It is clear that this term is negative for $p >7$ objects, i.e. D9-branes. It will also be negative for objects with negative tension, such as orientifold planes. Both D9-branes and O3-planes are stable in Type IIB.  In addition to satisfying (\ref{tension}), the sources must satisfy the integrated Bianchi identity
\begin{eqnarray*}
d \tilde F _{(5)} & = & H_{(3)} \wedge F_{(3)} + 2 \kappa_{10}^2 T_3 \rho_3^{loc} \\
\Rightarrow0 &= & \frac{1}{2 \kappa_{10}^2 T_3} \int_{{\cal M}_6} H_{(3)} \wedge F_{(3)} + Q_3^{loc},
\end{eqnarray*}
where $p^{loc}$ and $Q^{loc}$ are the D3 charge density and charge from localised sources. 

Thus (\ref{nogoeqn}) does not rule out warped flux compactifications in string theory as long as the required localised objects are present. In addition to giving a natural string realisation of hierarchies from warping, nonzero fluxes enter into the superpotential and stabilise the compactification moduli. 
In the special case that 
\begin{eqnarray}
\frac{1}{4} \left ( T_m^m - T_\mu ^\mu \right )^{loc} \geq T_3 \rho_3^{loc}
\end{eqnarray}
for all localised sources (this condition is satisfied by D3- and D7-branes and O3-planes), the global constraints determine all the moduli except for the radial modulus. While Klebanov-Strassler gave the local structure of a highly warped throat in a non-compact geometry, GKP \cite{Giddings:2001yu} gave a fully consistent compact embedding of such a throat (see also \cite{Dasgupta:1999ss}). This is their central result. For calculations in warped throats in the remainder of the thesis, we will generally use the local KS description. One should nevertheless keep in mind a picture in which this throat is glued onto a compact manifold such that the global constraints above are satisfied. In general more than one such throat will be present. Hierarchies between scales are given by the suppressed interections between the IR modes in different throats.

\newchapter{Axions in string theory}
\label{chapter:axion}
In this chapter I present a string theory realisation of a particle physics mechanism known as the Peccei-Quinn or QCD axion. The Peccei-Quinn axion was introduced as a dymanical explanation for the observed low value of the theta term in QCD \cite{Peccei:1977ur}. Fields which behave like the axion are not hard to find in string theory, but it has proved difficult to constrain them to physically acceptable behaviour \cite{Svrcek:2006yi}, making a string theoretic realisation of the axion a longstanding problem. The axion decay constant $f_a$ is strongly bounded by astrophysical and cosmological observations to a value below $10^{12}$ GeV, while typical string theory axions have decay constants of the order of $10^{16}$ GeV. As in the paper, ``On the Warped Heterotic Axion" with Keshav Dasgupta and Hassan Firouzjahi \cite{Dasgupta:2008hb}, we show that $f_a$ is sensitive to the mass scale of the throat in a warped compactification. This means an axion with allowable decay constant can be produced by appropriate engineering of the geometry. We construct suitable warped heterotic backgrounds and find that the question of obtaining $f_a$ within the allowed bounds is reduced to the question of constructing a throat with warped mass scale in this range. This provides a natural mechanism for realising the axion in heterotic string theory. 

\section{The Peccei-Quinn Axion}
\subsection{The strong CP problem}
In principle, the QCD langrangian can include a CP-violating interaction 
\begin{eqnarray}
\nonumber S_\theta & = & \frac{\theta}{8 \pi^2} \int {\bf tr} F \wedge F\\
\nonumber S_ \theta & = & \frac{\theta}{16 \pi^2} \int {\bf tr} F_{\mu \nu} \tilde F^{\mu \nu}\\
\label{theta} S_{\theta} &=& \frac{\theta}{32 \pi^{2}} \int d^{4} x \, \epsilon^{{\alpha \beta \gamma \lambda} } 
 {\bf tr} \, F_{\alpha \beta} F_{\gamma \lambda}  \, ,
\end{eqnarray}
where $F = \frac{1}{2} F_{\mu \nu} dx^\mu \wedge dx^\nu$, $\tilde F_{\mu \nu} = \frac{1}{2} \epsilon_{\mu \nu \alpha \beta} F^{\alpha \beta}$ and ${\bf tr}$ is a trace in the three-dimensional representation of SU(3). The gauge indices $a$ can be reinstated as follows:
\begin{eqnarray}
S_\theta & = & \frac{\theta}{64 \pi^2} \int d^4 x \epsilon^{\alpha \beta \gamma \lambda} F_{\alpha \beta}^a F_{\gamma \lambda}^a.
\end{eqnarray}
Here we are using the conventions of \cite{Svrcek:2006yi}, where the gauge fields are normalised such that the kinetic term is  $- \frac{1}{2 g^2} \int d^4 x \, {\bf tr} F_{\mu \nu} F^{\mu \nu} = - \frac{1}{4g^2}\int d^4 x F^a_{\mu \nu} F^{\mu \nu \, \, a}$.

 The interaction can also be written as $\theta N$, where $N$ is the winding number (used to label the degree of the mapping of the boundary of space to the space of vacua), necessarily an integer. This is sometimes referred to as an instanton number, since $N = N_1 - N_2$ gives the tunnelling amplitude of a transition from a configuration with winding number $N_1$ at the boundary (spatial infinity) to a configuration with winding number $N_2$ at the boundary. This is a nonperturbative process, with instantons interpolating between the vacuum configurations. It was 't Hooft who first showed that nonperturbative effects could give rise to this symmetry-breaking term \cite{'tHooft:1976up, 'tHooft:1976fv}.

Theta is an angular parameter. This can be seen by considering the change of variable \cite{Peccei:1977hh, Weinberg:1996kr}
\begin{eqnarray}
\label{rotation} \psi &\rightarrow& e^{\imath \gamma_5 \eta} \psi;\\
\nonumber \bar \psi & \rightarrow & \bar \psi e^{- \imath \gamma_5 \eta} ,
\end{eqnarray} 
which results in a change in the  fermion measure
\begin{eqnarray*}
{\cal D} \psi {\cal D} \bar \psi & \rightarrow e^{- [ \frac{\imath}{32 \pi^2} \int d^4 x \eta \epsilon^{\mu \nu \rho \sigma} F_{\mu \nu}^a\tilde  F_{\rho \sigma}^a]}{\cal D} \psi {\cal D} \bar \psi,
\end{eqnarray*}
equivalent to a shift 
\begin{eqnarray*}
\label{thetashift}
\theta &\rightarrow & \theta + 2 \eta.
\end{eqnarray*}
 One might conclude that $|\theta|$ could therefore take any value between $0$ and $2\pi$, but in fact it is subject to strong observational constraints.  Measurements of the neutron dipole moment (which is non-zero only for a non-zero value of $\theta$) give an upper bound for $\theta$:  the most recent analysis of the upper limit of the electric-dipole moment of the neutron\footnote{Obtained by measuring the Larmor frequency with which the neutron spin polarisation precesses about the field direction in an applied electric field. This is given by $h\nu = |2 \mu_n B + 2 d_n E|$ when the electric and magnetic fields are parallel, and $h \nu = |2 \mu_n B - 2 d_n E |$ when the fields are antiparallel, where $\mu_n$ is the magnetic moment and $d_n$ the electric dipole moment. The dipole moment $d_n$ is thus obtained by comparing the Larmor frequency for the two cases (fields parallel and antiparallel). 
 }is \cite{Baker:2006ts}, where the dipole moment limit is given as $|d_n| < 2.9 \times 10^{-26} e$ cm. Using the relation\footnote{This appears to have been given first by Witten in 1980 \cite{CrewtherErratum}, a correction to \cite{Crewther:1979pi}. See (for instance) \cite{Zioutas:2009bw, Srednicki:2007qs} for some discussion of the calculation.} $d_n \approx 3.6 \times 10^{-16} \,\theta\, e$ cm, this implies $|\theta| \approx 8 \times 10^{-11}$, i.e. an upper bound on $|\theta|$ of $10^{-10}$. Explaining this small observational value of the theta term is the strong (interaction) CP problem, so-called because the presence of the theta term  violates T and P invariance since the epsilon tensor involves one time and three space indices. Because CPT is a good symmetry, this means the term is both parity and CP-violating. 

\subsection{The Peccei-Quinn mechanism}
\subsubsection{Historical development}
As detailed in \cite{Svrcek:2006yi}, there are a few possible solutions to the strong CP problem. First, one must note that $\theta$ is not an independent parameter when the fermions (quarks) in the theory are massive. The mass terms in the Lagrangian can be written \cite{Srednicki:2007qs} (the fermion index is suppressed)
\begin{eqnarray}
{\cal L}_m & = & - |m| \bar \psi e^{- \imath \phi \gamma_5} \psi,
\end{eqnarray}
where $m = |m| e^{\imath \phi}$.  Then under the chiral rotation (\ref{rotation}), $\phi \rightarrow \phi + 2 \eta$. Since a change of variable cannot change the path integral, it cannot depend on $\theta$ or $m$ separately, but only on the combination $me^{- \imath \theta}$ or $ \Pi m_f e^{- \imath \theta}$ where $f$ is the fermion index. Thus a first possible solution is given by noting that the theta term would have no effect if any of the quark masses were zero. We now know that this is inconsistent with observation. The condition for P and T conservation is that $\theta = 0$ when the quark fields are rotated such that $m$ is real \cite{Weinberg:1977ma} (if they are not real, the chiral transformation needed to make the quark masses real will result in a nonzero $\theta$ angle).  

The Peccei-Quinn solution relies on postulating that (\ref{rotation}) is a symmetry of the system, called the $U(1)_{PQ}$ symmetry. Peccei and Quinn proposed that the quark masses arise from their coupling to a scalar field $\varphi$ \cite{Peccei:1977hh}, the vacuum expectation value of which is found by minimising their $\theta$-dependent potential. Symmetry under the transformation (\ref{rotation}) is thus broken by instanton effects which give $\varphi$ a VEV and the quark fields a mass. Peccei and Quinn found that the minimum of $V(\varphi)$ relates $\phi$ and $\theta$ above such that making the quark masses real sends the theta term to zero, i.e. $\phi = 0 \Rightarrow \theta = 0$. 

This spontaneous symmetry breaking gives rise to a pseudoscalar Goldstone boson which has zero bare mass, as was pointed out by Weinberg \cite{Weinberg:1977ma} and Wilczek \cite{Wilczek:1977pj}. This particle was named the \em axion\em, probably because, as pointed out by 't Hooft, it is the trace of the axial vector current which is associated with the instanton term:
\begin{eqnarray}
\partial_\mu J_\mu^5 & = & - \imath \frac{n g^2}{8 \pi^2} {\bf tr} F_{\mu \nu}\tilde F^{\mu \nu},
\end{eqnarray}
where 
\begin{eqnarray*}
J_\mu^5 & = & \sum_t  J_\mu ^{5, tt}
\end{eqnarray*}
and the traceless part of the axial vector current is
\begin{eqnarray*}
J_\mu^{5, st} & = & \imath \bar \psi^s \gamma_\mu \gamma_5 \psi^t. 
\end{eqnarray*}
$s$ and $t$ are the fermion indices. 

\subsubsection{The axion}
We can understand the PQ mechanism directly via the inclusion of the axion field $a$ from the start. It couples to $F \wedge F$, with action:
\begin{eqnarray}
\label{axionaction}
S_a = \int d^4 x \left ( \frac{1}{2}f_a^2 \partial_\mu a \partial^\mu a + r \frac{a}{32 \pi^{2}}  \, \epsilon^{{\alpha \beta \gamma \lambda} }
{\bf tr} \,  F_{\alpha \beta} F_{\gamma \lambda}  \right ) ,
\end{eqnarray} 
where $f_a$ has dimensions of mass. In terms of $a$, the PQ symmetry is now a shift symmetry ($a \rightarrow a + constant$) which is 
broken by the $F \wedge F$ term. This is clear since the chiral rotation (\ref{rotation}) implies a shift in theta (\ref{thetashift}) and $\theta$ can be absorbed by $a$. In effect, we are promoting  $\theta$ to a dynamical field so that
\begin{eqnarray}
\nonumber S_\theta & =&  \int d^4 x \, \left ( \frac{1}{2} ff_a^2 \partial_\mu \theta \partial^\mu \theta + \frac{\theta}{32 \pi^2} \epsilon^{\alpha \beta \gamma \lambda} {\bf tr}   F_{\alpha \beta} F_{\gamma \lambda}\right ),\\
\label{axion2} S_{\tilde \theta} & = &  \int d^4 x  \, \left ( \frac{1}{2} \partial_\mu \tilde \theta \partial^\mu \tilde \theta 
+  \frac{\tilde \theta}{32 \pi^2 f_a}   \epsilon^{\alpha \beta \gamma \lambda} {\bf tr}  F_{\alpha \beta} F_{\gamma \lambda} \right )
\end{eqnarray}
where  we have rewritten the action so as to write the kinetic term in canonical form. The physics is then independent of $\theta$ and can be investigated as a function of the axion $a = \theta f_a = \tilde \theta$. $f_a$ is called the axion decay constant, and gives the scale at which the PQ symmetry is spontaneously broken. 
In either case, this CP-violating term is then relaxed to zero as the axion is subject to an instanton-generated potential \cite{Conlon:2006tq} (see also \cite{Svrcek:2006yi, Srednicki:2007qs})
\begin{eqnarray}
\label{instpot}
V_{\mathrm instanton} & \sim & \Lambda^4_{QCD} \left (1 - \mathrm{cos} \left (\frac{a}{f_a} \right )  \right )
\end{eqnarray}
which is minimised at $a \sim 0$. Thus the theta term is relaxed to zero dynamically. 
\subsection{Constraints on the axion}
The potential (\ref{instpot}) implies a mass for the axion
\begin{eqnarray}
m_a & \sim & \frac{\Lambda_{QCD}^2}{\sqrt{2} f_a},
\end{eqnarray}
arising from the quadratic term in $a$. This relation between the axion mass and axion decay constant makes it clear that there could be observational bounds on the axion decay constant $f_a$ from known particle physics. Specifically, astrophysical and cosmological bounds constrain the value of $f_a$ from above and below (see \cite{Svrcek:2006yi, Sikivie:2006ni, Kim:1999ia, Kim:2007qa} and references therein). For small values of $f_a$, the axion couples strongly to matter. This would accelerate the evolution of stars such as red giants, by transporting their energies into the outer regions more efficiently and shortening their lifetimes \cite{Raffelt:1987yt, Turner:1987by, Mayle:1987as}. Similarly, values of $f_a$ which are too large are ruled out on cosmological grounds in order to avoid production of too much axionic dark matter (which could overclose the universe) \cite{Abbott:1982af, Dine:1982ah, Preskill:1982cy}. Thus experimentally acceptable values of $f_a$ must fall within the range
\begin{equation}\label{bound}
10^{9} \, {\rm GeV} < f_{a} < 10^{12}  \, {\rm GeV} \, .
\end{equation}
Note that in models of axion inflation, a subset of large field inflation models which generically give rise to observable tensor fluctuations, a super-Planckian value of $f_a$ is required (see for instance \cite{Kallosh:2007cc}). This is compatible with the bounds presented here because the axion responsible for axion inflation is not the QCD axion needed to solve the strong CP problem, but another pseudoscalar field with shift symmetry produced much earlier in the universe's history. However, it is still difficult to construct string models with $f_a > M_{Pl}$ \cite{Banks:2003sx}. Possible ways around this are presented in \cite{Kallosh:2007cc, Dimopoulos:2005ac}.
\section{Warped Heterotic Axions}
\subsection{Axions in String Theory}
PQ symmetries and axions arise naturally in string theory. As explained in \cite{Svrcek:2006yi}, the terms in the low-energy effective action that lead to anomaly cancellation in the Green-Schwarz mechanism \cite{Green:1984sg} also cause light string modes to behave as axions \cite{Green:1987mn, Kim:1986ax}. For other reviews see \cite{Choi:1985je, Choi:1985bz, Choi:2006qj, Conlon:2006tq,Svrcek:2006yi, Kim:2006aq}. However, 
axion construction in conventional string theory models typically results in an axion decay constant higher than the range of phenomenologically allowed values. This was extensively studied in \cite{Svrcek:2006yi} with the conclusion that for string scale $m_{s}$ comparable to $M_{P}$ the axion decay constant is generically of order $10^{16}$ GeV, which is too large to be allowed. One way out of this problem is to lower the string scale by employing an exponentially large compactification. Axion construction in very large compactification volumes was studied in \cite{Conlon:2006tq}. A very large compactification corresponds to a low-scale string theory. It is argued that up to
numerical factors of order unity, $f_{a} \sim m_{s} \sim 10^{11}$ GeV.  Other proposals range from anthropic arguments \cite{Linde:1987bx}, which predict significant abundance of axionic dark matter in the universe \cite{Wilczek:2004cr}, to modifying the usual cosmological assumptions about QCD \cite{Dvali:1995ce} or inflation \cite{Banks:1996ea, Banks:2002sd, Steinhardt:2004gk} (see \cite{Svrcek:2006yi}). It may be that future experimental data on axionic dark matter will rule out or confirm some of these suggestions. 

Here we consider whether it is possible to exploit the effects of warping to reduce the scale of $f_{a}$
in heterotic string theory.  Models of warped axions in the context of a five-dimensional Randall-Sundrum scenario were presented in \cite{Collins:2002kp, Choi:2003wr, Flacke:2006ad, Flacke:2006re}. We attempt a full string theoretic description of this set-up. We begin by showing the effect of warping on the axion decay constant in Section \ref{effect} and go on to give a full construction of the required heterotic background in Section \ref{hetconstruction}. In Section \ref{fa} we compute the axion decay constant in these models. This work was published in \cite{Dasgupta:2008hb}.

\subsection{The effect of warping on $f_a$}
\label{effect}
A warped geometry has the form 
\begin{eqnarray}
\label{metric}
ds^{2} =  h_{w}^{2}(y) \eta_{\mu \nu} dx^{\mu} dx^{\nu} + g_{mn} (y)dy^{m} dy^{n} ,
\end{eqnarray}
where $\mu, \nu = 0,1,2,3$ label co-ordinates in Minkowski space and $m, n = 4, ... 9$ label co-ordinates on the internal manifold ${\cal M}$. The warp factor $h_w$ can be a function of the internal dimensions $y$. To arrive at a 4-dimensional theory, any ten-dimensional starting point must be dimensionally reduced. The 4-dimensional axion is the Hodge dual of $B_{NS}$, the $NS-NS$ 2-form. $dB_2 = \star_4 d \phi$, and can therefore arise in two different ways: If $B_{MN}$ has no legs on the internal manifold, the resulting axion is said to be \em model independent\em. Conversely, if it wraps some cycle on the internal manifold, the axion is said to be \em model dependent\em. We shall see that for a warped heterotic compactification, the so-called ``model-dependent" axion does in fact depend on the details of the compactification. 

To see how this dependence arises, consider the ten-dimensional heterotic string action (in the Einstein frame):\footnote{This arises from converting (12.1.39) in \cite{Polchinski:1998rr} from string to Einstein frame. The relevant relations are $R^E = e^{\frac{\phi}{2}} R^S$ and $g_{\mu \nu}^S e^{- \frac{\phi}{2}} = g_{\mu \nu}^E$.}
\begin{eqnarray}
\label{10D}
\nonumber S_{het}^{(E)}=  \frac{1}{ 2 \kappa_{10}^{2}} \int d^{10} x\,  \sqrt{-g} \left( R - \frac{1}{2} \partial_{M} \phi \partial^{M} \phi-
\frac{e^{-\phi} }{2}  H_3 \wedge \star H_3
- \frac{\alpha'}{120} e^{-\frac{\phi}{2}}  \, {\rm Tr_A}~ F\wedge \star  F
\right)\\
\end{eqnarray}
Here $M,N$ are 10-dimensional indices, R is the Ricci scalar, $\phi$ the dilaton, $H = dB_2$ and F is the heterotic gauge curvature with trace in the adjoint representation. The trace in the fundamental representation is ${\bf tr}= {\rm Tr_A}/30$. 

We find the zero modes of the heterotic fields upon dimensional reduction separately, to evaluate the effect of warping. Beginning with the graviton, we keep only the four-dimensional part of the Ricci scalar $ R = R_4 + R_6 = g^{MN} R_{MN} = g^{ \mu \nu} R_{\mu \nu} + g^{mn} R_{mn}$ and pull out the warp factor dependence (leaving barred quantities):
\begin{eqnarray}
\label{gravitonzeromode}
\nonumber S_g & = & \frac{1}{2 \kappa_{10}^2} \int d^{10}x \sqrt{-g_{10}} R\\
\nonumber S_g^{(0)} & = & \frac{1}{2 \kappa_{10}^2} \int d^4 x d^6y  \sqrt{-g_4} \sqrt{g_6}R_4 \\
\nonumber S_g^{(0)} & = & \frac{1}{2 \kappa_{10}^2} \int d^6y h_w^2 (y) \sqrt{g_6} \int d^4 x \sqrt{-\bar g_4} \bar R\\
S_g^{(0)} & = & \frac{M_P^2}{2} \int d^4 x \sqrt{-\bar g_4} \bar R,
\end{eqnarray}
where we have defined the Planck mass as 
\begin{eqnarray}
\label{MP}
M_P^2 & = & \frac{1}{\kappa_{10}^2} \int d^6 y \sqrt{g_{6}} h_w^2(y).
\end{eqnarray}
The dependence on the warp factor arises from $R_4 = h_w^{-2} \bar R$ and $g_4 = h_w^8 \bar g_4$. Similarly,
\begin{eqnarray*}
\int  (H \wedge \star H)_4& = & \frac{1}{3!}\int d^4 x \sqrt{-g_4} \sqrt{g_6} H_{\mu \nu \rho} H^{\mu \nu \rho}\\
& = & \frac{1}{3!} \int d^4 x \sqrt{- \bar g_4}\sqrt{g_6} h_w^4(y) g^{\mu \mu'}g^{\nu \nu'} g^{\rho \rho'} H_{\mu \nu \rho} H_{\mu' \nu' \rho'}\\
& = & \frac{1}{3!} \int d^4 x \sqrt{-\bar g_4}\sqrt{g_6} h_w^4(y) h_w^{-6}(y) \bar H^2\\
& = & \frac{1}{3!} \int \sqrt{g_6} h_w^{-2}(y) \bar H \wedge \bar \star \bar H,
\end{eqnarray*}
where $\bar \star$ is constructed from $\bar g_{\mu \nu}$ and is independent of the warp factor. 
Then the zero mode of the NS-NS three-form is
\begin{eqnarray}
\label{NSzeromode}
\nonumber S_{NS}^{(0)} & = & - \frac{1}{4\kappa_{10}^2} \int \bar H \wedge \bar \star \bar H \int d^6 y \sqrt{g_6} e^{-\phi} h_w^{-2} (y)\\
S_{NS}^{(0)} & = & - \frac{\beta M_P^2}{4} \int \bar H \wedge \bar \star \bar H,
\end{eqnarray}
where $\beta$ is the ratio of normalizations of the graviton and NS-NS three-form \cite{Firouzjahi:2007dp}:
\begin{eqnarray}
\label{beta}
\beta= \frac{\int d^{6}y \,  \sqrt{ g_{(6)}} \,  e^{-\phi}
  h_{w}^{-2} (y)}{\int d^{6}y \,  \sqrt{g_{(6)}} \,  h_{w}^{2} (y)}.
\end{eqnarray}

Looking at (\ref{gravitonzeromode}) and (\ref{NSzeromode}), we see that in a flat background where $h_{w}= e^{-\phi}=1$, the zero modes of both the graviton and the NS--NS three-form are Planck suppressed. This is to be expected, since they both belong to the massless sector of the closed string theory in ten dimensions. However, as observed in \cite{Mukhopadhyaya:2002jn, Mukhopadhyaya:2007jn, Firouzjahi:2007dp}, they appear with different normalizations in a warped background.
This difference is parametrised by $\beta$.

The Bianchi identity for the gauge-invariant field $H$ is 
\begin{eqnarray} \label{bianchi}
d H= \frac{\alpha'}{4} \left( {\rm tr }~ R \wedge R - \frac{1}{30}{\rm Tr}~ F \wedge F \right). 
\end{eqnarray}
To incorporate the axion in our construction, we dualize the $B$-field in the four-dimensional action by a scalar field $a$, via  the following Lagrange multiplier for the Bianchi identity:
\begin{eqnarray}
\int   \, a \left[d H- \frac{\alpha'}{4} \left( {\rm tr }~ R \wedge R - {1\over 30}{\rm Tr}~ 
F \wedge F\right)\right] \, .
\end{eqnarray}
The action containing the $B$-field kinetic energy and the Lagrange multiplier is
\begin{eqnarray}
\label{actionL}
S=- \frac{ \beta  M_{P}^{2}}{4} \int d^{4} x   H \wedge  \bar \star   H  +  \int d^4 x \,
 a \left[d H- \frac{\alpha'}{4} \left( {\rm tr } R \wedge R 
- {1\over 30}{\rm Tr} F \wedge F \right)\right].
\end{eqnarray}
Integrating out $H$  in terms of the field $a$ one obtains
\begin{eqnarray}
\bar \star H= \frac{2}{\beta M_{P}^{2}} \,  d a \,.
\end{eqnarray}
This is equivalent to the statement that in four dimensions the axion is Hodge dual to the anti-symmetric
$B_{\mu \nu}$ field.
Substituting this into the action (\ref{actionL}) yields
\begin{eqnarray}
\label{axion3}
S(a)= \frac{2}{\beta M_{P}^{2}} \int d^{4} x  \, \left( -\frac{1}{2} \partial_{\mu} a  \partial^{\mu} a 
  \right) + \int a  \, \frac{\alpha'}{4} \left( {1\over 30}{\rm Tr}~ F \wedge F - {\rm tr}~ R \wedge R \right).
\end{eqnarray}
Upon rescaling the axion as in (\ref{axion2}) and noting that $2 \pi \sqrt {\alpha'} = m_{s}^{-1}$, we find
\begin{eqnarray}
\label{ourf}
f_{a} = \sqrt{\frac{1}{2 \beta}}  \, \frac{m_{s}^{2}}{M_{P}} .
\end{eqnarray}
In an unwarped compactification with $\beta =1$ and taking $m_{s} /M_{P} \simeq 1/18 $ in order to get the right GUT scale from string theory, one obtains $f_{a} \simeq 10^{16}$ GeV as in \cite{Svrcek:2006yi}, too big to be acceptable. However, $\beta$ can be significantly greater than one in a warped compactification. From (\ref{ourf}) we see that this can reduce $f_a$ to the range $10^{9}-10^{12}$ GeV. In subsequent sections we will provide specific warped examples where $\beta$ is found to be large enough such that  
$f_{a}$ falls within the desired window.


\section{Warped Heterotic Construction}
\label{hetconstruction}
\subsection{Heterotic Compactification on a non-K\"ahler Manifold}

First we shall review the axion construction in a warped heterotic background given by \cite{Kim:2006aq}. The model considered there is a heterotic
compactification on a non-K\"ahler manifold.\footnote{Some references on non-K\"ahler manifolds are \cite{Dasgupta:1999ss, Becker:2002sx, Becker:2003yv, Goldstein:2002pg, Becker:2003sh, Becker:2003gq, Becker:2003dz, Becker:2004qh, Alexander:2004eq, Becker:2005ef, Dasgupta:2006yd, Dasgupta:2006sg}.}The non-K\"ahler background
is a non-trivial $T^2$ fibration over a K3 base. In the Einstein frame the full ten-dimensional metric can be written:
\begin{equation}\label{nkmanifold}
ds^2 = e^{-{\phi\over 2}} \eta_{\mu\nu} dx^\mu dx^\nu + e^{-{\phi\over 2}}\left[(dx+\alpha_{1})^2 + (dy+\alpha_{2})^2\right]
+ e^{{3\phi\over 2}} ds^2_{\rm K3},
\end{equation}
where $x$ and $y$ are local coordinates such that $dx\,+\,idy$ is a holomorphic form on the $T^2$ fibers, and the
$\alpha_{i}$ are local one-forms on the K3 base.  
For this particular compactification we see that the dilaton is related to the warp factor
via $e^{-\phi} = h_{w}^{4}$. Substituting this into our expression for $\beta$ in (\ref{beta}), one finds
that $\beta=1$. Thus, as mentioned in \cite{Kim:2006aq}, the warping does not help to reduce $f_{a}$
for the model-independent axion in the above background. 

To avoid this cancellation of the warp factor with the dilaton, we construct a background with dilaton  independent of the warp factor. In this case, $\beta$ can be made sufficiently large. The backgrounds studied in Sections \ref{AdS} and \ref{constantcoupling} both satisfy this condition. 

\subsection{An AdS-type Background in Heterotic Theory}
\label{AdS}
As explained above, in order to get large enough $\beta$ we need to construct warped geometries where the dilaton is independent of the warp factor, which are given below. These can be constructed using sigma model identifications, given in detail in \cite{Dasgupta:2008hb}. In brief, we used sigma model identifications to move from torsional IIB backgrounds to backgrounds in the heterotic theory. The results of this analysis are that we can drag a given type IIB background to heterotic string theory provided that the original IIB background has (after U-dualising) non-trivial metric and NS-NS three-form, and no RR three-form. We also require a dilaton independent of the radial coordinate. The U-dualities consist of two T-duality transformations and an S-duality transformation. Sigma model identification is then used to correctly modify the Bianchi identity and construct the relevant vector bundles (see \cite{Dasgupta:2008hb} for details). We will show in Section \ref{fa} that in the resulting warped heterotic background, described below, $f_a$ can be lowered to values within the phenomenological window. In Section \ref{constantcoupling} we present another new heterotic background which has a warp-independent dilaton but non-trivial torsion.

A natural starting point is type IIB theory on $AdS_5$ space. However, the minimally supersymmetric $AdS_5$ background, i.e. $AdS_5 \times T^{1,1}$,  given by Klebanov-Witten \cite{Klebanov:1998hh} cannot be pulled to the heterotic side using our sigma--model identification. The non-trivial fibrations of the internal space $T^{1,1}$ create extra fluxes under U-duality which prohibit a heterotic dual for this background \cite{Dasgupta:2008hb}. We are therefore left with the other choices: $AdS_5 \times S^5$ and 
$AdS_5 \times {S^5\over {\bf Z}_n}$ with 
$n = 2, 3, 4, 6$. 

We now claim that in the heterotic theory we will have a background of the 
form
\begin{equation}\label{nhebag}
ds^2 = e^\phi ds^2_{AdS_5} ~ + ~ ds^2_{X^5},
\end{equation}
that satisfies all the requirements sketched above. Here 
$\phi$ is the dilaton that depends only on the coordinates of the internal space $X^5$ and not on the radial
coordinate $r$. This non-trivial dilaton will be supported by a background torsion $H$. 

To find it, we start with an $AdS_5 \times S^5$ background in type IIB string theory given (in units of $\alpha'$) by
\begin{eqnarray}
\label{Ads5}
ds^{2}= \frac{r^{2}}{R^{2}} \eta_{\mu \nu} dx^{\mu} dx^{\nu} + \frac{R^{2}}{r^{2}}~dr^2
+ R^{2} d\Omega_{5}^{2},
\end{eqnarray}
where $\mu, \nu = 0, 1, 2, 3$ are the spacetime directions and $R$ is the curvature radius of the AdS space given by
\begin{eqnarray}
\label{R}
R^{4} = 4 \pi g_{s} N \, .
\end{eqnarray}
Here $N$ is the quantised charge of the five-form $F_{5}$,
\begin{eqnarray}
\int_{S^{5}} F_{5} = (4 \pi^{2} \alpha')^{2} \, N ,
\end{eqnarray}
and we take $H_{NS-NS} = H_{RR} = \phi = 0$. In the absence of NS and RR three-forms the five-form $F_5$ can be written as
$F_{5}= d C_{(4)} + * d C_{(4)}$ with the RR four-form, $C_{(4)}$, given by
\begin{eqnarray}
\label{C4}
C_{(4)} = \frac{r^{4}}{   g_{s} R^{4}} \,  dx^{0} \wedge dx^{1} \wedge dx^{2} \wedge dx^{3} \, .
\end{eqnarray}
Finally, the metric of the five-sphere,  $d\Omega_{5}^{2}$, in (\ref{Ads5}) is 
\begin{equation}
\label{sfive}
d\Omega_5^2 = d\gamma^2 + {\rm cos}^2 \gamma ~d\varphi_3^2 + {\rm sin}^2\gamma 
\left(d\psi^2 + {\rm cos}^2 \psi ~d\varphi_1^2 + {\rm sin}^2\psi ~d\varphi_2^2\right),
\end{equation}
where $0 \leq \gamma, \psi \leq \frac{\pi}{2}$ and $ 0 \leq \varphi_i \leq 2 \pi$.
We see that there are three local isometries along the $\varphi_1$, $\varphi_2$ and 
$\varphi_3$ directions. We can choose $\varphi_1$ and $\varphi_2$ as the directions along which to perform our T-dualities, but we have to take care because there are no global one-cycles in the manifold. In fact, at the points
\begin{equation}\label{ncpoint}
\gamma ~ = ~ 0; ~~~~ \psi ~ = ~ 0; ~~~~ \psi ~ = ~ \frac{\pi}{2},
\end{equation}
the cycles all shrink to zero size and the U-dual manifold will be non-compact. To avoid these issues, we
will make our U-dualities away from the points \eqref{ncpoint}. We find the  
following background in heterotic theory: 
\begin{eqnarray}\label{hetbag}
\nonumber ds^2 &=& {1\over 2} R^2  {\rm sin}^2 \gamma ~{\rm sin} \, 2\psi \Big[{r^2 \over R^2} 
dx_\mu dx^{\mu} + {R^2 \over r^2} ~dr^2\Big]  \\
 \nonumber && + {1\over 2} R^4 {\rm sin}^2 \gamma ~{\rm sin} \, 2\psi \left[d\gamma^2 + {\rm cos}^2 
 \gamma d\varphi_3^2 
+ {\rm sin}^2 \gamma~d\psi^2\right] + {\rm tan}\psi \, d\varphi_1^2 + {\rm cot}\psi \, d\varphi_2^2 ; \\
 e^\phi~ &=& {1\over 2g_s}\left(R^2 ~{\rm sin}^2 \gamma ~{\rm sin}~2\psi\right)  \, ;  \\
 \nonumber H &=& {4r^3 \over R^4}  e^{2 \phi}~\ast \left(dx^0 \wedge dx^1 \wedge dx^2 \wedge dx^3 \wedge dr \wedge d\varphi_1  
\wedge d\varphi_2\right) + {\cal O}(\alpha')  \\
\nonumber &=& -\frac{4 R^{4}}{g_s^2} \sin^{3} \gamma \cos \gamma \sin \psi \cos \psi \,    d \gamma \wedge  d\psi 
\wedge  d \varphi_{3}      + {\cal O}(\alpha')        \,  ,
\end{eqnarray}
with an additional vector bundle given in \cite{Dasgupta:2008hb}. This  bundle has to satisfy the modified 
DUY equations which appear because of the background torsion \cite{Becker:2003yv, Becker:2003sh}. The Hodge star operation
is defined for a generic $p$-form as:
\begin{equation}\label{hodge}
\left(\ast \omega\right)_{\mu_1 \mu_2 ..... \mu_{10-p}} = \frac{\sqrt{-g}}{p!} ~
\epsilon_{\mu_1 \mu_2 ..... \mu_{10-p}}^{~~~~~~~~~~~~~~\nu_1 \nu_2 .....\nu_p} 
\omega_{\nu_1 \nu_2 .....\nu_p}.
\end{equation} 
Note that the new background on the heterotic side is not quite an $AdS_5$ background because of the unusual 
warp factors, although the radial dependence resembles that of the standard type IIB $AdS_5$ background. The internal 
space is also not an $S^5$ anymore. The metric has non-trivial warp factors that make the background non-K\"ahler. 
Furthermore, the dilaton is not a constant, and $H$ (the torsion) is in general more complicated than the 
standard form, although we expect a modified anomaly-cancelling Bianchi identity to hold and a suitable vector bundle to be defined (see \cite{Dasgupta:2008hb} for details). 

To read off physical quantities, we transform the metric into the Einstein frame via
$g_{MN}^{(E)}= e^{-\phi/2} g_{MN}^{(S)}$. After restoring the necessary factors of $\alpha'$ and rescaling $x^\mu$ 
($ g_{s}^{1/4} \alpha'^{1/2}\, x^{\mu} \rightarrow x^{\mu}$), \eqref{hetbag} 
in the Einstein frame is given by
\begin{eqnarray}
\label{Ein1}
ds^{2}&=& \sin \gamma \, \sqrt{\sin \psi \cos \psi} 
\Bigg[  \frac{r^2}{R } dx^{\mu} dx_{\mu}   + \alpha' \sqrt{g_s} R^{3} ( 
 \frac{dr^{2}}{r^{2}}  + d \gamma^{2} + \cos^{2} \gamma d \varphi_{3}^{2} + 
\sin^{2} \gamma d \psi^{2} )\Bigg]  \nonumber\\
&&+ \frac{\alpha' \sqrt{g_s \sin \psi}  }{R \cos^{\frac{3}{2}} \psi   \sin \gamma  } d\varphi_{1}^{2}
+\frac{  \alpha'\sqrt{g_s \cos \psi}  }{R \sin^{\frac{3}{2}} \psi   \sin \gamma  } d\varphi_{2}^{2},
\end{eqnarray} 
with $H$ given as in \eqref{hetbag}. Note that this geometry has the form of a warped metric (\ref{metric})
with warp factor 
\begin{eqnarray}\label{wfac}
h_{w}^{2} = \sin \gamma \sqrt{\sin \psi \cos \psi}  \,  \frac{r^2}{R }.
\end{eqnarray}
One can check that the background given by (\ref{Ein1}) or equivalently \eqref{hetbag}
is a consistent solution: with $H$ given as in \eqref{hetbag}, the equation of motion 
$d \star H = 0$  is trivially satisfied. The dilaton equation,
\begin{eqnarray}
\label{dileom}
\frac{1}{ \sqrt{-g} } \partial_{M} \left(   \sqrt{-g}     \partial^{M} \phi  \right) + \frac{e^{-\phi} }{12} H^{2}=0,
\end{eqnarray}
is also satisfied. Finally the Einstein equation, $G_{MN}= \frac{1}{2} T_{MN}$, where $G_{MN}$
is the Einstein tensor and $T_{MN}$ is the stress-energy tensor given by
\begin{eqnarray}
\label{Tmn}
T_{MN}=  \partial_{M} \phi \,   \partial_{N} \phi  
- \frac{1}{2} g_{MN} \partial_{P} \phi  \,  \partial^{P} \phi 
+ \frac{e^{-\phi} }{2 }  H_{MPQ} H_{N}^{\, \, PQ }
-\frac{e^{-\phi}}{ 12}   g_{MN}   H^2,
\end{eqnarray}
is also satisfied.

The components of the Einstein tensor for the background (\ref{Ein1}) are:
\begin{eqnarray} 
\nonumber && G_{00} ~ = ~ - G_{ii} =   
 \frac{ r^{2} \, (  1+1 2 \sin^{2} \psi \cos^{2} \psi \sin^{2} \gamma  )  }{ 4 \sqrt{g_{s}}  \alpha'   R^{2}\, 
 (\sin^{2} \gamma \cos^{2} \psi \sin^{2} \psi )    }; \\
\nonumber && G_{rr} ~ = ~ - \frac{R^{4}}{ r^{4}  }  \sqrt{g_{s}}  \alpha' \,      G_{00}; \\
\nonumber && G_{\gamma \gamma} ~ = ~
  \frac{  -1- 1 2 \sin^{2} \psi \cos^{2} \psi \cos^{2} \gamma   + 20  \sin^{2} \psi \cos^{2} \psi    }{4  \sin^{2} \gamma \cos^{2} \psi \sin^{2} \psi     }; \\ 
 &&  G_{\psi \psi} ~ = ~ (8- G_{\gamma \gamma} ) \sin^2 \gamma;\\
 \nonumber && G_{\phi1 \phi1}~ = ~ \frac{ \sqrt{g_{s}}  \alpha'   G_{ 00  } }{  r^{2} \sin^{2} \gamma \cos^{2}  \psi   }; \\
  \nonumber && G_{\phi2 \phi2}~ = ~ \frac{ \sqrt{g_{s}}  \alpha'  G_{ 00  } }{  r^{2} \sin^{2} \gamma \sin^{2}  \psi   }; \\
  \nonumber && G_{\phi3 \phi3}~ = ~ \frac{ \cos^{2} \gamma \,  ( -1+20 \sin^{2} \psi \cos^{2} \psi \sin^{2} \gamma )   }{4  \sin^{2} \gamma \cos^{2} \psi \sin^{2} \psi     };\\
   \nonumber && G_{\gamma \psi}~ = ~ \frac{ \cos \gamma(   \cos^{2} \psi -\sin^{2} \psi )  }{ \sin \gamma \sin \psi \cos \psi   }.
\end{eqnarray}
One can check that the off-diagonal component of the Einstein tensor is sourced by
$\partial_{\psi} \phi \partial_{\gamma} \phi$.
With these values for $G_{MN}$ along with $H$ and $\phi$ given as in (\ref{hetbag}), one can explicitly check that the Einstein equations are all satisfied \cite{Dasgupta:2008hb}. This demonstrates that the background  (\ref{Ein1})  is a genuine solution, giving us a powerful test of the consistency of our background. 

As mentioned above, the background 
\eqref{hetbag} is well defined away from the points \eqref{ncpoint}, where the cycles shrink to zero size. Our construction clearly fails 
at these points. It is then no surprise that the Ricci scalar for the metric (\ref{Ein1}), given by 
\begin{eqnarray}
{\cal R }= \frac{  1+ 4 \sin^{2} \psi  \cos^{2} \psi \sin^{2} \gamma}{2 \alpha' \sqrt {g_s}    R^{3} \sin^{3} \gamma
 \sin^{5/2} \psi   \cos^{5/2} \psi       } \, ,
\end{eqnarray}
diverges at the points \eqref{ncpoint} mentioned above. 
Such divergences can be cured by removing these points from the original $S^5$ \eqref{sfive}. Then the metric \eqref{hetbag} is a good description of the geometry away from these points, and the global six-dimensional manifold will be a compact non-K\"ahler manifold when we cut off the radial direction and replace it with a smooth cap. Physically, the quantity $g_sN$ corresponds to the integral of the three-form over a three-cycle of our non-K\"ahler manifold. As mentioned above, the four-dimensional spacetime is then a warped Minkowski spacetime with warp factor given by \eqref{wfac}. 

Once the singular points are smoothed out, the manifold will have a well-defined Riemann tensor globally. This will result in corrections to the torsion as expected \cite{Dasgupta:2008hb}. 

\subsubsection{Related AdS-type backgrounds}
Note finally that if we change the orientation of the three-form flux $H$ from $\varphi_1, \varphi_2$
to $\varphi_1, \varphi_3$, keeping other factors unchanged, we can generate a slightly different background that falls
in the same class as \eqref{hetbag}:
\begin{eqnarray}\label{hetbagtwo}
\nonumber  ds^2 & =&  {1\over 2} R^2 ~ {\rm sin}~2\gamma ~{\rm cos}~\psi \Big[{r^2 \over R^2} 
 dx_{\mu} dx^{\mu}+ {R^2 \over r^2} ~dr^2\Big]  \\
 \nonumber &&+ ~{1\over 2} R^4~ {\rm sin}~2\gamma ~{\rm cos}~\psi \left[d\gamma^2 + {\rm sin}^2 \gamma~d\psi^2 
+ {\rm sin}^2 \gamma~ {\rm sin}^2 \psi~
d\varphi_2^2\right]~  \\ \nonumber 
&&  ~+~  {\rm tan}~\gamma ~ {\rm cos}~\psi~d\varphi_3^2 + 
{\rm cot}~\gamma ~ {\rm sec}~\psi ~d\varphi_1^2; \\
e^\phi &= & {1\over 2g_s}\left(R^2 ~{\rm sin}~2\gamma ~{\rm cos}~\psi\right);\\  \nonumber
H&=& -\frac{4 R^{4}}{g_s^2} \sin^{3} \gamma \cos \gamma \sin \psi \cos \psi \,    d \gamma \wedge  d\psi 
\wedge  d \varphi_{2}      + {\cal O}(\alpha')        \,  .
\end{eqnarray}
This metric also has singularities. The Ricci scalar for this background in the Einstein frame scales like
${\cal R}  \sim  \sin^{-5/2} \gamma \,   \cos^{-5/2} \gamma \,   \sin^{-5/2} \psi  $, which diverges at
$\gamma=0$, $\gamma=\pi/2$ and $\psi=\pi/2$. Excising these points, and cutting off the radial direction to replace it with a finite cap, we can have a smooth non-K\"ahler manifold that has well-defined curvature forms. The full torsion for the manifold to higher orders in $\alpha'$
can now be easily computed following our earlier analysis \cite{Dasgupta:2008hb}. 

A third background, also falling in the same class, can be derived by changing the orientation of the three-form
from $\varphi_1, \varphi_2$ to $\varphi_2, \varphi_3$. This corresponds to taking
$\sin \psi \rightarrow \cos \psi$ in \eqref{hetbagtwo}:
\begin{eqnarray}
\label{hetbagthree}
 \nonumber ds^2&  =&  {1\over 2} R^2 ~ {\rm sin}~2\gamma ~{\rm sin}~\psi \Big[{r^2 \over R^2} 
dx_{\mu} dx^{\mu}+ {R^2 \over r^2} ~dr^2\Big]  \\ \nonumber
&&+ {1\over 2} R^4~ {\rm sin}~2\gamma ~{\rm sin}~\psi \left[d\gamma^2 + {\rm sin}^2 \gamma~d\psi^2 
+ {\rm sin}^2 \gamma~ {\rm cos}^2 \psi~
d\varphi_1^2\right]~  \\ \nonumber 
&&+  {\rm tan}~\gamma ~ {\rm sin}~\psi~d\varphi_3^2 + 
{\rm cot}~\gamma ~ {\rm cosec}~\psi ~d\varphi_2^2; \\  
e^\phi & =& {1\over 2g_s}\left(R^2 ~{\rm sin}~2\gamma ~{\rm sin}~\psi\right);\\  \nonumber
H&=& -\frac{4 R^{4}}{g_s^2} \sin^{3} \gamma \cos \gamma \sin \psi \cos \psi \,    d \gamma \wedge  d\psi 
\wedge  d \varphi_{1}      + {\cal O}(\alpha')        \,  .
\end{eqnarray}
This geometry also has singularities at $\gamma=0$, $\gamma=\pi/2$ and $\psi=0$, and, following the same procedure 
as before, we can deform it to give a smooth non-K\"ahler geometry with torsion and non-trivial vector bundles.


\subsection{The Axion Decay Constant}
\label{fa}
Having constructed specific warped AdS heterotic backgrounds we can calculate the normalization constant
$\beta$ from (\ref{beta}) and find the axion decay constant $f_{a}$ from (\ref{ourf}).
The AdS geometries presented in the previous section
should be considered as local warped regions
or throats, which are glued in the UV to the compactification bulk.
The warped throat is glued to the bulk at $r=L$, where for consistency we impose $R   \, \simeq L$ , with $ R = (4\pi g_{s} N)^{1/4} $ the AdS curvature radius  of the AdS geometry.
The overall size of the bulk of the compactification, $R_{6}$, is assumed to be much bigger
than the size of the throat, $R_{6}\gg L$, such that the bulk contains most of the volume of the compactification. Furthermore, it is assumed that the bulk is not warped.

The AdS geometry is also subject to an IR cut off when $r \rightarrow 0$.
There is a conical singularity at $r=0$ and we assume that the geometry near the tip
of the cone or throat is modified such that this singularity is smoothed out as in the 
Klebanov-Strassler (KS) background \cite{Klebanov:2000hb}. The IR geometry is cut off 
at $r=r_{0}$ and the value of the warp factor $h_0$ at $r_0$ (after integrating over the angular directions)
 is given by
\begin{eqnarray}
\label{h0}
h_{0} =\frac{r_{0}}{\sqrt R} \, .
\end{eqnarray}
As in the KS solution, there are corrections to $h_{0}$ due to IR modification of the throat.
We expect them to be sub-leading and that they will not play a significant role in our discussion.

Noting that $4 \pi \kappa_{10}^{2} = m_{s}^{-8}$, and defining the volume of the bulk to be 
$v_{6} = R_{6}^{6}/4\pi$, 
the four-dimensional gravitational coupling from (\ref{MP}) is
\begin{eqnarray}
\label{MP2}
M_{P}^{2}&= &4\pi m_{s}^{8}  \left(
 \frac{R_{6}^{6}}{4\pi} + \frac{1}{2}\pi^{3} R^{4} \alpha'^{3} g_{s}^{\frac{3}{2}} (1-\sin^{4} \gamma_{1}) 
(\sin^{2}\psi_{2} -  \sin^{2}\psi_{1} )(L^{2} -r_{0}^{2})
 \right) \nonumber\\
&\simeq& m_{s}^{8} R_{6}^{6} \left(  1+ 2 \pi^{4}  g_{s}^{\frac{3}{2}}  \frac{R^{4}   L^{2} \alpha'^{3}  }{  R_{6}^{6}  }
\right) \nonumber\\
&\simeq&   m_{s}^{8} R_{6}^{6} \, .
\end{eqnarray}
Here $\gamma_{1}$ and $\psi_{i}$ are the cut-off values for the 
angular variables $\gamma$ and $\psi$ at the singular points (\ref{ncpoint}).
In going from the first line to the second line above, it is assumed that $r_{0}\ll L$, 
$\gamma_{1}\rightarrow 0, \, \psi_{1}\rightarrow 0$ and $\psi_{2}\rightarrow \pi/2$.
To obtain the final answer, 
as mentioned before, it is assumed that the bulk contains most of the volume of the compactification, corresponding
to $R^{4} L^{2} \alpha'^{3} \sim L^{6} \alpha'^{3} \ll R_{6}^{2}$.

Similarly, we obtain
\begin{eqnarray}
\beta &=& \left(\frac{R_{6}^{6}}{4\pi} \right)^{-1}
\left[  e^{-\phi_{B}}  \frac{R_{6}^{6}}{4\pi} +  4 \pi^{3} R^{4} \alpha'^{3}  g_{s}^{\frac{5}{2}}
\left| \ln\left(   \frac{ \tan \psi_{2}  }{\tan \psi_{1}   }  \right)  \ln (\sin \gamma_{1}  ) \right|
\left(\frac{1}{r_{0}^{2}} -   \frac{1}{L_{0}^{2}}\right)  \right] \nonumber\\
&\simeq&    \frac{g_{s}^{\frac{5}{2}} }{4\pi^{2}} \frac{m_{s}^{2}}{ M_P^{2} }
 \left| \ln\left(   \frac{ \tan \psi_{2}  }{\tan \psi_{1}   }  \right)  \ln (\sin \gamma_{1}  ) \right|
  \frac{R^{4}}{ r_{0}^{2}  }.
\end{eqnarray}
In the first line, it is assumed that in the bulk the dilaton field does not change significantly from some bulk value $\phi_{B}$. To go from the first line to the second line it is assumed that $r_{0}/R^{2}<<1$, so the second term dominates over the first term. Physically, this means that the normalization of the zero mode of $B_{NS}$ gets its largest contribution from the highly warped throat \cite{Mukhopadhyaya:2002jn, Mukhopadhyaya:2007jn, Firouzjahi:2007dp}. This should be contrasted with the normalization of the graviton zero mode, which is insensitive to the warp factor \cite{Firouzjahi:2005qs}. The calculation of the four-dimensional gravitational coupling in (\ref{MP2}) reflects this.

Substituting this value of $\beta$ into (\ref{ourf}), we obtain the axion decay constant 
\begin{eqnarray}
\label{ourf1}
f_{a} \sim \frac{  \pi}{g_{s}^{\frac{5}{2}} R^{3/2}}
 \left|     \ln\left(   \frac{ \tan \psi_{2}  }{\tan \psi_{1}   }  \right)  \ln (\sin \gamma_{1}  )  \right|^{-1/2}
 (h_{0} \,  m_{s} ) \, .
\end{eqnarray}
The dependence on the cut-off angles $\gamma_{1}$ and $\psi_{1,2}$ is expected on physical grounds: the cut-off represents the deformation of the geometry near the singularities. These deformations will eventually show up in $\beta$ and $f_{a}$, when integrals over the non-singular compactification are performed. However, the axion decay constant is very insensitive to the angular coordinate cut-off; it  depends only logarithmically on $\gamma_{1}$ and $\psi_{1,2}$. As long as one is not exponentially fine-tuning the cut-off parameters to their singular values, the logarithmic expressions in $f_{a}$ will be of ${\cal O}(1)$. On the other hand, $R= (4 \pi g_{s} N )^{1/4}$. For parameters of
physical interest, one can assume $g_{s} \sim 0.1$ and
$1\lesssim     g_{s} N \lesssim 100$ 
such that $R \gsim  1$. Combining all these in $f_{a}$, we obtain the following expression for the axion decay constant:
\begin{eqnarray}
\label{ourf1b}
f_{a}  =   c\,   h_{0} m_{s} \, ,
\end{eqnarray}
where $c$ is a constant of order $1 - 10$, depending on the geometry of the throat. 
Recall that $h_{0} \, m_{s}$ is nothing but the physical mass scale at the bottom of the throat.
This indicates that the axion decay constant is controlled by the physical scale of the throat and is 
insensitive to the details of the bulk.

To obtain $f_{a}$ within the acceptable range, i.e. $10^{9}$ GeV $\lesssim f_{a}\lesssim 10^{12}$ GeV,
all one has to do is to construct a throat in the string theory compactification  with the physical
mass scale within this range. This is easily achieved in the light of recent progress in flux
compactifications \cite{Dasgupta:1999ss, Giddings:2001yu}.

The situation becomes more interesting in a multi-throat compactification.
One immediate conclusion of the result above is that in the multi-throat compactification,
the normalization of the axion field (or $B_{\mu \nu}$ field) zero mode is controlled by the longest throat
in the compactification.
To obtain axion decay of the right scale, one has to make sure that the physical mass scale of the
longest throat is within the range $10^{9}- 10^{12} $ GeV. In the case where the physical scales
of the throats are comparable, our formulation for calculating  $\beta$ indicates that
\begin{eqnarray}
\label{multi}
f_{a } =\left[ \sum_{i} c_{i}^{-2} h_{0 \, i}^{-2}  \right]^{-1/2}\,  m_{s}
\end{eqnarray}
where the sum is over all throats. Here
$h_{0\,  i}$ represents the warp factor at the bottom of the $i$-th throat and the
$c_{i}$ are constants of order unity depending on the construction of the corresponding throat. Thus, even in the situation where the physical mass scale for each throat is bigger than $10^{12}$ GeV, all throats contribute to the normalization of the axion
zero mode such that the sum in (\ref{multi}) can bring $f_a$ within the desired range.


\section{Constant Coupling Background}
\label{constantcoupling}
The previous examples we have studied give rise to large $\beta$ provided we impose a reasonable cut-off when compactifying the geometry. Two important aspects of our previous analysis were firstly that the dilaton remained  independent of the radial coordinate $r$, even though the background had non-trivial torsion, and secondly that the analysis of $f_a$ was insensitive to the cut-off. This was not surprising since local cut-offs in the geometry should not affect many of the global features of a system. However, it would be nice to construct a background with torsion in the 
heterotic theory that is compact from the beginning, and allows a dilaton that is (at least) independent of the radial coordinate $r$. A construction of such a background is sketched in \cite{Dasgupta:2008hb}, and although it was not explicitly verified that $f_a$ can be lowered to within the desired window for this background, it is believed that this should be possible.


\section{``Model-dependent" axions}
We have focussed on the construction of  a so-called ``model-independent'' axion. It is interesting to ask whether ``model-dependent'' axions with allowed values of $f_a$  can be constructed in our formalism. 
The answer seems to be affirmative,\footnote{We thank Joe Conlon for discussion on this issue.}but addressing the issue explicitly would require detailed knowledge of the cohomological and homological properties of the internal space.

To give an example, consider a model where the axion arises from the zero mode of the RR four-form potential $C_{(4)}$ in type IIB string theory (for example the background given by Ouyang \cite{Ouyang:2003df}).  To support such an axion we need a D7-brane on which
\begin{eqnarray}
\int C_{4} \wedge F\wedge F
\end{eqnarray}
is non-zero, where the integration is over the D7 worldvolume. It is assumed that  $F$ has legs along
the Minkowski coordinates, while $C_{4}(x^{\alpha}, y^m)$ has legs entirely along the compactified directions. $x^\alpha$ and $y^m$ denote the coordinates on the Minkowski and the internal spaces respectively.  This means that we are decomposing $C_4$ as 
\begin{equation}\label{cfour}
C_4 (x^\alpha, y^m) ~ = ~ \varphi(x^\alpha) \otimes h_4(y^m)
\end{equation}
where $h_4(y^m)$ is a harmonic four-form in the internal space and $\varphi(x^\alpha)$ is a scalar which will 
have axion-like couplings. Clearly, since the harmonic four-forms in the internal space are classified by the 
second Betti numbers $b_2$, there are $b_2$ axions from this decomposition. 
In the following we will choose $b_2 = 1$ to get a 
single axion for our case, but this is of course a model-dependent statement. 
It is easy to see that in terms of powers of the warp factor $h_w$, the kinetic energy
of the axion scales like
\begin{eqnarray}
\int \sqrt{g_{(6)}} \,  h_{w}^{2} \vert h_{4}\vert^{2} \, .
\end{eqnarray}
where $\vert h_{4}\vert$ is the magnitude of the harmonic four-form in the internal space. Using Hodge duality, this
can be mapped to the harmonic (1, 1) form, $h_2$, in the internal space. 
Now, computing $\beta$ using (\ref{beta}) we see that:
\begin{equation}\label{betnow}
\beta= \frac{\int d^{6}y \,  \sqrt{ g_{(6)}} \,
  h_{w}^{2} (y) \vert h_2\vert^2}{\int d^{6}y \,  \sqrt{g_{(6)}} \,  h_{w}^{2} (y)},
\end{equation}
which may be significantly bigger than one depending on the behaviour of the harmonic two-form in the internal space. 

Unfortunately, the exact form of $h_2$ for a CY (or non-CY) manifold is not known so we cannot make 
a more concrete statement.\footnote{To determine the harmonic form we need the metric of the internal space (say a CY). So far there is no known solution for the metrics of compact CY spaces.}To give a value of $\beta$ much greater than 1 we require the harmonic form to be peaked near the throat of our internal space, which again requires us to know the precise form for the $h_2$. However it seems clear that an internal space with the requisite $h_2$ forms could in principle be found.

This conclusion generalises to the zero modes of other ``model-dependent'' axions. The upshot is
that model-dependent axions look like scalars from a four-dimensonal point of view and, as one can easily check,  $\beta$ for a scalar is dependent on the warp factors as well as the magnitude of the harmonic forms in the internal space. Tuning these correctly could give a large 
$\beta$ for such compactifications although the result depends crucially on our knowledge of these harmonic forms (which 
is lacking at this stage). 

In contrast, the detailed topological properties of the internal space are not required for the model-independent axion. There is one and only one harmonic zero form in the internal space (which can be set to 1). As we have shown, it is relatively easy to construct compactifications for which $\beta$ is much greater than 1 in these models.  Thus for the model-independent axion there exist compactifications which result in values of the axion decay constant low enough to be allowed.

\section{Discussion and conclusions}
We have given a construction of the axion in warped heterotic backgrounds of the form
\begin{equation}\label{lajbat}
ds^2 = e^\phi ds^2_{AdS_5} ~ + ~ ds^2_{X^5},
\end{equation}
with non-trivial dilaton and torsion (where we are in the string frame). While other heterotic embeddings of the axion have been studied, they have been found to result in values of the axion decay constant $f_a$ outside the allowed range. Our backgrounds are warped Minkowski spacetimes with warp factor given by \eqref{wfac}. These are new non-K\"ahler manifolds, distinct from those studied in different contexts in \cite{Dasgupta:1999ss, Becker:2002sx, Becker:2003yv, Goldstein:2002pg, Becker:2003sh}.
Since we found that it is possible in these backgrounds to lower $f_a$ to within experimentally acceptable bounds, we can view these results as the construction of new phenomenologically motivated compact non-K\"ahler backgrounds in heterotic string theory. 

We gave one possible class of constructions of these backgrounds, which we believe to have smooth global metrics, using sigma-model identifications. It is interesting to note that the corresponding torsions can be expressed in powers of $\alpha'$ and satisfy the Bianchi identity in terms of the vector bundles and the torsional spin connections \cite{Dasgupta:2008hb}. A crucial feature of these backgrounds is that the dilaton is independent of the radial direction and therefore not quite proportional to the warp factor as in 
\cite{Strominger:1986uh, Dasgupta:1999ss, Becker:2002sx}.  A second class of non-K\"ahler manifolds with similar properties was also presented in \cite{Dasgupta:2008hb}, with the important difference that the torsion comes only from the Chern classes of the manifolds. These manifolds have been classified in the literature as 
either  ``conformally balanced''  or ``balanced'' manifolds. Our case was an example of a balanced 
manifold with a constant dilaton but non-trivial torsion. 

Our constructions focussed on the so-called ``model-independent" axion, constructed from
the zero mode of the NS-NS two-form potential $B_{\mu \nu}$. We have shown that the normalization of the zero mode of the $B_{\mu \nu}$ field gets most of its contribution from the highly warped regions of the  compactification. It is on this fact that our mechanism for lowering the axion decay constant $f_a$ to the desired range ($10^{9}- 10^{12} $ GeV) hangs. More specifically, we have shown that $f_{a}$ is given by the warped mass scale of the longest throat
in the compactification. Hence, the question of achieving an axion construction with a viable value of $f_a$ in this set-up is reduced to the question of constructing an AdS-type throat with mass scale in the desired $10^{9}-10^{12}$ GeV range. Heterotic compactification on any of the presented manifolds with appropriate torsion, dilaton and vector bundles should thus allow axions with permissible decay constants in a natural way. We have not checked that string or other non-QCD instantons in our models are suppressed  compared to QCD instantons. This condition must be met in order for the PQ mechanism to solve the strong CP problem, and can be a severe constraint on string models of the axion \cite{Svrcek:2006yi}.

 It may be possible to generalise our result to the case of  ``model-dependent'' axions, in order to give values of $f_a$ in the desired range in these models also. This would require precise knowledge of the harmonic two-forms  $h_2$ in the internal space, which would have to be peaked near the throat of our internal space.  Lacking such knowledge, we are unable to state whether or not this is the case. We leave the question of finding manifolds whose harmonic forms have the required properties for future work.

\newchapter{Cosmic Strings}
\label{chapter:CS}
\section{Introduction}
This and the following two chapters focus on string cosmology, and specifically on cosmic strings. Cosmic strings are macroscopic strings expected to have been produced during phase transitions in the early universe, which arise in grand unified theories as well as string theory. Macroscopic networks of strings are also predicted to be produced at the end of brane inflation. The tension of allowed cosmic strings is tightly constrained by measurements, so that viable string theoretic cosmic strings (cosmic superstrings) must have their tensions lowered somehow. This can be achieved in warped geometries, which is where brane inflation generally takes place. With Keshav Dasgupta and Hassan Firouzjahi, I studied the tension properties of 3-string or Y-shaped junctions in these geometries \cite{Dasgupta:2007ds}. These objects are extremely important in the evolution of cosmic string networks. This work is presented in Chapter \ref{lumps}.

Issues of stability for cosmic strings descending from string theory can be addressed by the use of wrapped higher-dimensional branes. In Chapter \ref{magneto} I present my investigation (with Stephon Alexander et al) of whether cosmic strings arising from string theory could generate primordial galactic magnetic fields \cite{Gwyn:2008fe} - fields that must have been present at the time of galaxy formation in order to explain the coherent magnetic fields of galactic scale observed today. We found that suitably wrapped M--branes, acting as strings in $3+1$-dimensions, can produce these fields, although the necessary construction is not generic.

This chapter is devoted to a review of cosmic strings and their properties, which should provide sufficient background material for Chapters \ref{lumps} and \ref{magneto}. The main reference is the book by Vilenkin and Shellard \cite{2000csot.book.....V}.
 
\section{Topological defects}
\subsection{Topological classification}
Cosmic strings are a type of topological defect formed during suitable phase transitions in the early universe. Topological defects arise in a range of physical systems upon spontaneous symmetry breaking to a non-trivial set of degenerate ground states.

With the physics of the early universe described by a theory which undergoes spontaneous symmetry breaking, the universe is expected to have gone through various phase transitions as it cooled. Cosmologically, domain walls and other topological defects are expected to have formed during these phase transitions in the particle physics regime of the early universe. The type of defect that forms is determined by the topology of the vacuum manifold, as pointed out by Kibble \cite{Kibble:1976sj} and summarised as follows.

We assume that the universe is correctly described by a gauge theory which undergoes spontaneous symmetry breaking (SSB) of its symmetry group G at some critical temperature $T_C$. In the hot Big Bang model of the universe, this phase transition will necessarily occur at some point in the very early universe as the temperature drops below $T_C$. 

The field $\phi$ undergoing SSB can be taken to have a minimum at $\phi = 0$ in the high temperature phase. Upon SSB, $\phi$ takes on a vacuum expectation value (vev) $\langle \phi \rangle$ in some orbit of G. $\langle \phi \rangle$ is invariant under transformations in $H \subset G$, the remaining symmetry group, so that the orbit may be identified with the coset space $ {\cal M} = \frac{G}{ H}$. ${\cal M}$ is the vacuum manifold: the space of degenerate low-energy states which break the original symmetry. 

During the phase transition, $\phi(\bf{x})$ will take a vev $\langle \phi \rangle$ in ${\cal M}$, which will be uncorrelated in different regions of space provided they are far enough apart, forming domains. For energetic reasons (because a spatial derivative term appears in the Hamiltonian) a constant or slowly varying $\langle \phi \rangle$ is preferred, and so many of these domains will join smoothly and average out their values of the vev.  However, depending on the topology of ${\cal M}$, some boundaries may survive between domains, where $\phi = 0$. These are defects: configurations of the higher-energy state which survive because they are topologically stable. Why this is so will be discussed in more detail for each type of defect. In brief, the dimension of allowed defects is determined by the nontrivial homotopy groups of ${\cal M}$, as pointed out by Kibble \cite{Kibble:1976sj}.\footnote{The homotopy groups of a manifold give information about its topology, via the types of cycles which can be defined on it. See for instance \cite{nak} for a thorough mathematical introduction.}This is summarised in the following table:

\begin{table}[htbp]
  \begin{center}
    \TableCaption{Classification of topological defects}
    \begin{tabular}{l | lll}
      \hline
      Defect & Dimension & Nontrivial homotopy group & ${\cal M}$\\
      \hline
      Domain walls & 2 & $\Pi_0({\cal M})$ & disconnected \\
      Cosmic strings & 1 & $\Pi_1({\cal M})$ & not simply connected  \\
      Monopoles & 0 & $\Pi_2({\cal M})$ &  \\
      \hline
    \end{tabular}
  \end{center}
\end{table}

We will use the general example given in \cite{Kibble:1976sj}. Consider an N-component real scalar field $\phi$ with Lagrangian
\begin{eqnarray}
\label{example}
{\cal L} & = & \frac{1}{2} ( D_\mu \phi)^2 - \frac{1}{8} \lambda^2 (\phi^2 - \eta^2)^2 + \frac{1}{8} {\mathrm Tr} (F_{\mu \nu} F^{\mu \nu} ),
\end{eqnarray}
where 
\begin{eqnarray*}
D_\mu \phi & = & \partial_\mu \phi - e A_\mu \phi;\\
F_{\mu \nu} & = & \partial_\nu A_\mu - \partial_\mu A_\nu + e [A_\mu, A_\nu].
\end{eqnarray*}
The Lagrangian is invariant under $O(N)$, and the number of vector fields $A_\mu$ is $\frac{1}{2} N ( N- 1)$. $O(N)$ is spontaneously broken to $O(N-1)$ at low temperatures, when $\phi$ acquires an expectation value\footnote{The temperature-dependent corrections to the potential, arising from one-loop diagrams, are given in \cite{Kibble:1976sj}:
\begin{eqnarray*}
V(\phi) & = & \frac{1}{8} \left ( \phi^2 - \eta^2 \right )^2 + \frac{1}{48} \left [ (N + 2) g^2 + 6 (N-1) e^2\right] T^2 \phi^2.
\end{eqnarray*} 
For $T>T_C$ the minimum is at $\phi = 0$. The critical temperature is given by cancellation of the mass term
\begin{eqnarray*}
T_C & = & \eta \left (\frac{N+2}{12} + \frac{N-1}{2} \frac{e^2}{g^2} \right) ^{-\frac{1}{2}},
\end{eqnarray*}
and below this temperature the minimum is given by
\begin{eqnarray*}
\phi^2 & = & \eta^2 \left [ 1 - \frac{T^2}{T_C^2}\right].
\end{eqnarray*}}
\begin{eqnarray}
\phi^2 & = & \eta^2 \left [ 1 - \frac{T^2}{T_C^2}\right].
\end{eqnarray}
The manifold of degenerate vacua is therefore an $N-1$-sphere:
\begin{eqnarray}
{\cal M} & = & \frac{O(N)}{O(N-1)} \, = \, S^{N-1}.
\end{eqnarray}
\subsection{Domain walls}
\begin{figure}[ht]
\centering
\subfigure{
\includegraphics[scale=0.4]{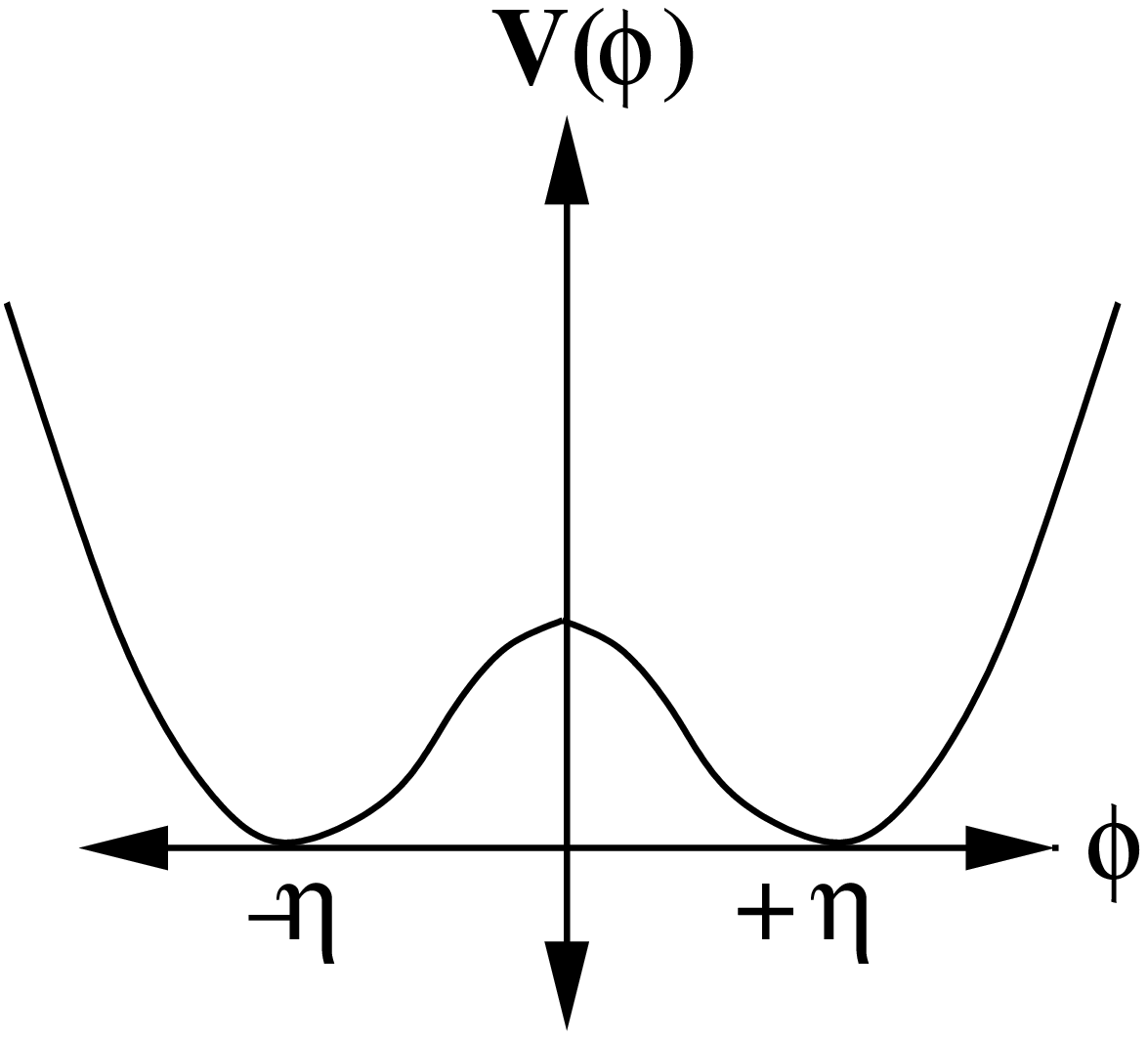}
\label{DW1}
}
\subfigure{
\includegraphics[scale=0.3]{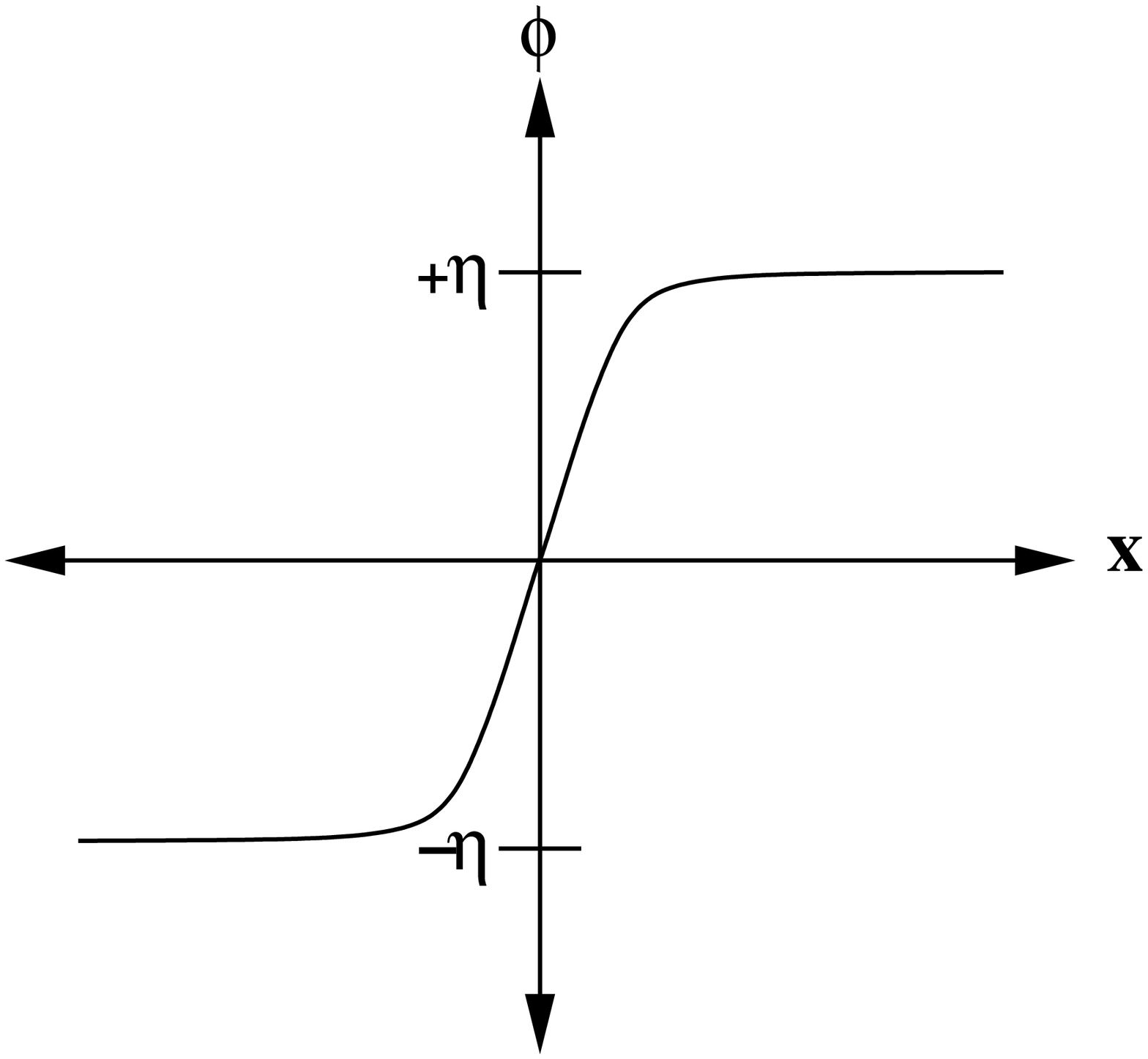}
\label{DW2}
}
\caption[Domain wall vacuum manifold]{The potential and vacuum manifold for $N=1$ in (\ref{example}), which results in a domain wall.}
\label{DW}
\end{figure}

Domain walls are two-dimensional defects, which are formed when $\Pi_0({\cal M})$ is nontrivial. $\Pi_0({\cal M})$ is not strictly speaking a homotopy group, representing merely the number of connected components of ${\cal M}$. In other words, domain walls form when the vacuum manifold is disconnected, which is the case for (\ref{example}) when $N=1$. In that case the Lagrangian only possesses a discrete reflection symmetry, and the vacuum manifold ${\cal M} = S^0$ is made up of two points: $\langle \phi \rangle \,  =  \, \pm \eta$ (see Figure \ref{DW}). 

$\langle \phi \rangle$ may take on either of these values in different regions of space. Consider two such regions, idealised as (three-dimensional) cells, characterised by $\langle \phi \rangle \,  = \, \phi_1$ and $\langle \phi \rangle  \, = \, \phi_2$ respectively, with a common boundary surface (here we generalise the $ \langle \phi \rangle \, = \, \pm \eta$ vacuum to a case with more than two points). There is a spatial derivative term in the energy so if there were a continuous path in ${\cal M}$ joining $\phi_1$ and $\phi_2$ then it would be energetically favourable for the width of the boundary to expand, leaving a smoothly varying $ \langle \phi \rangle $ rather than a sharp boundary. However, when $\phi_1$ and $\phi_2$ are disconnected no smooth transition is possible. As shown in Figure \ref{DW} one cannot pass from $ \langle \phi \rangle  \, = \, \eta$ to $\langle \phi \rangle  \, = \, - \eta$ without passing through a region where $\langle \phi \rangle  = 0$. A wall of this high-energy phase will thus be formed between the two cells and is stable against decay for topological reasons. This solution is effectively a soliton.

\begin{figure}[htp]
\centering
\subfigure[Schematic of magnetic domains in a ferromagnet.]{
\includegraphics[scale=0.28]{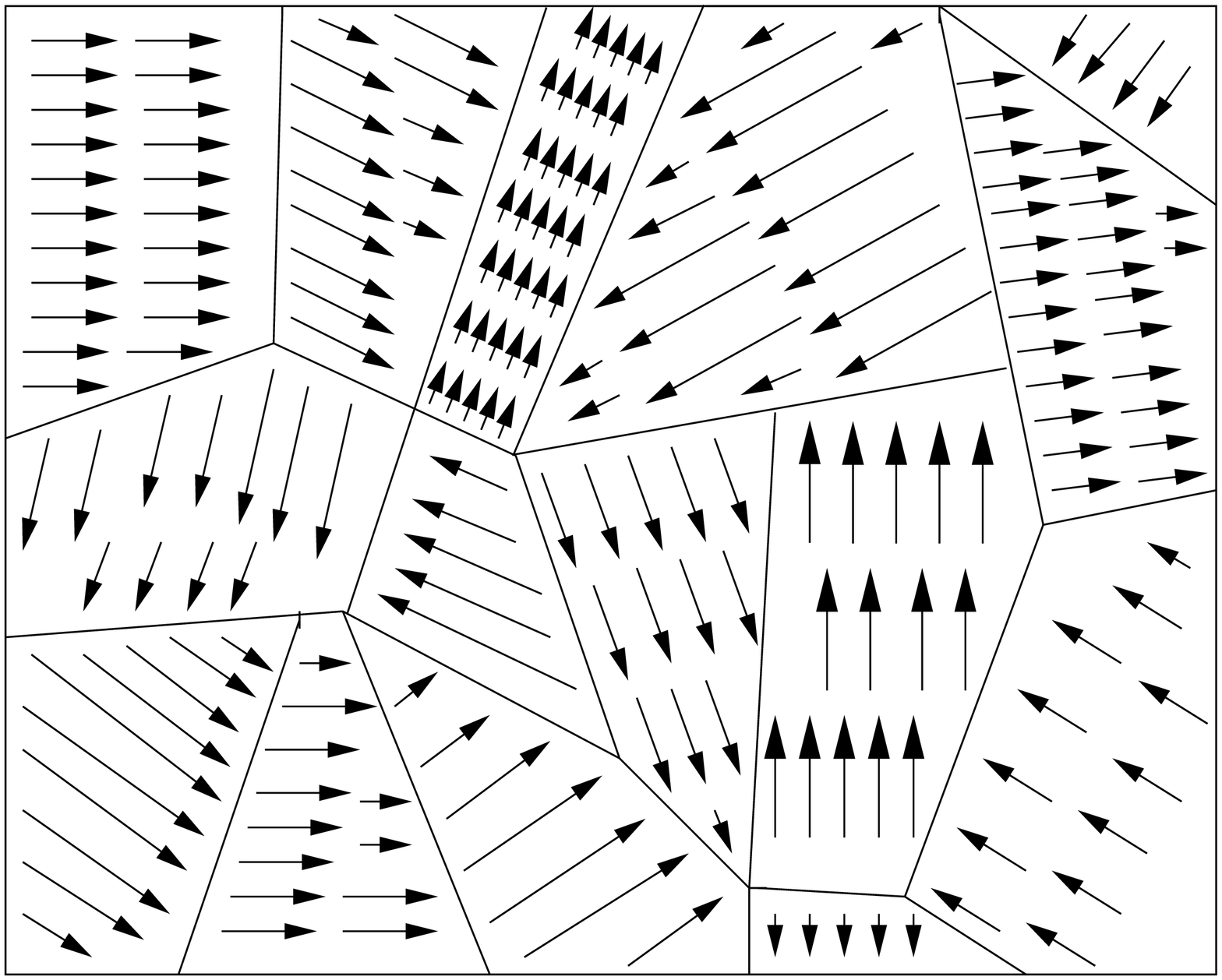}
\label{fig:dw}}
\subfigure[Photomicrograph of NdFeB, showing the magnetic domain structure.]{
\includegraphics[scale=0.5]{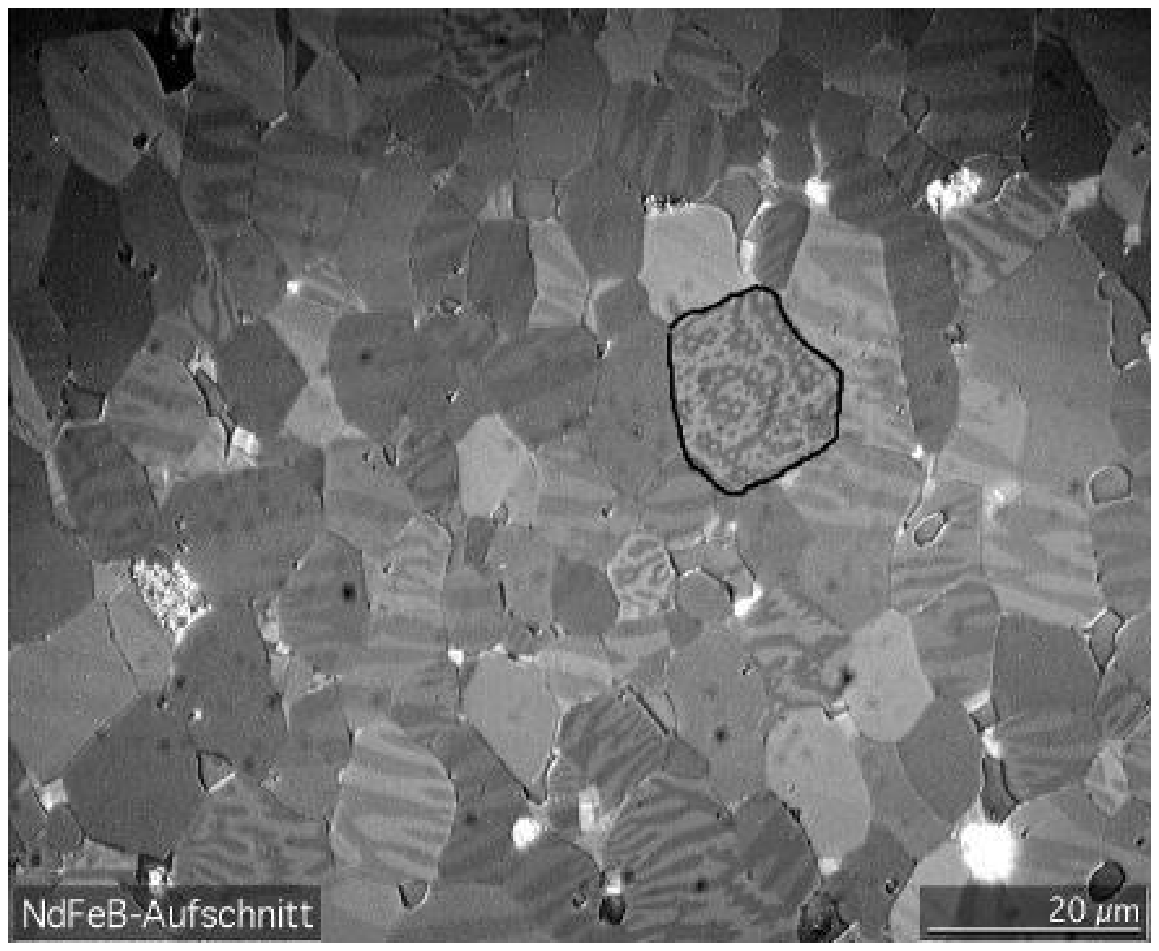}\label{wiki}}
\caption[Domain walls in a ferromagnet]{The domain walls in a ferromagnet are the boundaries of magnetic domains. Figure \ref{wiki} is from the Wikimedia commons and is used under the GNU Free Documentation License.}
\label{dw}
\end{figure}

A familiar example of domain wall formation is given by ferromagnets. In a ferromagnet, the magnetic dipole moments of individual electrons are randomly oriented at temperatures above $T_C$, the Curie temperature. As the ferromagnet is cooled below this temperature, domains of alignment are formed, breaking the rotational symmetry of the unordered state. Domain walls appear at the boundaries between these domains, as shown in Figure \ref{fig:dw}.\footnote{It must be kept in mind however that there are differences between condensed matter and cosmological domain walls. For instance, it is energetically favourable (in the absence of an applied magnetic field) for a large ferromagnet to break up into domains with different alignments because of the long-range dipole-dipole interaction. No such constraint applies cosmologically: cosmological domain walls occur because of non-equilibrium effects and the limits on correlation length set by causality \cite{Kibble:1976sj}.} 

The cosmological evolution of a domain wall is determined by its surface tension. The structure will grow with time until it is comparable to the Hubble scale, leading to huge inhomogeneities in the background radiation which are not observed. If domain walls are predicted by a model to be produced in the early universe they must  disappear by recombination, otherwise the model is ruled out by observation \cite{Kibble:1976sj, Zeldovich:1974uw}.

\subsection{Vortices - Cosmic strings}
Consider three or more domains which meet at an edge (this is just the smallest number of domains needed to define an edge). Each corresponds to some value of $\langle \phi \rangle$; we can characterise the domains by $\phi_1, \phi_2$ and $\phi_3$ respectively. We take $\phi_1, \phi_2$ and $\phi_3$ as belonging to the same connected component of ${\cal M}$, so that no pair of cells is separated by a domain wall. We can then define a smooth path as we go around the edge, from $\phi_1$ to $\phi_2$ to $\phi_3$ and back to $\phi_1$.

If this path in ${\cal M}$ can be continuously deformed to a point, all three domains can fuse such that $\langle \phi \rangle$ over the whole region is smoothly varying. However, if there exist closed paths on ${\cal M}$ which cannot be shrunk to a point, a vortex will be defined, as in a superfluid or a superconductor. This is called a cosmic string: cosmic strings are one-dimensional defects which form when the first homotopy group of the vacuum manifold $\Pi_1({\cal M})$ is non-trivial. $\Pi_1({\cal M})$ gives the homotopy classes of closed paths in ${\cal M}$. 

\begin{figure}[htp]
\centering
\subfigure{
\includegraphics[scale=0.35]{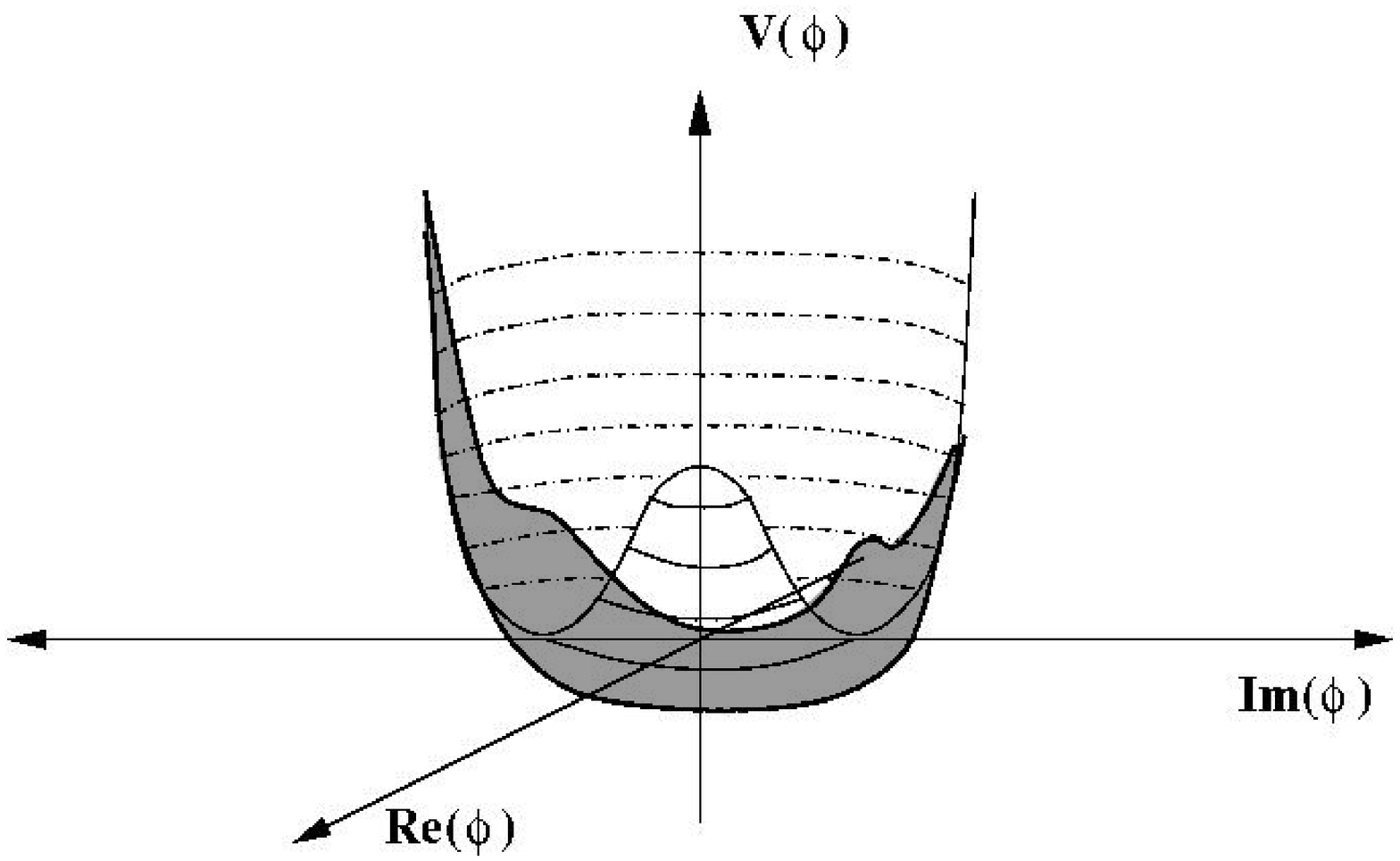}
\label{mexhat}
}
\subfigure{
\includegraphics[scale=0.25]{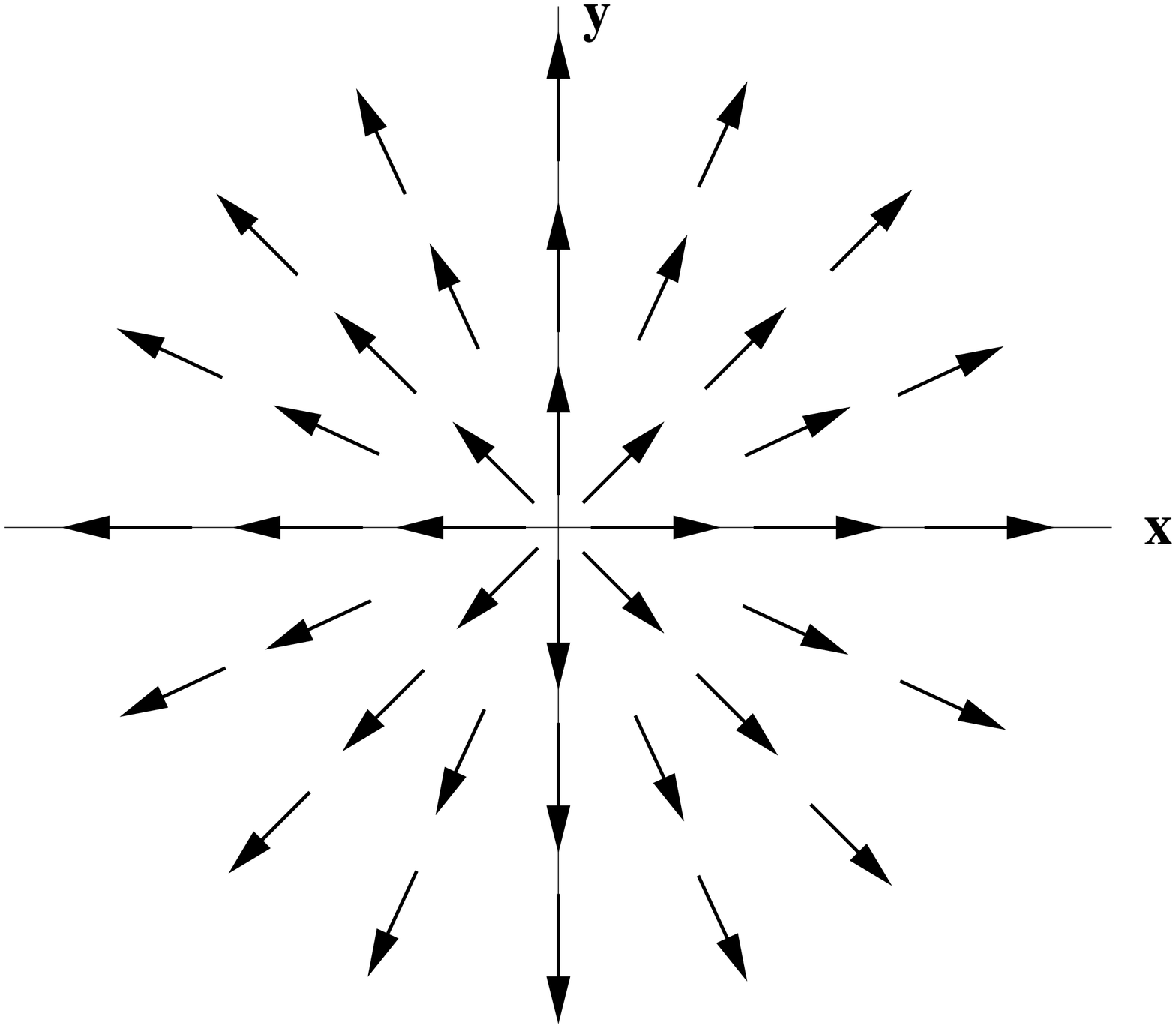}
\label{axsym}
}
\label{CS}
\caption[Cosmic string vacuum manifold]{The potential and vacuum manifold for $N=2$ in (\ref{example}), which results in a cosmic string.}
\end{figure}

This corresponds to the $N=2$ case in (\ref{example}), in which ${\cal M} = S^1$. The potential that results is often called the Mexican hat potential, and is shown in Figure \ref{CS}. We see that cosmic strings form when an axial or cylindrical symmetry is broken. The Nielsen-Olesen string \cite{Nielsen:1973cs} is an example of a vortex solution. 


\subsection{Monopoles}

Monopoles are pointlike defects which form when the second homotopy group of the vacuum manifold is nontrivial, i.e. when $\Pi_2({\cal M}) \neq 1$. We can visualise this topology by considering a vertex at which four cells meet, characterised by vevs $\phi_i$, where $i = 1,..,4$. Again we assume that there are no domain walls between them, so that a  closed surface around the vertex can be constructed without leaving ${\cal M}$. If this surface can be shrunk to a point, $\Pi_2({\cal M}) $ is trivial and monopole defects are not possible. If not, a pointlike defect will form at the vertex. This corresponds to the $N=3$ case and the 't Hooft-Polyakov monopole solution \cite{'tHooft:1974qc, Polyakov:1974ek}. Thus monopoles form when a spherical symmetry is broken.

Monopoles, like domain walls, are ruled out in cosmological theories for observational reasons. They are extremely massive and are usually predicted to be produced in such numbers that they would overclose the universe.\footnote{This is the case for local monopoles, which can form when a gauge symmetry is broken \cite{Preskill:1979zi}. By contrast, global monopoles have a total energy which increases with distance, so that long-range correlations are possible. Monopole pairs can find each other and annihilate and global monopoles are thus cosmologically safe \cite{Bennett:1990xy}.} Inflation can help to solve this problem by exponentially diluting the monopole population. 

\section{Cosmic strings}
\subsection{Phase transitions in the early universe}
In a cosmological context, one has to consider symmetry breaking in a grand unified theory at some point in the early universe, whether in string theory or some other theory. The basic premise of grand unified theories is that the known symmetries descend from a larger (not yet known) group $G$. The grand unified group $G$ will be broken down somehow to the $SU(3) \times U(1)_{\mathrm{em}}$ subgroup of our world:
\[G \rightarrow H \rightarrow ... \rightarrow SU(3) \times SU(2) \times U(1) \rightarrow SU(3) \times U(1)_{\mathrm{em}}. \]

This chain of phase transitions will generically result in the production of different topological defects at different times. Monopoles  and domain walls might overclose the universe and be inconsistent with Big Bang cosmology, but cosmic strings are likely to have formed and may have played an important role in cosmological evolution. 

 Cosmic strings are produced as a string network via the Kibble mechanism. Upon cooling past some critical temperature $T_c$ (corresponding to a time $t_c$) the field undergoing spontaneous symmetry breaking at a point ${\bf x}$ in space relaxes to the minimum of its potential, acquiring a vacuum expectation value $\langle\phi(\bf x)\rangle = v$ in the vacuum manifold $\cal M$. The phase will be chosen randomly for points in space separated  by more than some correlation length $\xi$, where $\xi$ is bounded above by the Hubble radius at time $t_c$ (by causality) but is typically much smaller. Specifically, if matter is in thermal equilibrium before the transition, then the initial  correlation length at the time of the phase transition is a microphysical  scale, the so-called {\it Ginsburg length} \cite{Kibble:1976sj, Kibble:1981gv}. When the vacuum manifold is not simply connected, cosmic strings will be produced in a network with correlation length $\xi$. The evolution of the string network characterised by this parameter is discussed below. 

\subsection{Network evolution}

During a phase transition, a network of cosmic strings will form with a characteristic length scale comparable to $\xi$. This correlation length gives both the typical curvature radius of the strings as well as the typical distance between neighbouring strings. As the universe expands, so will the correlation length $\xi(t)$. When long strings intersect themselves or each other, they will intercommute, producing loops. Thus the string network can be separated into the so-called infinite strings (strings with 
curvature radius larger than the horizon at the time $t$) and a distribution of string loops with radii $R$ smaller than $t$. Both loops and long strings emit gravitational radiation. Detection of this radiation could provide evidence of their presence, and is discussed below in Section \ref{gravrad}. Sufficiently small loops will radiate all their energy away and decay. This means that an initially dense string network will be diluted as the strings chop each other up and the resulting loops decay. 

Dilution of the network via loop production and decay will result in an increase in correlation length $\xi(t)$, However, by causality the correlation length $\xi$ can never grow larger than $t$, since this would imply the presence of correlations in the position of the field in the vacuum manifold over lengths greater
than the distance light could have travelled. The rate at which the strings can chop each other off into loops is thus limited by the speed of light. This means that either the network approaches a scaling solution in which $\xi$ remains a fixed fraction of $t$ or it grows more slowly, in which case ${\xi(t)}/{t}$ decreases. In the latter case strings would eventually come to dominate the total energy density of the universe.

\begin{figure}[htp]
\centering
\includegraphics[scale=0.5]{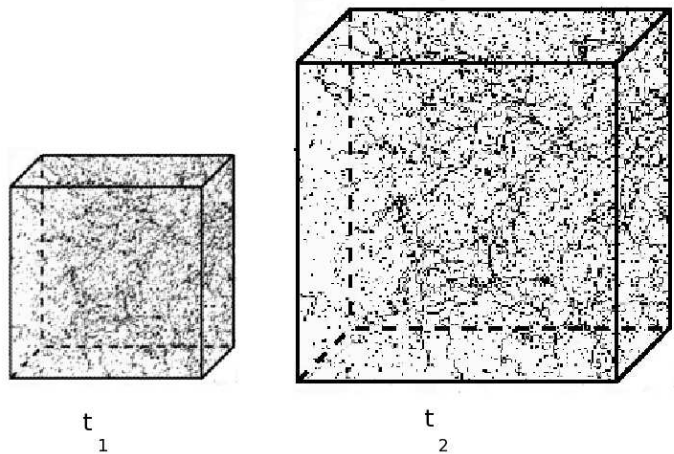}
\caption[Scaling solution]{The string network will reach a scaling solution in which $\xi(t) \sim t$.}
\label{network}
\end{figure}

It has been verified using numerical string network evolution simulations \cite{Albrecht:1984xv, Albrecht:1989mk, Bennett:1987vf, Bennett:1989yp, Allen:1990tv} that instead the distribution of infinite strings 
will converge to a so-called scaling solution in which ${\xi(t)}/{t}$ is independent of time and the string density is constant relative to the rest of the radiation and matter energy density in the universe. It is called a scaling solution because, scaled to the Hubble radius, the string network looks the same at all times. The string properties, such as $\xi(t)$, are proportional to the time passed \cite{2000csot.book.....V,Hindmarsh:1994re,Brandenberger:1993by}, as shown schematically in Figure \ref{network}.

\section{Cosmological implications of cosmic strings}
\subsection{Structure formation}
\label{struct}
Cosmic strings carry energy and can seed gravitational instabilities, the starting points for structures like galaxies to form \cite{Battye:1997hu, Avelino:1997hy}.
Density fluctuations that can give rise to such instabilities are not hard to generate in the early universe, since they will be produced by any phase transition. However, it is difficult to produce fluctuations on sufficiently large scales because of the limit set by causality in the early universe. Cosmological phase transitions are predicted to occur very early in the radiation era ($t < 10^{-4}$ seconds \cite{Kibble:1976sj}) and can only affect comoving scales up to a few parsecs, so any fluctuations produced during phase transitions are generally insufficient to explain fluctuations on megaparsec scales. Inflation is one way out of this difficulty: all perturbations (quantum fluctuations of the inflaton) originate in a causally connected region in this model. Until they were ruled out, cosmic strings were also under active investigation as sources for structure formation, since the scaling property of cosmic string networks avoids the causality constraint.

Structure formation from cosmic strings was first suggested in 1980 and 1981 by Zeldovich \cite{Zeldovich:1980gh} and Vilenkin \cite{Vilenkin:1981iu}.\footnote{See Vilenkin and Shellard \cite{2000csot.book.....V} for a detailed discussion of the prospects for structure formation from cosmic strings.}The key was the use of GUT scales, which gave rise to the right galactic mass scales and galactic evolution. The critical temperature $T_C$ at which the phase transition occurs is determined by the symmetry-breaking scale $\eta$, which is related to the mass per unit length of the resulting cosmic string by $\mu \sim \eta^2$. Then 
\begin{eqnarray}
\label{gravint}
G \mu & \sim & \frac{\eta^2}{M_{Pl}^2}
\end{eqnarray}
is the dimensionless constant characterising the strength of gravitational interaction of the strings. G is Newton's constant. GUT-scale strings  with $\eta \sim 10^{16}$ GeV give $G\mu \sim 10^{-7}$, which is of the right order for seeding density perturbations. 

Investigations of structure formation by cosmic strings in the early to mid-eighties assumed the `old picture' of string evolution, in which strings were believed to be smooth on the horizon scale, with loops being chopped off on this scale and having small initial velocities.\footnote{This picture was verified by an early simulation \cite{Albrecht:1984xv}.}This model assumed that the energy in the string network was primarily carried by loops, which were treated as stationary accretion centres. The metric around a static straight string is locally flat, which means that test particles in the region around the string will experience no gravitational force, while in contrast a closed loop will act like any other object with mass ($M(R) \approx 2 \pi R \mu$ where $R$ is the loop radius) and attract a test particle gravitationally. The loops decay via gravitational radiation, leaving clumps of matter that can act as seeds for galaxies and galactic clusters. 

This model looked promising, and was explored in several papers during the nineteen eighties \cite{Vilenkin:1983jv, Turok:1983sp, Turok:1985zz, Turok:1985tt, Sato:1985rt}. 
It was shown that not only did GUT string loops have the right mass to  explain galaxy formation, but they could also produce the observed galaxy-galaxy correlation function \cite{Turok:1984cn, Turok:1985tt}. Loops were shown to be in one-to-one correspondence with bright galaxies.

\begin{figure}[htp]
\centering
\includegraphics[scale=0.55]{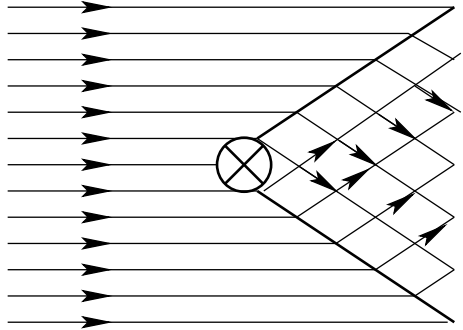}
\caption[Matter accretion onto wakes]{Matter accretes onto the wake formed behind a moving string \cite{2000csot.book.....V}.}
\label{wake}
\end{figure}

However, detailed simulations towards the end of the eighties \cite{Bennett:1987vf, Bennett:1989ak, Albrecht:1989mk, Bennett:1989yp, Allen:1990tv} made it clear that several of the model's assumptions were flawed: strings are neither straight nor motionless, with long strings possessing a wiggly structure that results in far more loops being chopped off than was previously assumed.\footnote{This effect is more marked during the radiation than the matter era, because the rate of expansion is comparatively slower in the radiation era and strings are stretched out less. A scaling solution exists in both cases. On the level of numerical simulation, this means that a one-parameter model (that parameter being the correlation length $\xi$) of the string network is no longer sufficient. We will not go into any detail on the simulations used here.}  Loops are formed on scales much smaller than the horizon, with larger initial velocities and larger number densities than previously thought. Fast-moving loops result in elongated structures which are unlikely to seed galaxies and, moreover, the energy density in the string network was shown to be primarily in the long strings. The accelerated decay mechanism means that more long strings must be present in order to preserve a scaling solution. It had already been pointed out that wakes will form behind long strings with non-negligible velocities \cite{Silk:1984xk}, with consequences for structure formation since matter will accrete onto the planar wakes of moving strings, as shown in Figure \ref{wake} \cite{Vachaspati:1986gf, Stebbins:1987cy}.

Thus it was realised that any contribution to structure formation by cosmic strings must arise from this accretion process in the wakes behind them \cite{Stebbins:1987cy}. Shorter and thinner wakes produced by strings with a lot of small-scale structure are termed filaments \cite{Brandenberger:1993by}.  

Since string network evolution is a nonlinear numerical problem, it soon became clear that making detailed predictions was difficult, although a scaling solution is still believed to exist.  This fact is what makes cosmic strings a compelling candidate for primordial magnetogenesis, as discussed in Chapter \ref{magneto}. It must be noted that the arguments for a scaling solution are quite general, and independent of which of the above string network scenarios is applicable. 

Although the number of papers on cosmic strings waned during that decade, cosmic strings remained viable candidates for generators of the primordial density perturbations needed to seed structure until the late nineties. They were then ruled out as a primary source of these perturbations by the data from COBE, BOOMERanG and WMAP, which showed a gaussian scale-invariant power spectrum consistent with adiabatic perturbations from inflation being the dominant source of primordial density fluctuations \cite{Contaldi:1998qs, Bouchet:2000hd, Durrer:2001cg, Peiris:2003ff, Komatsu:2003iq, Pogosian:1999np,  Pogosian:2003mz, Komatsu:2008hk}.\footnote{A review of the rise and fall of cosmic strings as potential generators of primordial density perturbations is given in \cite{Perivolaropoulos:2005wa}. All the key references can be found there.}Density fluctuations produced by cosmic strings are strongly non-gaussian. It was shown that cosmic strings (or other topological defects) cannot account for more than 10\% of the density perturbations in the early universe.  Specifically, the CMB temperature power spectrum can be modelled as a superposition of adiabatic gaussian perturbations arising from inflaton and cosmic string perturbations \cite{Perivolaropoulos:2005wa}:
\begin{eqnarray*}
C_l & = & A C_l^{\mathrm{adiabatic}} + BC_l^{\mathrm{strings}},
\end{eqnarray*}
where the predicted forms of $C_l^{\mathrm{adiabatic}}$ and $C_l^{\mathrm{strings}}$ are known. The observed data is well fit by a purely adiabatic spectrum \cite{Jaffe:2000tx, Spergel:2003cb}, and gives $B=0$ at 90\% confidence levels \cite{Pogosian:2003mz}. A small contribution from strings is permitted, but $B\leq 0.09$ at the 99\% confidence level \cite{Pogosian:2003mz, Pogosian:2004ny, Wu:2005apb}. 

Thus what initially seemed a promising place to look for signs of the role of cosmic strings in the universe's evolution led to a strong constraint on cosmic strings once the data was improved. Cosmic strings did \em not \em play a large role in structure formation, but there remain other possible signals of their presence, discussed below. 
\subsection{Gravitational waves}
\label{gravrad}
Having ruled out cosmic strings as the primary model for structure formation, interest in the subject waned considerably. However, it remains possible, even likely, that cosmic strings with small $G\mu$ were formed in the early universe. Detection of them would be interesting because of the light they could shed on the particle physics regime of the early universe, and so other signals of their presence are worth exploring.

The first is the production of gravitational radiation by cosmic strings. A cosmic string network should result in a stochastic gravitational wave background \cite{Vilenkin:1981bx, Hogan:1984is}, which could interfere with pulsar timing measurements. Such interference has not yet been observed, and gives rise to a constraint on the tension of cosmic strings, given in Section \ref{constraints} below.
\subsection{Gravitational lensing}
The space around a cosmic string is conical (although the metric is locally flat, as mentioned above) with deficit angle $\Delta = 8 \Pi G \mu$. This means that the cosmic string, if aligned correctly, will act as a gravitational lens, providing the earth-bound observer with a double image of visible objects behind the string \cite{Vilenkin:1981zs, Vilenkin:1984ea, Vilenkin:1986cb, 1985ApJ...288..422G}. Massive objects which can act as gravitational lenses, such as galaxies and black holes,  produce an odd number of distorted images of unequal brightness. In contrast, images due to a straight string (or a straight segment of a string) should be identical. The typical angular separation between the images is comparable to the conical deficit angle. 

Recently there was some excitement over a report of two seemingly identical galaxies observed not far from each other \cite{Sazhin:2003cp}. These were later shown to be a pair of similar galaxies rather than images of the same galaxy \cite{Agol:2006fb, Sazhin:2006fe}. However, the possibility remains that gravitational lensing events by cosmic strings will one day be observed.

\subsection{Observational constraints}
\label{constraints}
For convenience we collect here the bottom line of various observational constraints on cosmic strings. These can all be expressed as constraints on $G\mu$. The more detailed CMB data which ruled out  density fluctuations produced by cosmic strings as the primary seeds for large scale structure formation also gave a stringent constraint on the string tension \cite{Wyman:2005tu, Fraisse:2005hu, Bevis:2007gh, Battye:1998xe}, using WMAP and SDSS (Sloan Digital Sky Survey) data. 
\begin{eqnarray}
\label{CMB}
G \mu & \leq & 2 \times 10^{-7}.
\end{eqnarray}
This constraint is tightened by pulsar timing measurements, which give an upper bound \cite{Vachaspati:1984gt, Lommen:2002je, Jenet:2006sv}
\begin{eqnarray}
G \mu \leq 10^{-8},
\end{eqnarray}
although this result is highly sensitive to the distribution of cosmic string loops which is not known exactly.

\section{Cosmic strings from string theory}
\label{csstring}
In the early literature it was assumed that macroscopic cosmic strings arising as topological defects in grand unified field theories and microscopic strings arising in superstring theory were entirely distinct. Most obviously, the relevant energy scale for cosmic strings is the GUT scale or less, while for fundamental strings it should naively be the Planck scale, leading to an inadmissibly high tension for these strings (using (\ref{gravint})). Furthermore, they were believed to be unstable either to breakage or confinement by domain walls (this is the axionic instability discussed in greater detail in Section \ref{axion}).
In fact, either in warped geometries \cite{Randall:1999ee} or in models with large extra dimensions \cite{ArkaniHamed:1998rs, ArkaniHamed:1998nn}, the tension can be lowered substantially. This became clear after the construction of viable brane inflation models \cite{Dvali:1998pa, Burgess:2001fx, Dvali:2001fw, Jones:2002cv, Kachru:2003sx, Iizuka:2004ct} in string theory. In these models, inflation is driven, for instance, by the attraction between a brane-antibrane pair.  The inflaton is provided by the distance between the branes, which feel a gravitational attraction, so that inflation ends when the branes collide (annihilating in the case of brane-antibrane systems). At the same time, a tachyonic field takes on a vacuum expectation value, randomly chosen over causally disconnected regions. This results in the formation of topological defects \cite{Jones:2003da, Dvali:2003zj, Sarangi:2002yt, Pogosian:2003mz, Majumdar:2002hy} via a mechanism explained in \cite{Bergman:2000xf}. As argued by \cite{Sarangi:2002yt}, the defects can only be cosmic strings. Brane inflation and annihilation take place in the IR region of a Klebanov-Strassler (KS) throat \cite{Klebanov:2000hb}, which is where the strings are produced. Their tensions are compatible with observation \cite{Firouzjahi:2005dh, Wyman:2005tu, Shandera:2006ax}. Moreover, stable or metastable cosmic strings arising from F-strings and D-strings or wrapped D-, NS- and M-branes have also been constructed \cite{Copeland:2003bj, Leblond:2004uc}.
In fact, the cosmic strings of string theory come in many more varieties than those of a GUT and with a richer spectrum \cite{Copeland:2003bj, Polchinski:2005bg, Firouzjahi:2006vp}, as we will discuss in Chapter \ref{lumps}.

This opened the possibility of a new observational angle on string theory models, and led to a renewal of interest in the subject - see, for instance, Kibble \cite{Kibble:2004hq}. The annual count of papers on Spires-HEP with ``cosmic strings" in the title rose to 2000 levels again in 2005 and has remained there at around 30. (In the decade between 1986 and 1995 the number was around 50.) The spectrum and production rate of different cosmic strings are highly model-dependent, so any cosmological input from observation might potentially constrain viable string models, in the same way that observational input to inflation is doing.  
In the following chapter we explore the properties of cosmic superstring networks in these set-ups.

\newchapter{Cosmic strings in warped geometries}
\label{lumps}
\section{Introduction}
As discussed in Section \ref{csstring}, cosmic superstring production is predicted to take place at the end of brane inflation in string theory models. Several kinds of strings are possible in string theory, from the fundamental F-strings to D-strings, otherwise known as D1-branes. The latter can also be achieved by wrapping higher dimensional Dp-branes on $(p-1)$-cycles in the internal manifold. In addition, $(p,q)$-strings can form. These are bound states of $p$ F- and $q$ D-strings which do not always intercommute but instead can form junctions.  Nontopological defects called semilocal strings are also possible. A spectrum of possible string tensions results. It is clear that such richness adds a great deal to the resulting cosmic string networks, and requires careful study in order for predictions and comparisons with observation to be made. Some steps have been taken towards studying $(p,q)$-strings in warped backgrounds, the natural habitat of brane inflation models. In \cite{Dasgupta:2007ds} we took this further, investigating 3-string junctions, beads and semilocal strings in Klebanov-Strassler (KS)-type throats. We were able to describe all of these objects using D3-branes wrapped on cycles of varying dimension.

This chapter follows the presentation of \cite{Dasgupta:2007ds}. We begin by reviewing the basic properties of F-, D-, $(p,q)-$ and semilocal strings. We then discuss in turn $(p,q)$-strings, three-string junctions, cosmic beads and necklaces, all in the throat. It turns out that all these solitons can be studied using wrapping modes of D3-branes. For example, when a D3-brane with appropriate electric and magnetic fluxes on its worldvolume wraps a two-cycle in a KS throat, it will appear as a $(p,q)$-string in 4-dimensional Minkowski spacetime. A three-string junction in the throat can similarly be constructed using a combination of three suitably wrapped D3-branes. Beads, monopoles on which strings can end, can be constructed using D3-branes wrapping the three-cycle in the KS throat. These can therefore also be used in the construction  of cosmic necklaces.  We end with a discussion of the cosmological implications of these results. 

\section{Strings produced in brane inflation}
\subsection{F- and D-strings}
Brane inflation \cite{Dvali:1998pa, Burgess:2001fx, Dvali:2001fw, Jones:2002cv, Kachru:2003sx, Shandera:2003gx, Iizuka:2004ct} has been suggested as a string realisation of early universe inflation. The canonical example is a brane-antibrane pair in some realistic vacuum (for example the KKLT vacuum \cite{Kachru:2003aw}). The inflaton is given by the brane separation, which decreases because of an attractive potential between the branes. Inflation ends upon collision and annihilation of the brane-antibrane pair. During this process both F-strings (fundamental string theory strings) and D-strings (one-dimensional D1-branes) are formed \cite{Sarangi:2002yt, Majumdar:2002hy, Jones:2003da, Dvali:2003zj, Pogosian:2003mz}.

The D-string is simply the D1-brane, with tension
\begin{eqnarray*}
\mu_D & = & \frac{\mu_F}{g_s},
\end{eqnarray*}
where $g_s$ is the string coupling, $g_s = e^{\phi}$. In cases where the dilaton is not constant, the tension is given by this relation with the value of the dilaton at the position of the brane. This chapter and most of the literature on cosmic superstring networks produced at the end of brane inflation focus on Type IIB string theory, in which both the F- and D-strings exist, as well as strings arising from suitably wrapped odd-dimensional D-branes and NS5-branes. Wrapped NS-branes are classical solitons like the GUT cosmic strings of the previous chapter.  In Chapter \ref{magneto} we explore heterotic strings arising from wrapped M-branes in heterotic M-theory. 
\subsection{$(p,q)$-strings}
$(p,q)$-strings are bound states of $p$ F-strings and $q$ D-strings. They form when F-strings and D-strings intersect, and are produced as part of a network of strings at the end of brane inflation. They are generically expected because F- and D-strings are not independent, and are not produced independently. In brane-antibrane annihilation, it is the breaking of the $U(1) \times U(1)$ gauge group of the branes that gives rise to the strings. The tachyon stretching between the branes couples to one $U(1)$ factor only (a linear combination of the original $U(1)$s), generating abelian strings (D-strings) when it takes a symmetry-breaking vev. The remaining $U(1)$ factor becomes confining (since no $U(1)$ remains after the branes annihilate), and the resulting flux tubes can be identified as F-strings \cite{Tye:2005fn, Bergman:2000xf}. In fact, because of the $SL(2, \mathbb{Z})$ symmetry of Type IIB string theory \cite{Hull:1994ys, Witten:1995ex}, it does not make sense to consider only one or the other. $B_2$, the NS-NS two-form under which the fundamental F-strings are charged, and $C_2$, the Ramond-Ramond 2-form under which the D-strings are charged, transform as a doublet under $SL(2,\mathbb{Z})$. In other words F- and D-strings are S dual to each other. The more general object is thus the $(p,q)$-string, which has $p$ units of F-string charge and $q$ units of D-string charge \cite{Schwarz:1995dk, Schwarz:1995du}.

Furthermore, the $(p,q)$-strings do not necessarily intercommute, but can form new bound states (depending on the energetics) when they intersect. This results in a rich network of different strings, with different tensions. The tension of a $(p,q)$-string in a flat background is given by \cite{Schwarz:1995dk, Schwarz:1995du}
\begin{eqnarray}
\label{flatjxn}
T_{(p,q)} & = & \frac{T_{F1}}{g_S} \sqrt{(p - C_0q)^2 g_s^2 + q^2},
\end{eqnarray}
where $C_0$ is the axion. In this notation, F-strings are $(1,0)$-strings, and D-strings are $(0,1)$-strings. $(p,q)$-strings are stable when $p$ and $q$ are coprime \cite{Schwarz:1995dk, Schwarz:1995du}; see also \cite{Witten:1995im}.\footnote{Two integers are coprime if they contain no common factor other than 1.}The sign of $p$ or $q$ refers to the orientation of the string, and by convention is positive for a string pointing into a vertex.  
Three $(p,q)$-strings can meet at a junction (in flat space) provided \cite{Jackson:2004zg}
\begin{eqnarray}
p_1 + p_2 + p_3 & = & q_1 + q_2 + q_3 = 0;\\
\cos \, \theta_{ij} \, =  \,\cos \, \theta_{\mathrm{crit}}& = & \frac{e_i \cdot e_j} {|e_i| |e_j|},\label{flat}
\end{eqnarray}
where $\theta_{ij}$ is the angle between strings $(p_i, q_i)$ and $(p_j, q_j)$, and $e_i = ([p_i - C_0 q_i] g_s, q_i)$. (\ref{flat}) is just the condition that the three strings lie on a plane, i.e. that there is no unbalanced force on the vertex. 

\begin{figure}[htp]
\centering
\includegraphics[scale=0.8]{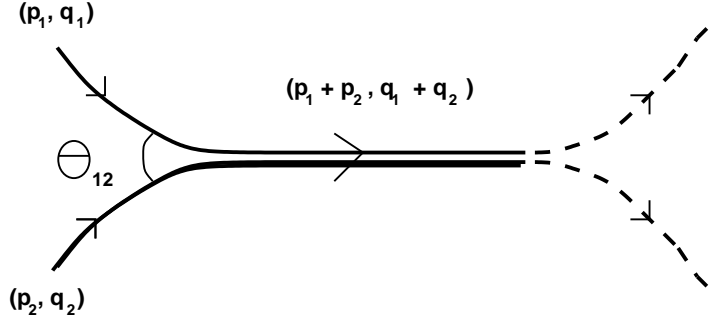}
\caption{Formation of a $(p_1 + p_2, q_1 + q_2)$-string upon intersection of two $(p,q)$-strings.}
\label{int}
\end{figure}

When two strings collide, they will form a bound state with tension depending on the size of $\theta_{12}$. If it is small enough (smaller than $\theta_{\mathrm{crit}}$), the bound state with $p = p_1 + p_2$ and $q = q_1 + q_2$ can form, is energetically favourable, and is the final state. One can think of this process as being due to a force to the left on the three-string junction - see Figure \ref{int}.

\begin{figure}[htp]
\centering
\includegraphics[scale=0.6]{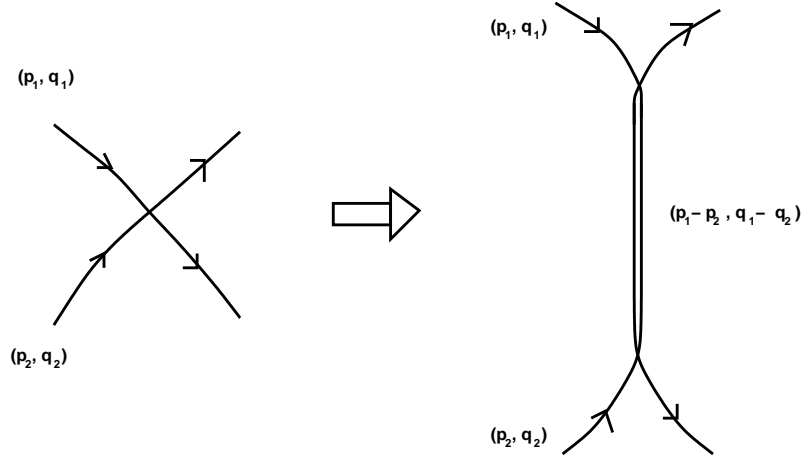}
\caption{Formation of a $(p_1 - p_2, q_1 - q_2)$-string upon intersection of two $(p,q)$-strings.}
\label{minus}
\end{figure}

In the case that $\theta > \theta_{\mathrm{crit}}$, the $(p_1, q_1)$-string and $(p_2, q_2)$-string will pull away from each other, and a $(p_1 - p_2, q_1 - q_2)$ bound state will form instead, as in Figure \ref{minus}. 

What results is a multi-tension string network with 3-string junctions.\footnote{It is also possible that 4-string and higher vertices could form. These can be constructed from wrapped branes and are referred to as baryons because of their role in the gauge/gravity dictionary. Their production is suppressed, however, because of their high mass \cite{Jackson:2004zg}.}

\subsection{Semilocal strings}
Semilocal strings \cite{Vachaspati:1991dz, Achucarro:1999it} are non-topological defects that are nevertheless stable for a long period of time. They form when both local and global symmetries are present, in particular when $\Pi_1 \left( \frac{\mathrm{G}_l}{\mathrm{H}_l} )\right ) $ is nontrivial (the subscript $l$ referring to only the gauged parts of the initial and final symmetries) but $\Pi_1 \left ( \frac{\mathrm{G}}{\mathrm{H}} \right )$ is trivial.\footnote{This also means that a sudden emergence of global symmetries may render an unstable defect stable for some time. For details on this see \cite{Preskill:1992bf, Preskill:1992ck}.}

In 2004 it was shown that semilocal strings can be produced at the end of brane inflation \cite{Urrestilla:2004eh, Dasgupta:2004dw}. Their cosmological properties are discussed in \cite{Achucarro:1998ux, Achucarro:1997cx, Urrestilla:2001dd}. These strings satisfy the current CMB constraints and appear not to pose any cosmological problems. In fact the formation rate of semilocal strings at weak gauge coupling is generically almost one third that of cosmic strings \cite{Achucarro:1997cx}, since they are short and end on monopoles. Only when they combine to form a long string can they have large-scale density perturbations; otherwise they will shrink to zero size. 

Semilocal strings can also be constructed in the throat. In subsequent sections we present constructions of $(p,q)$-strings, three-string junctions and monopoles in the throat using wrapped D3-branes. To construct a 4-dimensional $(p,q)$-string, a D3-brane with appropriate fluxes must be wrapped on a 2-cycle in the warped compactification geometry, while to construct a bead on which strings can end a suitable D3-brane must wrap a 3-cycle. Semilocal strings can be constructed by wrapping D3-branes on zero-cycles in the throat, giving rise to D3s in 4-dimensions. In the presence of D7-branes, semilocal defects can form on these D3s, given certain conditions. This construction was discussed in detail in \cite{Dasgupta:2007ds} but is not discussed any further here. 
\section{$(p,q)$-strings in the throat}
In most models of it, brane inflation takes place in a highly warped region of the compactification geometry, often called the throat. In order to give a realistic embedding of brane inflation, one needs a complete compactification scheme which stabilises all the moduli and allows the relevant branes to exist in the compactification space. This was first achieved by KKLMMT \cite{Kachru:2003sx}; see also \cite{Firouzjahi:2003zy} and \cite{Burgess:2004kv}. As was shown in \cite{Firouzjahi:2006vp}, this warping will affect the tension of the resulting $(p,q)$-strings. For our purposes we will use the warped Klebanov-Strassler throat \cite{Klebanov:2000hb}, reviewed in Chapter \ref{throat}.

 The metric in the tip can be written as in (\ref{metric0}):
 \begin{eqnarray}
 ds^2 & = &  h^2\, \eta_{\mu \nu} dx^{\mu} dx^{\nu}+
  g_s M \alpha'(d\psi^2 +\sin^2 \psi\, d \Omega_2 ^2),
 \end{eqnarray}
 where \[h = H ^{-\frac{1}{4}}( \tau \rightarrow 0) = \epsilon^{\frac{2}{3}} 2^{-\frac{1}{6}} a_0^{- \frac{1}{4}}(g_s M \alpha')^{-\frac{1}{2}}.\] $\psi$ is the usual polar co-ordinate in an $S^3$, and ranges from $0$ to $\pi$.

M gives the number of units of Ramond-Ramond fluxes $F_3$ turned on inside this $S^3$.
The two-form associated with $F_3$ is given by (\ref{C2}) \cite{Firouzjahi:2006vp}:
\begin{eqnarray}
C_2=  M \alpha' \,  \left(\psi-\frac{\sin (2\psi)}{2} \right) \sin \theta\, d \theta \, d\phi \, .
\end{eqnarray}

\subsection{D3-brane construction of $(p,q)$-strings in the throat}
\label{pq}
The $(p,q)$-string can be constructed using a D3-brane with $p$ units of electric and $q$ units of magnetic flux, wrapped on an $S^2$ inside the $S^3$ at the bottom of the throat \cite{Firouzjahi:2006vp}.\footnote{These D3 branes do not slip out of the three-cycle because of the stabilising RR background fluxes that, via the underlying Myers effect \cite{Myers:1999ps}, give a finite radius at the stable value $\psi_{\mathrm{min}}$: $r^2 \sim \sin^2 \psi$ - see \cite{Thomas:2006ud, Bachas:2000ik}. A derivation of the tension $T_{(p,q)}$ using the dielectric brane method is given in \cite{Firouzjahi:2006xa}.}The position of the $S^2$ is given by $\psi$; the D3 can be taken to be extended along the $x^0$ and $x^1$ directions and wrapping $\theta$ and $\phi$ in the throat. To the 4-dimensional observer it will appear as a 1-dimensional object extended along $x^1$. 

The action of the D3-brane is
\begin{eqnarray}
\label{action1}
S_{D3} &= &
  -\frac{\mu_3}{g_s}\int d^4x\, \sqrt{ -|g_{ab} + {\cal F}_{ab} | } + \mu_3 \int \left ( C_2\, \wedge {\cal  F}  + \frac{1}{2} \lambda^2 C_0\, {\cal F} \wedge{\cal F} \right).
\end{eqnarray}
Here $\mu_3$ is the D3-brane charge, $\lambda= 2\pi \alpha'$, and ${\cal F}_{ab} = B_{ab} + \lambda F_{ab}$. The metric is given by (\ref{metric0}). In the following we will take $C_0 = 0$, since it is zero to leading order in the KS solution.

To induce the necessary units of flux for a $(p,q)$-string in 4-dimensions, we require the flux configuration 
\cite{Firouzjahi:2006vp}
\begin{eqnarray}
\label{charges}
F_{\theta \phi} & = & \frac{q}{2};\\
\nonumber \tilde F^{01}& = & -\frac{p}{4\pi}, 
\end{eqnarray}
where $\tilde F^{01}=  \frac{\delta S}{\delta F_{01}}$ is dual to $F_{01}$.

This can be seen by recalling that the D-string charge comes from the equation of motion for $C_2 = C_{01}$ (see \cite{Firouzjahi:2006vp}): 
\[ Q_{D1}^{\mathrm{induced}}  =  \int_{S^2} \frac{\delta S_{D3}}{\delta C_2} = \mu_3 \int_{S^2} {\cal F},\] and noting that $2 \pi \lambda \mu_3 = \mu_1$, the D1-charge. As argued in \cite{Firouzjahi:2006vp}, the induced charge will be quantised if the pullback $B_2$ is either quantised or vanishes on $S^2$, which we take to be the case. Similarly, the F-string charge is given by integrating the EOM for $B_2 = B_{01}$: \[Q_{F1}^{\mathrm{induced}}  = \int_{S^2} \frac{\delta S_{D3}}{\delta B_2} = \int_{S^2}\frac{1}{\lambda}\frac{\delta S_{D3}}{\delta F_{01}} = - \int_{S^2} \frac{1}{\lambda} \tilde F^{01} = p.\]
Note that the dilaton has been set to zero so $\mu_1 = \mu_{F1} = \frac{1}{2 \pi \alpha'} = \frac{1}{\lambda}$. Using (\ref{charges}) as the only non-vanishing components of F and integrating out $\theta$ and $\phi$ gives the 2-dimensional D-string action
\begin{eqnarray}
\label{action2}
S_{(p,q)}= \int dt\, dx^1\, \left(-\Delta \sqrt{ h^4 - \lambda^2 F_{01}^2} +
\Omega\, F_{01} 
\right),
\end{eqnarray}
where $\Omega$ and $\Delta$ are defined by \cite{Firouzjahi:2006vp}
\begin{eqnarray}
\label{Omega}
\Omega \equiv \lambda \mu_3 \int_{S^2} C_{2} = \frac{M}{\pi}\, (\psi - \frac{1}{2} \sin\, 2\psi  )
\end{eqnarray}
and
\begin{eqnarray}
\label{delta}
\Delta &\equiv& \mu_3 g_s^{-1} \int_{S^2} \sqrt{g_{\theta \theta} g_{\phi \phi} + \lambda^2 F_{\theta \phi}^2 }\nonumber\\
&=&\lambda^{-1} \sqrt{\frac{ M^2}{\pi^2}\, \sin^4 \psi +\frac{q^2}{g_s^2} },
\end{eqnarray}
and $F_{01}, F_{\theta \phi}$ are now scalars. 
(\ref{action2}) is thus the action of a $(p,q)$-string extended along $x^1$. The D-string charge $q$ is encoded in $\Delta$ and the electric charge $p$ is given by the U(1) gauge field on the D-string worldvolume ($F_{01}$).

The Hamiltonian of the cosmic string is easily found, and gives its tension:
\begin{eqnarray}
\label{H}
\nonumber {\cal H} & = & pF_{01} - {\cal L}\\
\nonumber {\cal H} & = & \frac{h^2}{\lambda} \sqrt{ (p - \Omega)^2 + \Delta^2 \lambda^2}\\
{\cal H}& = &  \frac{h^2}{\lambda}\, \sqrt{ \frac{q^2}{g_s^2} + \frac{ M^2}{\pi^2}\, \sin^4 \psi 
+ \left( p - \frac{M}{\pi} (\psi -\frac{\sin 2\psi}{2}) \right)^2 }.
\end{eqnarray}
In going from the first to the second line we used the fact that $ p = \frac{\Delta \lambda^2 F_{01}}{\sqrt{h^4 - \lambda^2 F_{01}^2}} + \Omega$. 
Minimising the tension with respect to the string position gives the stable configuration:
\begin{eqnarray}
\label{psi}
\frac{\delta {\cal H}}{\delta \psi } = 0 & \Rightarrow & \psi= \frac{\pi\, p}{M},
\end{eqnarray}
 with the Hamiltonian at this point given by
\begin{eqnarray}
\label{E}
\left . {\cal H} \right |_{\frac{\delta {\cal H}}{\delta \psi } = 0}= \frac{h^2}{\lambda}\, \sqrt{ \frac{q^2}{g_s^2} + \frac{ M^2}{\pi^2}\,
 \sin^2 \left(  \frac{\pi p}{M}\right) }.
\end{eqnarray}
This was the result of \cite{Firouzjahi:2006vp}. The formula reduces to other known expressions in the appropriate limits: 
\begin{itemize}
\item When either $p$ or $q$ is taken to zero, (\ref{E}) reduces to the correct formula for F- or D-strings in the warped deformed conifold background. The relevant expressions are given in \cite{Herzog:2001fq, Hartnoll:2004yr, Gubser:2004qj}:\footnote{In fact $T_{F1}$ should have an additional prefactor of $b=0.93$. Here and elsewhere we have made the approximation $b \approx 1$.}
\begin{eqnarray*}
T_{F1} & \approx &\frac{h^2}{2 \pi \alpha'} \frac{M}{\pi} \sin \left (\frac{\pi p}{M} \right ); \\
T_{D1} & \approx & \frac{h^2}{2 \pi \alpha'} \frac{q}{g_s}.
\end{eqnarray*}
\item In the limit $ M \rightarrow \infty$, the radius of the finite $S^3$ goes to infinity and we should regain the flat background in which $h=1$. Then we expect to recover (\ref{flatjxn}), which is indeed the case.
\item The F-string tension vanishes when $p=M$. This is consistent with the interpretation of the F-strings in the dual $SU(M)$ gauge theory as flux tubes connecting quarks and antiquarks. M quarks make up a baryon in this theory, on which it is possible for F-strings to terminate. Their tension should then vanish, as implied by (\ref{E}). This means the fundamental strings are charged in $\mathbb{Z}_M$, which has interesting cosmological consequences (discussed below).  
\end{itemize}
\section{Three-string junctions in the throat}
\subsection{Three-string junctions}
\begin{figure}[htp]
\centering
\includegraphics[scale=0.5]{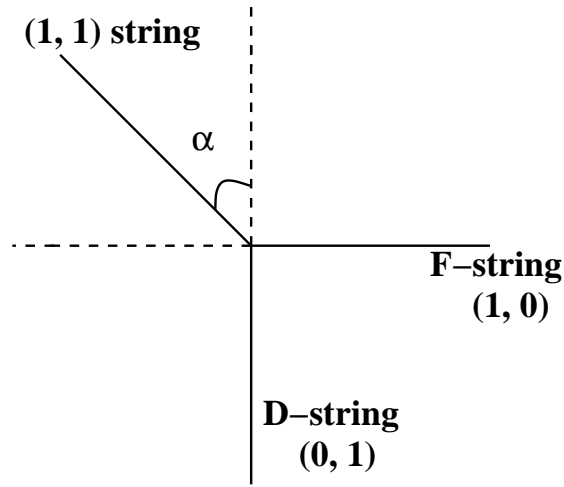}
\caption{An F-string ending on a D1-brane results in a 3-string junction.}
\label{DM}
\end{figure}

The three-string junction in a flat background was studied in \cite{Dasgupta:1997pu, Rey:1997sp, Sen:1997xi} by analysing the SUSY conditions for an F-string to end on a D1-brane. The three-string junction predicted by Schwarz \cite{Schwarz:1995dk, Schwarz:1995du} and shown in Figure \ref{DM} results, with tension given by (\ref{flatjxn}) and
\begin{eqnarray}
\tan \alpha & = & {g_s},
\end{eqnarray}
where $\alpha$ is the angle between the D-string and the bound state as shown in the figure.

In the following we construct general $(p,q)$-string junctions in the throat from a different perspective, using wrapped D3-branes with appropriate fluxes. It would be interesting to do the analogous SUSY analysis in this case also. 

\subsection{D3-brane construction of a three-string junction in the throat}
\begin{figure}[htp]
\centering
\subfigure[A three-string junction as seen by a 4-dimensional observer.]{
\includegraphics[width=2in]{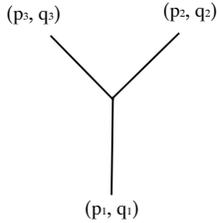}
\label{fig:subfig1}
} \hspace{2cm}
\subfigure[The three-string junction from the wrapped-brane viewpoint.]{
\includegraphics[width=2in]{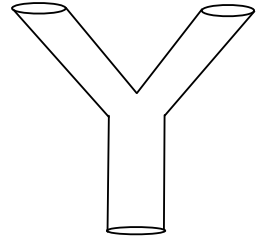}
\label{fig:subfig2}
}
\label{string}
\caption[Schematic of the three-string junction]{A schematic picture of the three-string junction. Figure \ref{fig:subfig2} shows the higher-dimensional construction.}
\end{figure}

Following the construction of $(p,q)$-strings using wrapped D3-branes with appropriate fluxes given in Section \ref{pq}, we construct a three-string junction using three such strings, as shown in Figure \ref{string}.

We assume that the three strings are coplanar in the $(X^1,X^2)$-plane and that the junction is stationary. This means neglecting time-dependent quantities in the action and is only a valid assumption under certain conditions, which we shall derive. Each string is a wrapped D3-brane with suitable fluxes labelled by $(p_i, q_i)$, $i = 1,2,3$ and worldvolume action given by  
\begin{eqnarray}
\label{actionD3}
S_{D3} &= &
  -\frac{\mu_3}{g_s}\int d^4x\, \sqrt{ -|g_{ab} + {\cal F}_{ab} | } + \mu_3 \int \left ( C_2\, \wedge {\cal  F}  + \frac{1}{2} \lambda^2 C_0\, {\cal F} \wedge{\cal F} \right),
\end{eqnarray}
where once again we take $C_0 = 0$. All three strings wrap an $S^2$ parametrised by $\theta$ and $\phi$ and whose position in $S^3$ is given by the angle $\psi$. For now, we assume that they wrap the same $S^2$, i.e. that the $\psi_i$ are identical. We consider the more general case in Section \ref{necklace} below. 

Each string has worldsheet coordinates  $(\tau, \sigma)$ and spatial coordinates $X_i^s$ in the $(X^1, X^2)$-plane, where $s=1,2$ and $i$ is the string index. We may choose $\tau=X^0$, where $X^0$ is the space-time time coordinate, and $-\infty < \sigma < \infty$. It is assumed that the strings do not perturb the flat background metric and fields.

Proceeding as before, with $a,b$ in (\ref{actionD3}) replaced by $0$ and $\sigma$ and using the formalism of \cite{Copeland:2006if, Copeland:2006eh}, we find
\begin{eqnarray}
\label{S1}
S= \sum_i \int  d \tau \, d \sigma \,
 \left(-\Delta_i \sqrt{ h^4 X_i'^2 - \lambda^2 { F_{0\sigma}^i} ^2} + \Omega_i\, F_{0\sigma}^i  \right)
\theta(-\sigma)
  \nonumber\\
+ \int d \tau\, \left[  f_i . (X_i-\bar X) + g_i (A_i  -\bar A)  \right],
\end{eqnarray}
where prime denotes a derivative with respect to $\sigma$.
This is found by evaluating the determinant
\begin{eqnarray}
|g_{ab} + \lambda F_{ab}| & = & \left | \begin{array}{cc}- h^2 & \lambda F_{0\sigma} \\ - \lambda F_{0 \sigma} & h^2 X_i'^2 \end{array} \right |.
\end{eqnarray}
The $\theta$ function implies that the strings are extended for $-\infty < \sigma <0$, meet at $\sigma=0$ and vanish at $X=\bar X$ for $\sigma >0$. The last two terms in the above action contain Lagrange multipliers $f_i$ and $g_i$ which enforce the constraints that the strings meet at $X=\bar X$, with equal gauge field potential 
$A_i=\bar A$. In this notation $F_{i\,  \tau \sigma}= \partial_\tau A_{i\,  \sigma}- \partial_\sigma A_{i\, \tau} $. We choose the gauge where $A _ {i\, \sigma} =0$ and to simplify the notation we set the convention 
$A_{i\,  \tau} \equiv  A_i$. The definitions of $\Omega_i$ and $\Delta_i$ are given by (\ref{Omega}), (\ref{delta}) and (\ref{psi}), replacing $q$ by $q_i$ and $\psi$ by $\psi_i$. 

\subsubsection{Charge conservation}
To find the charge constraints on the three-string junction, we evaluate the equations of motion. 
The equation of motion for the fields $X_i^s$ is
\begin{eqnarray}
\label{eom1}
\partial_\sigma \left(  \frac{ h^4\Delta_i {X^s_i}' }{  \sqrt{h^4 {X_i'}^2 - \lambda^2 {A_i^{\prime} }^2 } } \theta(-\sigma) 
\right)= -f_i^s \, \delta(\sigma),
\end{eqnarray}
which yields the following boundary condition at $\sigma=0$:
\begin{eqnarray}
\label{bc1}
f_i^s =  \frac{ h^4 \Delta_i {X^s_i}' }{ \sqrt{ h^4{X_i'}^2 - \lambda^2 {A_i^{\prime} }^2 } }. 
\end{eqnarray}
Similarly the equation of motion for the $A_i$ is
\begin{eqnarray}
\partial_\sigma \left(  \frac{\lambda^2 \Delta_i A_i' }{ \sqrt{ h^4{X_i'}^2 - \lambda^2 {A_i^{\prime} }^2 } } 
\theta(-\sigma) - \Omega_i \theta(-\sigma)  \right)= g_i \delta(\sigma),
\end{eqnarray}
with the boundary condition
\begin{eqnarray}
\label{bc2}
 g_i= -\frac{\Delta_i  A_i' \lambda^2 }{ \sqrt{h^4 {X_i'}^2 - \lambda^2 {A_i^{\prime} }^2 } } 
 +\Omega_i.
 \end{eqnarray}
Variation with respect to the Lagrange multipliers $f_i$ and $g_i$ imposes the conditions  $X_i=\bar X$ and $A_i=\bar A$ at $\sigma=0$. Variation with respect to $\bar X$ and $\bar A$
respectively implies that 
\begin{eqnarray}
\label{fg}
\sum_i f_i= \sum_i g_i = 0\, .
\end{eqnarray}

Note that the electric charge $p_i$ on the worldvolume of each $(p,q)$-string is defined
by $p_i =\delta {\cal L}/ \delta F_{i\, \tau \sigma}$ which gives 
\begin{eqnarray}
\label{p}
p_i=g_i =-\frac{\Delta_i  A_i' \lambda^2 }{ \sqrt{h^4 {X_i'}^2 - \lambda^2 {A_i^{\prime} }^2 } } 
 +\Omega_i \, .
 \end{eqnarray}
Thus (\ref{fg}) is just the condition for electric charge conservation at the junction ($\sigma= 0$):  $\sum_i p_i=0$.

Similarly, the magnetic charges $q_1$ are conserved at the junction, To see this, consider switching on  a hypothetical background $C_{2}$ field along the $(\tau,\sigma)$ directions. This induces a Chern-Simons term $C_2 \wedge {\cal F}$ for each string and we obtain the following extra contribution to the action
\begin{eqnarray}
S_{CS}=  \sum_i  q_i\int  d \tau \, d \sigma \,  \partial_\sigma X_i^s  C_{\tau s} \theta(-\sigma), 
\end{eqnarray}
where $\partial_\sigma X_i^s$ arises because of a co-ordinate transformation taking $C_{\tau, \sigma}$ to $C_{\tau, s}$, $s$ being the direction of alignment of the $i$th string in the $(1,2)$-plane. 
In ten dimensions, which one may consider as the bulk,  
the theory is invariant under the gauge transformation 
$C_{2\, \tau \sigma} \rightarrow C_{2\, \tau \sigma} +
 \partial_\sigma \Lambda$, where $\Lambda$ is a scalar. For a string with no boundary, i.e. closed strings or infinite strings, the string action is invariant under this gauge transformation. 
 However, in our model
 the strings terminate at $\sigma=0$. Upon this gauge transformation 
 on $C_{2}$, the string actions gain a surface term 
 \begin{eqnarray}
\delta S_{CS}= \sum_i q_i \int d\tau d\sigma \partial _\sigma X_i^s \partial_s \Lambda
 = 2  \sum_i q_i \int d\tau \Lambda(\sigma=0)\, .
 \end{eqnarray}
To keep the action invariant under this gauge transformation we therefore require that
$\sum_i q_i=0$, which is the D-string charge conservation condition.

 \subsubsection{Tension constraints}
The Hamiltonian of the system is
\begin{eqnarray}
\label{H1}
{\cal H } &=& \sum_i p_i F_{i\, \tau \sigma} - {\cal L}\nonumber\\
&=& \sum_i f_i^s {X^s_i}',
\end{eqnarray}
which gives the energy of the system when integrated over $\sigma$: $E=\int d\sigma {\cal H}$ . In terms of the string tensions $T_i^s$, the energy of the system is given by  $E= \sum_i \int dX^s  T_{i\, s} $. Using the above expression for the Hamiltonian, one obtains 
\begin{eqnarray}
\label{Ti}
T_i= f_i= \frac{ h^4 \Delta_i\, {X_i}' }{ \sqrt{ h^4{X_i'}^2 - 
\lambda^2 {A_i^{\prime} }^2 } }  \, .
\end{eqnarray}
Furthermore, from (\ref{fg}) we conclude that 
\begin{eqnarray}
\label{balance}
\sum_i T_i=0\, .
\end{eqnarray}
Geometrically, this means that the vector
sum of the string tensions vanishes. This is indeed the necessary condition for the string junction point to be stationary \cite{Rey:1997sp}.

 \subsubsection{Geometrical Interpretation}
An interesting geometrical interpretation of these results is found by defining each string's direction by the unit vector  $\hat{n}_i=\frac{X_i'}{|X_i'|}$ and substituting the expression (from (\ref{p}) for the gauge field, 
\begin{eqnarray}
\lambda A_i'= \frac{h^2(p_i- \Omega_i ) | X_i'|}{ \sqrt{ (\lambda  \Delta_i)^2 +  (p_i- \Omega_i )^2 } },
.\end{eqnarray}
into (\ref{Ti}) to rewrite the string tensions as
\begin{eqnarray}
\label{Ti2}
 T_i 
= \frac{h^2}{\lambda}\, \sqrt{ \frac{q_i^2}{g_s^2} + \frac{ M^2}{\pi^2}\,
 \sin^2 \left(  \frac{\pi p_i}{M}\right) }\, \hat{n}_i ,
\end{eqnarray}
which is in agreement with (\ref{E}). Now consider the coordinate system where $p$ and $q$ represent the horizontal and the vertical axes respectively as in Figure \ref{figconstraints}:
\begin{figure}[htp]
\centering
\includegraphics[scale=0.6]{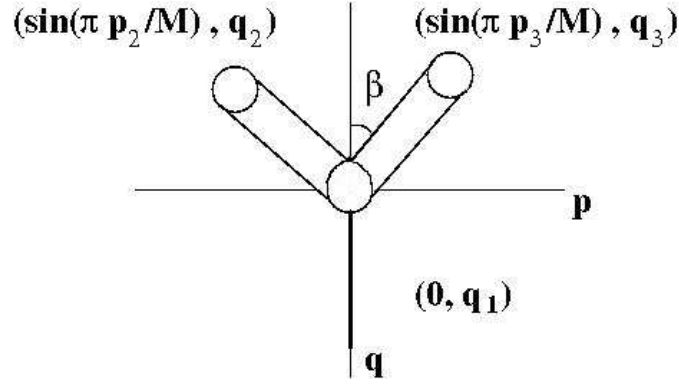}
\caption[Constrained three-string junction]{Our constrained three-string junction; (\ref{const}) implies that one string must be a D- or $(0,q)$-string.}
\label{figconstraints}
\end{figure}
Then each string's direction in this coordinate system is given by\footnote{Some constants omitted in \cite{Dasgupta:2007ds} have been restored.}
\begin{eqnarray}
\label{ni}
\hat{n}_i= \frac{h^2}{\lambda |T_i|} \left(  \frac{ M}{\pi}\,  \sin \left (  \frac{\pi p_i}{M} \right)  \, , \, \frac{q_i}{g_s} \right).
\end{eqnarray}
Defining $\beta_i$ as the angle between the $i$-th string and the $q$-axis, one obtains
\begin{eqnarray}
\label{angle}
\tan \beta_i = \frac{ M g_s  }{\pi q_i}\,   \sin \left(  \frac{  \pi p_i}{M} \right)  .
\end{eqnarray}
In the limit where $M\rightarrow \infty$, this reduces to the result for the flat background
\cite{Dasgupta:1997pu}.
For the junction point to be stationary  
one obtains from Eqs (\ref{balance}), (\ref{Ti2}) and (\ref{ni}) that 
\begin{eqnarray}
\sum_i \sin ( \pi p_i/M ) = 0 \quad \quad ; \quad \quad \sum_i q_i =0
.\end{eqnarray}
The second equation is automatically satisfied due to D-string \ conservation. Combining the first equation with the electric charge conservation $\sum_i p_i=0$ implies that 
\begin{eqnarray}
\label{const}
\prod_i  \, \sin \left(  \frac{\pi p_i}{M} \right) =0.
\end{eqnarray}
This means that one of the strings should be a D-string, i.e. a $(0,q)$-string, while
the remaining two strings are $(p,q)$-strings with opposite $p$ charges. In other words, the system contains $(0,q_1), (p_2, q_2)$ and $(-p_2,q_3)$
strings. This configuration is shown in Figure \ref{figconstraints}. These constraints are not surprising since, as mentioned above, the $|\psi_i|$ must be equal or zero in order to cover the boundaries properly. This means that the two D3-branes which represent the $(p_2,q_2)$ and $(-p_2, q_3)$-strings wrap the same $S^2$ inside $S^3$, but with opposite orientations. The D3-brane corresponding to the $(0,q_1)$-string, on the other hand, shrinks to a point on this $S^2$ labelled by $\psi=0$.  
This construction, pictured in Figure \ref{figconstraints}, corresponds to a limited class of three-string junctions. It would be an interesting exercise to consider the general $(p,q)$-string junction with the boundary issue properly addressed. In some situations this may mean that the D3-branes end up on a spherical D3-brane. We study this case briefly in the next section. 
\section{Beads in the throat}
\label{necklace}
\subsection{Beads}
Beads are pointlike charged objects or monopoles on which strings can end. They are also known as dyons \cite{Hindmarsh:1985xc}, because they carry both electric and magnetic charge $(P,Q)$. With the charges of strings ending on a dyon labelled by $(p_i, q_i)$, the charge conservation conditions
\begin{eqnarray}
\sum_i p_i \, = \, P &\,\,;\,\,& \sum_i q_i \, = \, Q 
\end{eqnarray}
apply. In the context of a string network, the existence of beads in the spectrum can lead to so-called cosmic necklaces being formed, whose cosmological evolution is studied in \cite{Berezinsky:1997td, Siemens:2000ty, Leblond:2007tf, Leblond:2009fq} and discussed in Section \ref{beadevol}.
There are several ways in which one can understand the appearance of monopoles as beads in cosmic string networks. In general they can form upon phase transitions of the form 
\[ G \, \rightarrow \, H \times U(1) \, \rightarrow \, H.\]
Monopoles can form during the first phase transition, while their flux can be confined in the second, threading them onto strings. If inflation dilutes these monopoles away, they can still be formed by strings breaking, and if their mass is low enough they could give a strong gravitational wave signature because their interactions are ultrarelativistic \cite{Berezinsky:1997td}. In the context of $(p,q)$-string networks, note that the F-strings are charged under $\mathbb{Z}_M$ and are thus non-BPS (their charge is not additive because it appears as the argument of sine). They can thus end on monopoles, or baryons in the dual gauge theory. When $|p_1 + p_2| >M$ at the intersection of two $(p,q)$-strings, a bead must be formed \cite{Leblond:2007tf}.
\subsection{D3-brane construction of beads in the throat}
\begin{figure}[t]
\vspace{-3cm}
  \centering
   \hspace{2.1cm}
   \includegraphics[width=4in]{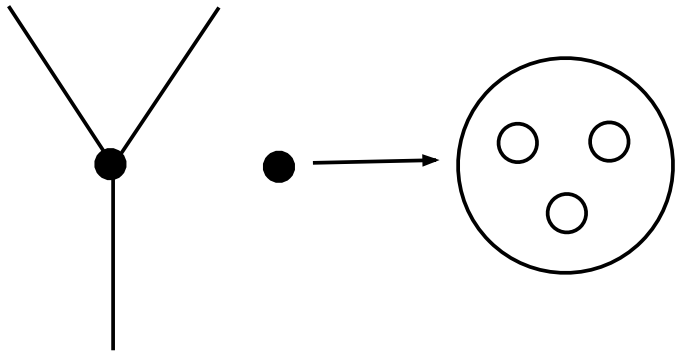} 
   \vspace{-5cm}
\caption[$(p,q)$-strings ending on a bead]{D3-brane construction of $(p,q)$-strings ending on a bead.
The figure on the left hand side represents the junction of three strings on a 
bead or dyon, pictured on the right hand side as a D3-brane wrapped on an $S^3$ with three holes.
Each hole represents an $S^2$ that the D3-branes corresponding to the $(p,q)$-strings can wrap. \label{bead}
}
\vspace{1cm}
\end{figure}

Using wrapped D3-branes, we can present a microscopical realization of  $(p,q)$-strings which end on point charges. As hinted at above, a bead can be constructed by wrapping a D3-brane on the $S^3$ in the throat.  This was suggested in \cite{Leblond:2007tf}, and is shown schematically in Figure \ref{bead}, where the D3-brane is represented as an $S^3$ with 3 holes, each of which is an $S^2$. Wrapping D3-branes with appropriate fluxes on these $S^2$s gives rise to $(p,q)$-strings, as in the previous sections, with these strings now ending on a bead. 

The bead is a D3-brane with $P$ units of electric and $Q$ units of magnetic flux, and wraps the entire $S^3$ at the bottom of the throat. For a four-dimensional observer, this wrapped D3-brane looks like a point-like charged object, as desired. A $(p,q)$-string can wrap around the $S^2$ holes, so all boundaries are covered properly. From a four-dimensional point of view this is just a three-string junction on a dyon. Similarly, one can also construct the configuration where two $(p,q)$-strings end on a bead or a bead and an anti-bead are trapped in a closed $(p,q)$-string loop.

In our construction, the mass of the bead is given by the tension of D3-brane wrapped around $S^3$:
\begin{eqnarray}
\label{beadmass}
M_b&=& 8 \pi\, h\,  (g_s M \alpha')^{3/2}\,  T_3 \nonumber\\
&=&\frac{M^{3/2}g_s^{1/2}}{\pi^{2} \alpha'^{1/2}}\, h \sim h\, m_s.
\end{eqnarray}
In this expression, $h$ stands for the warp factor at the bottom of the throat.
This just indicates that the physical mass is redshifted inside the warped throat. In the second line $m_s$ represents the scale of string theory
where $m_s \sim \alpha'^{-1/2}$. 
The implications of this result are discussed in Section \ref{beadevol}.
\section{Cosmological implications}
\subsection{Evolution of $(p,q)$-string networks}
 Unlike usual cosmic strings, which have a reconnection probability of 1 \cite{Shellard:1987bv, Moriarty:1988em}, fundamental strings can have a much lower reconnection probability, which goes like $g_s^2$, the string coupling. 
In \cite{Jackson:2004zg} it was found that F-strings have reconnection probabilities  $10^{-3} < P< 1$ and D-strings have reconnection probabilities $10^{-1} < P<1$. This could lead to an increased density
 of strings, and so to the enhancement of certain string network signatures \cite{Jackson:2004zg}. Given this, it is possible that such networks would never approach scaling, but might instead oscillate indefinitely or approach scaling only asymptotically. Another possibility is that the network might evolve into a three-dimensional structure and freeze \cite{Copeland:2003bj}, in which case it would dominate the energy density of the universe. 
However, these cosmic superstring networks have been numerically shown to reach a scaling solution, because of the energy loss mechanism provided by formation of $(p,q)$-string bound states \cite{Tye:2005fn}. In this case one can use certain results from the GUT string network simulations to make predictions for cosmic superstring networks.
 Note that if the resulting $(p', q')$-string in Figure \ref{int} or Figure \ref{minus} is such that $p'$ and $q'$ are not coprime, it will decay into a set of strings with lower $(p,q)$-values \cite{Tye:2005fn}. Specifically, if $p'$ and $q'$ are not coprime, then we can write them as $p' = N\tilde p$ and $q' = N\tilde q$, and
\begin{eqnarray*}
(N\tilde p, N\tilde q) & \rightarrow & N (\tilde p,\tilde q).
\end{eqnarray*} 
The $(p,q)$-string will decay into $N$ $(\tilde p, \tilde q)$-strings which move on a common classical trajectory \cite{Jackson:2004zg, Tye:2005fn}.
Thus  higher tension strings are difficult to form and will in general break up into lower tension strings. In fact, \cite{Tye:2005fn} found a steep power-law dependence of string number density on tension: $n \propto \mu^{-n}, n \sim 10$. String decay is a dissipative process which limits the average tension of the network and is a large enough energy-loss mechanism for scaling to be reached even when $P = 0$ (i.e. no loops form) \cite{Tye:2005fn}.

The authors of \cite{Tye:2005fn} adapted the velocity-dependent one-scale model of Martins and Shellard \cite{Martins:1996jp, Martins:1995tg, Martins:2000cs} to the case of a multitension cosmic string network with self-intersection and string-string binding interactions. They found that the $(p,q)$-string network rapidly approaches a scaling network, that scaling is achieved even when the reconnection probability is as low as zero and that the total density of the network is comparable to that of a cosmic string network. Copeland et al examined further the reconnection probabilities of $(p,q)$-string networks, finding the condition for intercommutation to occur (a function of the incoming velocities, angle of approach and string tensions involved). Their results confirm that junctions will serve to remove heavier strings from the network \cite{Copeland:2006eh, Copeland:2006if}.

Even if a scaling solution is reached, $(p,q)$-string networks with junctions can still have important observable properties that could potentially distinguish cosmic superstring networks from cosmic string networks. These are
\begin{itemize}
\item A spectrum of string tensions. Evidence of a multi-tension network could be from gravitational waves or lensing measurements, either of a large variation between two lensing angles or a variation among a larger number of events greater than that likely to arise from random string orientations and velocities. 
\item An enhancement of lensing events. 
\begin{itemize}
\item The interaction rate is expected to be reduced by the presence of extra dimensions \cite{Tye:2005fn}.
\item A strong lensing signal is also predicted for binding events, since strings near a binding site are relativistic and the gravitational lensing due to moving strings is enhanced \cite{Shlaer:2005gk}.
\end{itemize}
\item Junctions. The Y-shaped junction itself, formed during string binding, could be detected via the Kaiser-Stebbins effect, by gravitational lensing or through gravitational waves. These signals will not arise in abelian string networks, but it should be noted that similar junctions can be found in nonabelian string networks \cite{Spergel:1996ai, McGraw:1997nx}.
\begin{itemize}
\item The Kaiser-Stebbins effect refers to the fact that the signal of a light source undergoes a discontinuous change in frequency when a string comes between it and the observer \cite{2000csot.book.....V}.  The three patches resulting from a Y-shaped junction moving at a constant velocity and oriented correctly would be a very distinctive signal \cite{Tye:2005fn}. As shown by \cite{Brandenberger:2007ae}, line discontinuities in the CMB would be the distinctive signature of a junction, with details depending on its orientation and direction of motion with respect to the line of sight.
\item Each leg of a Y-shaped junction will lens exactly like a straight string \cite{Shlaer:2005ry}, resulting in a triple image (with partial obstruction) in the case when the plane of the junction lies between us and the object being lensed. Similarly N-string junctions with the strings coplanar will result in N identical overlapping images \cite{Sen:1997xi, Brandenberger:2007ae}. Observation of such an image would be an unmistakable sign of the presence of a string network which allows junctions. 
\item Gravitational radiation by excitations on the cosmic strings forming cosmic string junctions was studied in \cite{Brandenberger:2008ni}, where it was observed that the presence of the junction leads to the mixing of left and right moving excitations which is necessary for the emission of gravitational waves. They found that the resulting gravitational radiation is independent of the polarisation of the incoming wave and has magnitude proportional to the frequency of that wave.
\end{itemize} 
\end{itemize}

\subsection{Evolution of cosmic necklaces}
\label{beadevol}
Beads in cosmic superstring networks are believed to be cosmologically safe and to reach a scaling solution if the loop size is sufficiently large in the scaling regime and bead production is sufficiently suppressed in the (pre-scaling) transient regime \cite{Leblond:2007tf}. This is true when the beads' contribution to the energy density is negligible; this contribution is parametrised by the dimensionless quantity \cite{Berezinsky:1997td}
\begin{eqnarray}
r=\frac{M_b}{\mu\, d},
\end{eqnarray}
where $M_b$ is the mass of the bead, $\mu$ is the tension of the cosmic string and 
$d$ is the typical interbead distance along the string. Thus the energy per unit length of a bead-carrying string is $(r + 1)\mu$. If $r <<1$ during the network evolution, then one may safely neglect the effect of beads and the web of cosmic necklaces effectively follows the evolution of the web of $(p,q)$ cosmic strings. On the other hand, if $r>>1$, then the web is mostly dominated by beads and the network becomes a web of
massive monopoles connected by light strings. It is well known that massive (local) monopoles will overclose the universe quickly \cite{Preskill:1979zi, Guth:1979bh}. In the regime where $r \sim 1$, beads are cosmologically safe and allowed, and could at the same time give rise to observable effects. Bead pair production upon cosmic string breaking is an ultrarelativistic process, expected to give rise to an extremely scale-invariant gravitational wave spectrum \cite{Berezinsky:1997td, Leblond:2009fq}. 

Using (\ref{beadmass}) for the mass of the bead, we obtain
\begin{eqnarray}
r \sim  (h\, m_s\, d)^{-1}
\end{eqnarray}
in our construction, where the relation $\mu \sim h^{2} m_s^2$ for the cosmic string tension  (from (\ref{Ti2})) has been used. The quantity $h\, m_s$ represents the physical mass scale of the throat where the junction is formed. It is the same throat where brane inflation takes place. 
In brane-antibrane inflation, point-like defects are not copiously produced \cite{Sarangi:2002yt}.  
This means that the interbead distance $d$ is typically bigger than the Hubble radius at the end of inflation. Suppose $T_r$ and $H_r^{-1}$ are the temperature of the universe and the Hubble radius at the end of reheating, respectively. We have $H_r \sim  T_r^2/M_P$, where $M_p$ is the Planck mass. Having $d\ge H_r^{-1}$, as argued above, implies that
\begin{eqnarray}
 r \le \left(\frac{T_r}{h\, m_s}\right) \left(\frac{T_r}{M_P}\right).
\end{eqnarray}
In the above equation, the first bracket is less than unity since the reheating temperature is smaller than the scale of inflation $h \, m_s$. The second bracket is also much smaller than unity for $T_r$ not bigger than the GUT scale.

One can readily see that $r$ is many orders of magnitude smaller than unity at early stages of the network evolution. This is a direct consequence of the facts that (a) in this model the monopoles are not copiously produced and (b) we have only one mass scale: $h\, m_s$. In conventional models when both monopoles and cosmic strings could be present, the mass of the monopoles and the tension of the cosmic strings have different origins. This is due to different symmetry breaking which happens at different energy scales in the early universe. This results in different mass scales for monopoles and cosmic strings and in those models $r$ can be bigger than or close to unity.

\subsection{Conclusions}
In this chapter we have presented the results of \cite{Dasgupta:2007ds} for cosmic superstring networks in warped geometries. In particular we looked at the properties of $(p,q)$-strings, three-string junctions and beads in the throat, making use of wrapped D3-branes on appropriate cycles in the higher dimensional theory. This can also be extended to semilocal strings, as discussed further in \cite{Dasgupta:2007ds}. 

$(p,q)$-string networks and their components have some distinctive cosmological features and possible observational signatures which make their study in a realistic compactification, particularly one which supports an inflationary process that can produce them, extremely important.  A detailed study of these solutions may give us insight into the cosmological evolution of our universe and string theory itself.

A question we were unable to address fully in this investigation was the BPS nature of $(p,q)$-string junctions in the KS throat. Motivated by \cite{Dasgupta:1997pu, Rey:1997sp} and \cite{Callan:1997kz},
we expect that the three-string junction in the throat should be supersymmetric, with the requirement that it be supersymmetric leading to the constraints obtained from the equations of motion. This is left for future work. 
\newchapter{Primordial magnetic fields from superconducting heterotic cosmic strings}\label{magneto}
\section{Introduction}
Large-scale galactic magnetic fields are observed today to be coherent on scales of up to a megaparsec, with strengths of a few microgauss \cite{Beck:1995zs}.  Their continued presence despite dissipative losses can be explained by the dynamo mechanism, which amplifies an existing field using turbulence effects in the interstellar medium. The dynamo mechanism also explains the observed configuration of the galactic magnetic fields, but needs a seed field to act on. These seed fields would have to have been present and coherent at the time of galaxy formation. The origin of the seed fields required at galaxy formation is still not settled. Because of their scaling properties, cosmic strings are compelling candidates for producing coherent seed fields, as pointed out in \cite{Brandenberger:1998ew}. 

This chapter discusses an attempt at a string theoretic realisation of this process, focussing on heterotic cosmic strings - cosmic strings in a 4-dimensional theory descending from a heterotic M--theory construction. Such a realisation would provide a natural explanation for the existence of large-scale magnetic fields observed in the universe today. Heterotic cosmic strings stemming from M--branes wrapped around 4 of the compact internal directions are stable \cite{Becker:2005pv}, avoiding a problem pointed out by Witten in 1985 \cite{Witten:1985fp}.

As we shall see, it is possible to generate large enough seed magnetic fields from heterotic cosmic strings. We require that the strings support charged zero modes, which forces us into a more general heterotic M--theory picture, in which the moduli of a large moduli space of M--theory compactifications are time dependent and evolve cosmologically. These results were published in \cite{Gwyn:2008fe}.\footnote{The presentation in this chapter follows closely the original version of \cite{Gwyn:2008fe}, which was shortened substantially for publication. An extended abstract of this work was also published in \cite{PiC}.}We found that suitably wrapped M--branes, acting as strings in $3+1$ dimensions, can produce these fields, although the necessary construction is not generic. If strings are responsible for these fields, the relevant string constructions are very tightly constrained. 

We begin in Section \ref{mag} by reviewing the astrophysics of the problem: the galactic-scale magnetic fields observed today and the dynamo mechanism, believed to explain their amplitude, configuration and continued presence despite dissipative losses. In Section \ref{CSsuit} we explain why cosmic strings are viable candidates for generators of the primordial magnetic fields needed to seed those observed today, and review the superconduction properties of different cosmic strings. We review the construction of stable heterotic cosmic strings by Becker, Becker and Krause \cite{Becker:2005pv} in Section \ref{hetero}. In Section \ref{spercon} we analyse the conditions required for these strings to be superconducting and show using fermionisation that this is indeed possible. We go on to discuss stability of these strings in Section \ref{stab}, and conclude with an analysis of the resulting magnetic fields in Section \ref{magnetogenesis} and a discussion of the cosmological implications of these results in Section \ref{magconc}.

\section{Galactic magnetic fields}\label{mag}
\subsection{Primordial Magnetic Fields}
It was Fermi who first proposed the existence of a large-scale magnetic field in our galactic disc.\footnote{See \cite{Fermi:1949ee}; the story is related by Parker \cite{1979cmft.book.....P}.}He argued that such a field was needed to confine cosmic rays to the galaxy; it would have to have a strength of $10^{-6}$ to $10^{-5}$ G. From measurements of synchrotron emission, Faraday rotation, Zeeman splitting and the polarisation of optical starlight, it is now known that the gaseous disc of the galaxy contains a general azimuthal (toroidal) magnetic field with a strength of $3 \times 10^{-6}$ G and which is coherent on galactic scales of up to a megaparsec \cite{1979cmft.book.....P, Turner:1987bw, Beck:1995zs}. This field is not only necessary for confinement of cosmic rays, but is responsible for a crucial step in stellar formation\footnote{The galactic magnetic field transfers angular momentum away from protostellar clouds, which is necessary for their collapse into stars. See \cite{1978ppim.book.....S}, cited by \cite{Kulsrud:2007an}.}and plays an important role in the dynamics of other objects like pulsars and white dwarfs \cite{Turner:1987bw}.
Moreover, such fields have been detected in many other galaxies, wherever the appropriate measurements have been made, and it is believed that they are ubiquitous in galaxies and galactic clusters. Whereas in spiral galaxies like ours the magnetic field is generally coherent on scales comparable to the visible disc, in elliptical galaxies the coherence length is much smaller than the galactic scale and the fields more random. There has been no detection of purely cosmological fields (fields not associated with any gravitationally bound structure) \cite{Widrow:2002ud}.

There are no contemporary sources for galactic-scale magnetic fields, so they must either be primordial or descended from primordial magnetic fields. These seed fields would have been present at galaxy formation, and can be reasonably supposed to have condensed along with matter from the original diffuse gas clouds which contracted to form galaxies. However, there are severe observational problems with the hypothesis that these primordial fields are the ones measured today. 

Firstly, the gaseous disc of the galaxy rotates non-uniformly, with an angular velocity dependent on the distance $r$ from the axis of rotation. This non-uniform rotation would shear the lines of force of the field 
into many filaments of alternating signs, contrary to observation. In addition, these fields could not have survived to be observed today. Large-scale magnetic fields in a turbulent medium can escape through various effects which result in a characteristic decay time of $10^8$ years, to be contrasted with the galactic lifetime of $10^{10}$ years \cite{1971ApJ...163..279P}. 

If the original fields could not have survived to present times, we must conclude that the fields we observe are not primordial. In order for fields still to be present at late times despite losses, there must be 
some process that generates galactic flux continually. This is the turbulent galactic dynamo, discussed below.

\subsection{The dynamo mechanism}\label{dyn}
The turbulent galactic dynamo consists of electrically conducting matter moving in a magnetic field in such a way that the induced currents amplify and maintain the original field. Here we give a schematic review; the classic texts on magneto-hydrodynamics and dynamo theory are \cite{1978mfge.book.....M} and \cite{1979cmft.book.....P}, among others. Parker showed in a series 
of papers \cite{1955ApJ...122..293P, 1970ApJ...162..665P, 1971ApJ...163..255P, 1971ApJ...163..279P} that the gaseous disc of the galaxy is a dynamo, and the formal equations on the matter are 
contained therein and in his 1979 book \cite{1979cmft.book.....P}. Parker's derivation was independent of and equivalent to the mean field dynamo theory developed by Steenbeck, Krause and R\"adler \cite{1966ZNatA..21..369S, 1980opp..bookR....K} and applied to the galactic dynamo by Vainshtein and Ruzmaikin \cite{1971AZh....48..902V}; see \cite{Kulsrud:2007an}. There exist many papers on the subject of the galactic magnetic field and its origins (see e.g. \cite{1987QJRAS..28..197R, 1991MNRAS.248..677M}). Widrow's review \cite{Widrow:2002ud} is especially lucid and contains the key references.

The interstellar medium is turbulent because of stellar winds, supernova explosions and hydro-magnetic instabilities. This turbulence is rendered cyclonic by the non-uniform rotation of the gaseous disc of the galaxy, which means that it gains a net helicity. These two effects, cyclonic turbulence and non-uniform rotation, are the key ingredients of what is called the $\alpha \omega$ dynamo, shown by Parker \cite{1971ApJ...163..255P, 1971ApJ...163..279P} to be responsible for the magnetic field of the galaxy. The dynamo mechanism also provides an explanation for the specific field configurations observed in spiral galaxies \cite{Beck:1995zs}. It is now thus the accepted explanation\footnote{Criticisms of the model and its assumptions are reviewed by Kulsrud \cite{Kulsrud:1999bg, Kulsrud:2007an}; the author concludes that although some issues merit closer examination (particlarly the assumption in some treatments that the ISM is horizontally homogeneous), none are serious enough to cast doubt on the dynamo as the most likely generator of galactic fields.}for regeneration and amplification of the magnetic field in spiral galaxies. (Elliptical galaxies and clusters are non-rotating or slowly rotating, and coherent large-scale fields are not observed in them, an observation which provides further support for the dynamo explanation. Only small-scale local dynamos can operate in these systems \cite{Widrow:2002ud}.)

\begin{figure}[htp]
\centering
\includegraphics[scale=0.2]{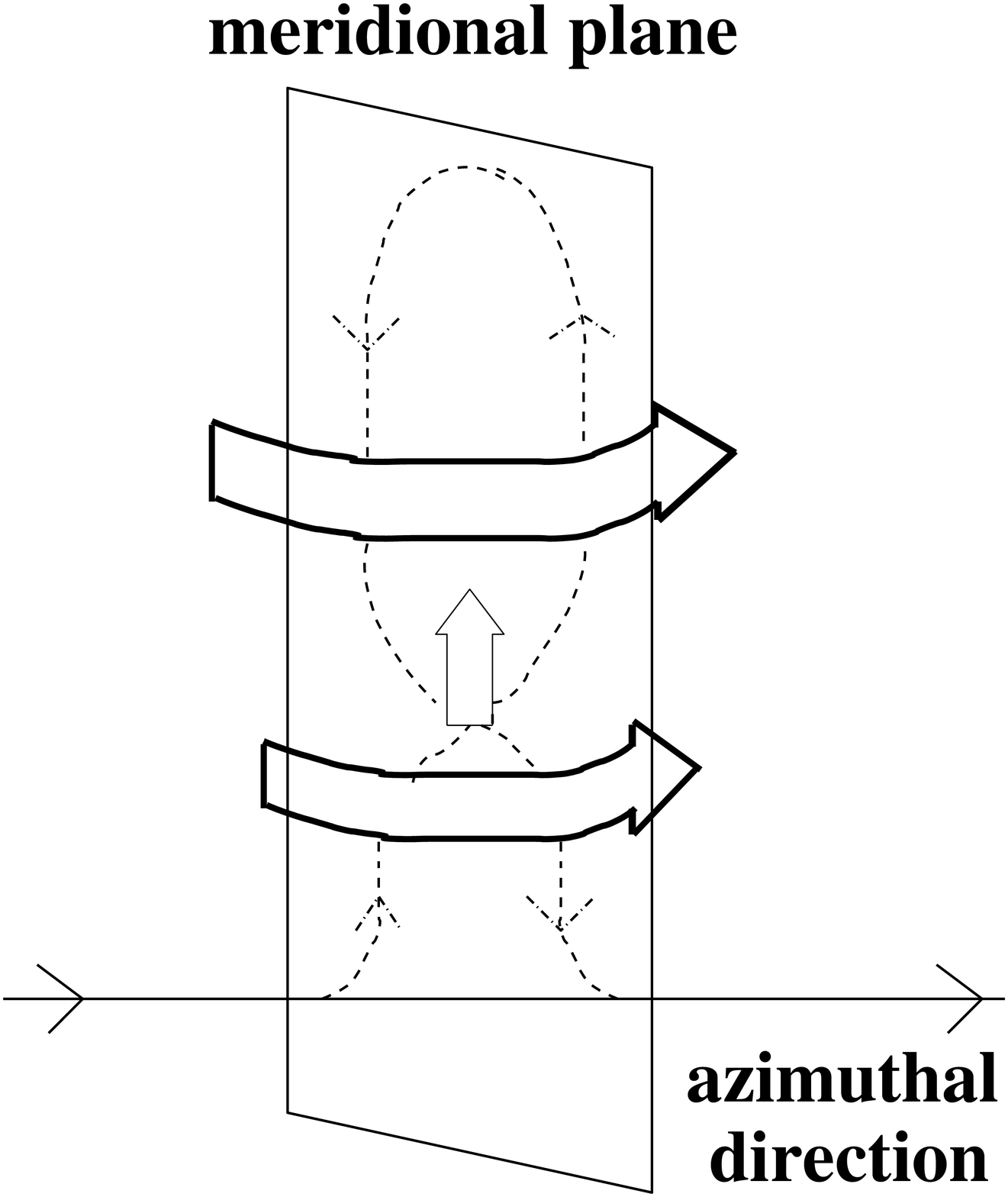}
\caption[Schematic of a cyclonic convective cell]{A cyclonic convective cell distorts and twists a magnetic field 
line in the azimuthal direction (solid black) into the meridional plane, 
generating a poloidal field line (dashed).}
\label{fig:cell}
\end{figure}

\begin{figure}[htp]
\centering
\includegraphics[scale=0.5]{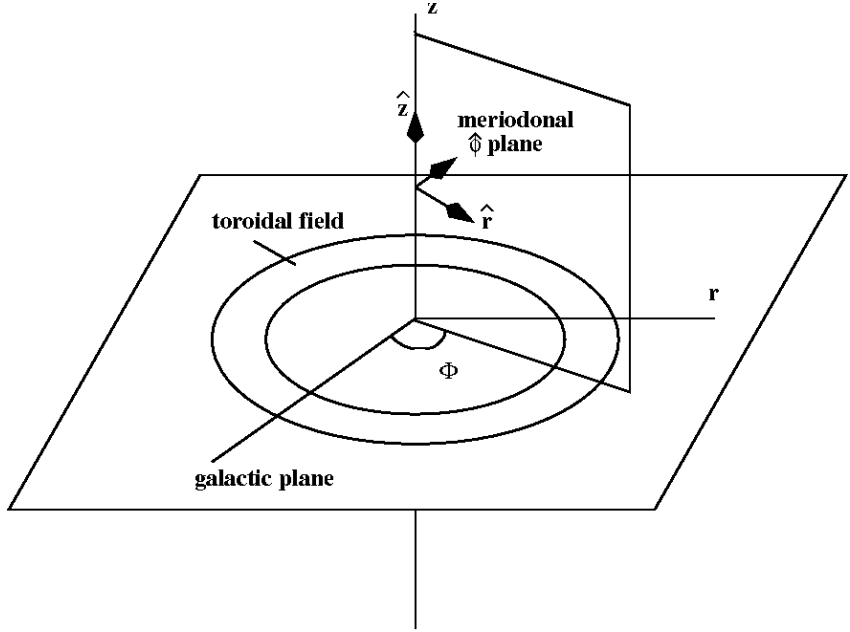}
\caption[Geometry of the galactic plane]{The galactic magnetic field lies in the plane of the galaxy, while the convective cell shown in Figure \ref{fig:cell} lies in the meridional plane perpendicular to it.}
\label{galdiag}
\end{figure}

The dynamo mechanism can be explained heuristically as follows: any poloidal field (in the meridional plane, which lies perpendicular to the plane of the galactic disc) will generate field lines in the azimuthal (toroidal) direction thanks to the non-uniform rotation. At the same time, cyclonic motion produces poloidal field from azimuthal field. This process is shown schematically in Figure \ref{fig:cell}, adapted from \cite{1971ApJ...163..279P}, where a cyclonic cell is shown raising and twisting the azimuthal field $B_\phi$ into a loop with non-vanishing projection in the meridional plane. Such loops are produced on scales comparable to the size of the largest turbulent eddies and then mixed and smoothed by general turbulence until they coalesce into a general poloidal field, from which toroidal field lines will be produced. Thus a feedback loop allowing amplification is set up. The cyclonic cell is in the meriodonal plane perpendicular to the plane of the galaxy in which the toroidal galactic magnetic field lies, as shown in Figure \ref{galdiag}. The twist that makes the convective cell cyclonic is supplied by the Coriolis force \cite{1971ApJ...163..279P}.

Formally, solution of the dynamo equations in a slab of gas, representing the galactic disc, produces regenerative modes in the azimuthal direction (for boundary conditions allowing magnetic flux to escape from the slab). Diffusion within the slab and diffusive escape from the surface of the slab are both essential to the operation of the dynamo because they permit the escape of reversed fields which would otherwise cause active degeneration \cite{Kulsrud:1999bg, 1955ApJ...122..293P, 1971ApJ...163..255P}.

These effects can be seen from the hydro-magnetic equation
\begin{eqnarray}
\label{induction}
\frac{\partial \vec{B}} {\partial t} & = & 
\vec{\nabla} \times \left ( \vec{v} \times \vec{B} \right )  + \vec \nabla \times \vec \epsilon\\
& = & \vec{\nabla} \times \left ( \vec{v} \times \vec{B} \right )  
+ \vec{\nabla} \times (\alpha \vec{B} + \eta \vec \nabla \times \vec B),
\end{eqnarray}
which governs the large-scale behaviour of magnetic fields. The approach in hydromagnetics is to treat matter as a fluid with density, velocity, pressure and current density fields.  $\vec \nabla \times \vec \epsilon$ is the inductive term: the loop in the meridional plane sketched in Figure \ref{fig:cell} will produce an electromotive force 
\begin{eqnarray}
\label{emf}
\epsilon_i & = & \alpha_{ij}  B_j + 
\eta_{ijk} \frac{\partial B_j}{\partial x_k}.
\end{eqnarray}
The first term corresponds to the helical part of the turbulence (labelled by $\alpha$) and the second term to the diffusion. $\eta$ is called the resistivity or diffusion coefficent. $\alpha$ and $\eta$ are determined by the local properties of the turbulence and are functions of position. 
The dynamo equations then follow from (\ref{induction}) and are given by \cite{1970ApJ...162..665P}
\begin{eqnarray}
\label{amp}
\left ( \frac{\partial}{\partial t} - \eta \nabla^2\right ) A_\phi & = & \Gamma(\vec{r}, t) B_\phi;\\
\nonumber \left ( \frac{\partial}{\partial t} -\eta \nabla^2 \right ) B_\phi & = &B_z \frac{\partial v_\phi}{\partial z} +  B_r \left ( \frac{\partial v_\phi}{\partial r} - \frac{v_\phi}{r} \right ),
\end{eqnarray}
where $\Gamma$ is a measure of the mean strength of the cyclonic motion and $v_\phi$ is the rotational velocity. $A_\phi$ is the vector potential of the poloidal field, the components of which are $B_r$ and $B_z$ (see Figure \ref{galdiag}). These equations show the feedback loop by which the dynamo can amplify and maintain existing fields (although a full treatment is much more complicated and nuanced - see \cite{Kulsrud:1992rk, Widrow:2002ud} or the series of papers by Parker for instance). The $\alpha \omega$ dynamo can operate in any differentially rotating body, and is accepted as the primary mechanism for the maintenance of magnetic fields in the sun and the galaxy \cite{Widrow:2002ud}.

\subsection{Seed fields and the coherence length}

We have seen that the galactic magnetic fields observed today cannot be primordial and that the dynamo effect provides a mechanism for continual generation of flux. However, seed magnetic fields which are primordial are still required.\footnote{In fact, large-scale dynamo action in a galaxy is preceded by a small-scale dynamo that prepares the seed fields for the former \cite{Beck:1995zs}.}This can be seen by considering the hydro-magnetic equation (\ref{induction}), which is linear and homogeneous in $\vec{B}$ and contains no source term. Seed fields must therefore have been present to be amplified by the dynamo mechanism. To determine the strength of the seed field required in order to obtain magnetic fields of order $10^{-6}$ G today, two effects must be considered. Firstly, magnetic fields will be amplified during galaxy formation by the stretching and compression of field lines that occur during the collapse of gas clouds to form galaxies. In spiral galaxies these processes can amplify a primordial field by several orders of magnitude \cite{Widrow:2002ud}. Amplification after galaxy formation is via the dynamo mechanism and is given by $\Gamma$, the growth rate for the dominant mode of the dynamo.  The amplification factor  ${\cal A}$ by which the magnetic field grows between times $t_i$ and $t_f$ after galaxy 
formation is then
\begin{eqnarray}
{\cal A}  \, = \, \frac{B_f}{B_i} \, =  \, e^{\Gamma ( t_f - t_i )}.
\end{eqnarray}
The maximum amplification factor is given in \cite{Widrow:2002ud} as ${\cal A} =10^{14}$, implying that a seed field with strength of at least $10^{-20}$ G is required. However, it must be noted that this 
minimum could increase. Observations of microgauss fields in galaxies at a redshift of 2 shorten the time available for dynamo action and lead to a seed field as large as $10^{-10}$ G \cite{Widrow:2002ud}. Similarly, imperfect escape of field lines may allow only a limited amplification of the mean field \cite{Kulsrud:1999bg}. On the other hand, the amplification will be greatly magnified in more general models where the cosmological constant is greater than zero \cite{Davis:1999bt}.

\begin{figure}[htp]
\centering
\includegraphics[scale=0.5]{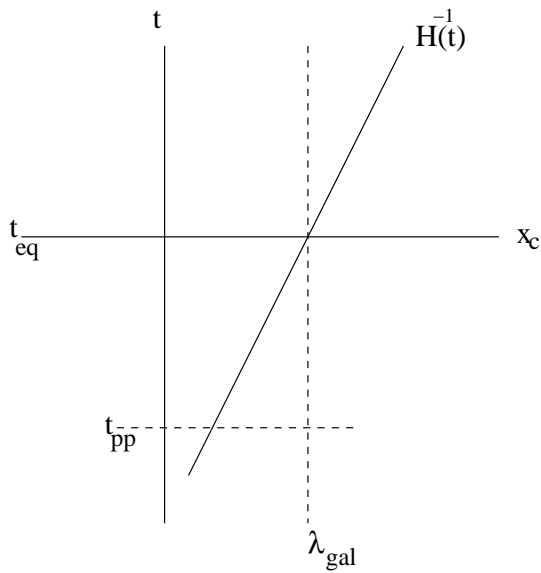}
\caption[The coherence problem]{The coherence problem. The scale of the Hubble radius at particle physics times $t_{pp}$ is much smaller than the scale of galaxy structures at the time of their formation.}
\label{fig:coherence}
\end{figure}

Various mechanisms  for generating the seed magnetic fields have been suggested, but coherence over the lengths required is not easily explained unless one makes use of a scaling string network. The challenge is the following: the seed magnetic fields need to be coherent on galactic scales at the time of galaxy formation, with a coherence length of over $100$ pc after protogalactic collapse.\footnote{See \cite{Davis:1999bt}, who cite \cite{zeldovich}, and \cite{Kulsrud:1992rk}. One might think that the dynamo could smooth out any existing fields, but in fact if the coherence length is too small the dynamo will be destabilised \cite{Davis:1999bt}. $100$ pc is the comoving scale of the largest turbulent eddy in the ISM, so that $\xi \geq 100$ pc is required for the dynamo to operate.}The comoving distance corresponding to the mean separation of galaxies has a physical size $\lambda_{\mathrm{gal}} \sim 100$ kpc similar to the Hubble radius $H(t)^{-1}$ at the time $t_{eq}$ of equal matter and radiation, which is the time when structures on galactic scales can start to grow by gravitational
instability. This is a very late time from a particle physics perspective (see Figure \ref{fig:coherence}). 

Typical particle physics processes will create magnetic fields whose coherence length is limited by the Hubble radius at the time $t_{pp}$ when the processes take place (i.e. in the very early universe). In fact, the coherence scale is typically microscopic even at that time. For instance, the comoving scale of the horizon at electroweak symmetry breaking is just $10^{-3}$ pc \cite{Dimopoulos:1997df}. Even if the coherence scale expands with the cosmological expansion of space, it will be many orders of magnitude smaller than the Hubble radius at $t_{eq}$ since the coherence length scales with $t^{1/2}$ whereas the Hubble radius grows linearly in $t$. 

Thus, explaining the coherence of the seed magnetic fields at the time corresponding to the onset of galaxy formation is a major challenge for attempts to generate seed magnetic fields using ideas from particle physics. A particle physics source that will scale appropriately so as to avoid this problem is given by cosmic strings. 

\section{Magnetic fields from cosmic strings}\label{CSsuit}
\subsection{General arguments}
Many candidate mechanisms have been suggested as generators of primordial magnetic fields. Most research has focussed on mechanisms during phase transitions or inflation, which result in highly incoherent seed fields \cite{Davis:2005ih}. Various suggestions are reviewed in \cite{Kulsrud:2007an, Widrow:2002ud} .

As we saw in Chapter \ref{chapter:CS}, cosmic strings were ruled out as a primary source of the primordial density perturbations needed to seed structure in the early years of this century. However, the results from cosmic network simulations in their heyday allow for the possibility that primordial magnetic fields were generated by cosmic strings, or even cosmic superstrings. This is precisely because of the scaling properties of these networks: whether the networks are dominated by loops or long strings they will scale such that any magnetic fields they produce are of galactic scales at the time of galaxy formation and could therefore suffuse the entire protogalactic cloud with a single coherent field. In the original models of cosmic string network evolution, loops seed galaxies and the galactic magnetic field is a remnant of the magnetic field of the string loop which seeded the galaxy. In the picture suggested by more recent simulations, long strings dominate with matter accreting in their wakes. Galaxy formation will occur in these wakes, and thus the galaxies will inherit the magnetic fields present in them. The coherence scale of these  fields is then comparable to or larger than the size of the regions which collapse to form galaxies. Thus galaxies will also inherit  coherent magnetic fields in this scenario. 

The possibility of seeding dynamos with magnetic fields produced by superconducting cosmic strings was first mentioned by Witten \cite{Witten:1984eb} and examined in detail by Vachaspati and Vilenkin \cite{Vachaspati:1991tt}, who argued that vorticity in the wakes of long wiggly strings could give rise to magnetic fields via the Harrison-Rees effect \cite{1970MNRAS.147..279H, 1987QJRAS..28..197R}. This mechanism has been invoked many times since - see for instance \cite{Avelino:1995pm, Dimopoulos:1999dn, Battefeld:2007qn, Battefeld:2007fj}.\footnote{See \cite{Hollenstein:2007kg} for a discussion of the difficulties of using the Harrison mechanism to create magnetic fields from topological defects.}In this picture the coherence scale of the resultant magnetic fields is set by the scale of wiggles on the string. For wakes created at $t_{\mathrm{eq}}$ it can be up to $100$ kpc, for fields with magnitudes of $10^{-18}$ Gauss \cite{Dimopoulos:1997df}, which is more than adequate. 

In 1997, Dimopoulos \cite{Dimopoulos:1997df} took up again (after Witten \cite{Witten:1984eb}) superconducting cosmic strings as candidates for magnetic field generation, emphasising the importance of the coherence length. Brandenberger and Zhang  \cite{Brandenberger:1998ew} considered a different mechanism in which charge is drawn into global strings by their anomalous interaction with electromagnetism \cite{Kaplan:1987kh} and showed that these strings could lead to the required field strength and coherence length for the seed fields. This set-up is reviewed below. 

\subsection{Pion strings}
Pion strings arise as global vortex line solutions of the effective QCD Lagrangian below the chiral symmetry breaking scale of $T_c \sim$ 100 - 200 MeV \cite{Zhang:1997is}. These pion strings couple to electromagnetism via anomalous axionic ($F \wedge F$) interactions \cite{Brandenberger:1998ew}. Using the results of Kaplan and Manohar \cite{Kaplan:1987kh} for such a coupling, it can then be shown that pion strings could generate seed magnetic fields greater than $10^{-20}$ G and coherent on comoving scales of a few kiloparsec, as required, provided the strings reach scaling soon enough.

\subsubsection{Anomalous coupling to electromagnetism: the Kaplan-Manohar mechanism}
In \cite{Kaplan:1987kh} the authors considered a theory with a single Dirac fermion $\psi$ coupled to a complex neutral scalar field $\phi$. The Lagrangian is
\begin{eqnarray}
\label{2.1}
{\cal L} & = & \imath \bar \psi \slashed{D} \psi + 
\left |\partial_\mu \phi \right |^2 - g \phi \bar \psi_L \psi_R - 
g \phi^* \bar \psi_R \psi_L \nonumber \\
& & - \frac{\lambda}{2} \left ( \left | \phi \right |^2 - f^2\right ) ^2 - 
\frac{1}{4} F_{\mu \nu}F^{\mu \nu},
\end{eqnarray}
and the theory has a local U(1) electromagnetic symmetry (where $\psi$ has charge 1 and $\phi$ is neutral) and a global U$_A$(1) symmetry under which
\begin{eqnarray*}
\psi_L & \rightarrow & e^{\imath \alpha} \psi_L,\\
\psi_R & \rightarrow & e^{- \imath \alpha} \psi_R,\\
\phi & \rightarrow & e^{2 \imath \alpha} \phi.
\end{eqnarray*}
The symmetry is spontaneously broken by the vacuum expectation value $\langle\phi\rangle \, = \, f$.  Then we are left with a massive scalar with mass $m_s = \sqrt{\lambda}f$, a massive fermion with $m_e = gf$, a massless photon and a massless pseudo-scalar Goldstone boson, termed the axion $a$.  

Because of the $U_A(1)$ anomaly, the axion couples to photons via the Adler-Bell-Jackiw triangle diagram. At low enough energies, only the massless particles are important, as in the low-energy effective Lagrangian obtained by integrating out the heavy particles:
\begin{eqnarray}
\label{wedge}
{\cal L} & = & \frac{1}{2} (\partial_\mu a)^2 - F \wedge \star F 
- \frac{e^2}{32 \pi^2} \left ( \frac{a}{f} \right ) F \wedge F \, .
\end{eqnarray}
Then the equation of motion for the electromagnetic field is
\begin{eqnarray}
\label{Maxwell}
dF \, = \, - \frac{\alpha}{\pi} d \left ( \frac{a}{f} \right ) \star F \, ,
\end{eqnarray}
which makes manifest the coupling between the axion and the photons.

The model given by (\ref{2.1}) has vacuum manifold ${\cal M} = S^1$, which has first homotopy group $\Pi_1({\cal M}) = \mathbb{Z}$ and hence admits vortex (cosmic string) solutions given by
\begin{eqnarray} \label{vortex}
\phi(r, \theta) \, = \, f(r) e^{\imath \theta} \, ,
\end{eqnarray}
where $f(r) \rightarrow 0$ as $r \rightarrow 0$, and $f(r) \rightarrow f$ when $r$ is much greater than the radius of the string core: $r >> r_0$. In the above, $r$ and $\theta$ are the polar coordinates in the plane perpendicular to the vortex, and $r = 0$
corresponds to the centre of the vortex. The vortex solution (\ref{vortex}) corresponds to the axion varying as we rotate about the vortex. Thus, via (\ref{Maxwell}), the vortex is coupled to the photons. Specifically, if the vortex carries a current, then the axionic coupling leads to a magnetic field circling the string.

To find the electromagnetic fields arising from this vortex configuration with current flowing along the vortex, we solve Maxwell's equations (\ref{Maxwell}) in the presence of the vortex string. This is accomplished by taking $\frac{a}{f} = \theta$ in (\ref{Maxwell}). One finds two static z-independent solutions \cite{Kaplan:1987kh}:
\begin{eqnarray} \label{field}
E_r & = & c_+ r^{-1-\frac{\alpha}{\pi}} + c_-r^{-1+\frac{\alpha}{\pi}},\\
\nonumber B_{\theta} & = & c_+ r^{-1-\frac{\alpha}{\pi}} - 
c_- r^{-1 + \frac{\alpha}{\pi}}.
\end{eqnarray}
 From the fermionic zero modes, found by solving the Dirac equation for $\psi$ in the vortex background, Kaplan and Manohar were 
able to solve for $c_\pm$, finding
\begin{eqnarray}
E_r \, = \, - B_\theta \, \sim \, r^{-1 + \frac{\alpha}{\pi}},
\end{eqnarray}
so the fall-off is slower than expected classically ($\frac{1}{r}$). Since $B_\theta = \pm E_r$, the solutions have the Lorentz transformation properties expected if the vortex were to carry a light-like current 4-vector $j^\mu = (\lambda, 0, 0, \pm \lambda)$.  This indicates that the charge carriers move along the vortex at the speed of light. The decay rate depends on the strength of the anomalous coupling. In the construction presented in \cite{Gwyn:2008fe} and below, we will apply these results to the case $\alpha = 0$.

\subsubsection{Anomalous pion strings}
 
The Lagrangian (\ref{2.1}) was generalised in \cite{Brandenberger:1998ew} to the case of the low energy nonlinear $\sigma$ model for QCD with two species of massless quarks. The model has two complex scalar fields: the first containing the charged pions $\pi^\pm$, the second the neutral pion 
$\pi^0$ and the $\sigma$ field. It is convenient to write the fields 
in an $SU(2)$ basis as
\begin{eqnarray}
\Phi \, = \, \sigma \frac{\sigma^0}{2} + \imath {\vec{\pi}}\cdot \frac{{\vec \tau}}{2} \, ,
\end{eqnarray}
where $\sigma^0$ is the unit matrix and the ${\tau_i}$ are the Pauli matrices. The bosonic part of the Lagrangian is
\begin{eqnarray}
{\cal L}_{\Phi}  \, =  \, 
{\rm tr} \left[(\partial_\mu \Phi)^{+} \partial^{\mu} \Phi \right ]  
- \frac{\lambda}{2} \left[{\rm tr}( \Phi^{+}\Phi) - f^2\right]^2 \, .
\end{eqnarray}
In addition the Lagrangian will contain the standard kinetic terms for the left- and right-handed fermion SU(2) doublets $\Psi_L$ and $\Psi_R$. The Yukawa coupling term takes the form
\begin{eqnarray}
{\cal L}_I \, = \, g {\bar \Psi}_L \Phi \Psi_R + \mathrm{h.c.} \, ,
\end{eqnarray}
where h.c. stands for Hermitean conjugate.

After spontaneous symmetry breaking, there are 3 Goldstone bosons, the massless pions $\vec{\pi}$, and a massive $\sigma$ particle:
\begin{eqnarray}
\phi \, = \, \frac{\sigma + \imath \pi^0}{\sqrt{2}}; \, \, 
\pi^\pm \, = \, \frac{\pi^1 \pm \imath \pi^2}{\sqrt{2}} \, .
\end{eqnarray}
As shown in \cite{Zhang:1997is}, this model admits vortex solutions, but not of the stable type since the vacuum manifold is ${\cal M} = S^3$ and hence has trivial first homotopy group. The vortex solutions of this model are of the embedded type. They are obtained by setting $\pi^{\pm} = 0$ and considering the vortex solution (\ref{vortex}) of the reduced two-field system where only $\phi$ is allowed to be non-vanishing. The resulting vortex solution is called the pion string. Pion strings are unstable in the vacuum since the winding of $\phi$ can disappear by $\pi^\pm$ being excited. However, as was argued in
\cite{Nagasawa:1999iv, Nagasawa:2002at}, electromagnetic plasma effects in the early universe 
will create an effective potential which drives $\pi^\pm$ to zero while not affecting $\phi$ (to leading order). Thus, one can apply the usual topological and dynamical arguments for defect formation to the
pion string model and conclude that after the QCD phase transition a network of pion strings will form which will be stabilized by the electromagnetic plasma until the time of recombination. It is then possible to apply the Kaplan-Manohar mechanism and one finds that pion strings can generate coherent seed magnetic fields greater than $10^{-20}$ G provided the strings reach scaling before the temperature $T_d$ at which they decay, and that $T_d < 10^{-2}$ MeV. 

\subsection{Superconduction mechanisms}
A superconducting string is defined as such by its response to an electric field \cite{2000csot.book.....V}. In the presence of an electric field, a superconducting string will develop an electric current:
\begin{eqnarray*}
\frac{dJ}{dt} & \sim & \left ( \frac{ce^2}{\bar h}\right ) E.
\end{eqnarray*}
Superconducting strings might be observable via synchrotron emission or relativistic jets from cusps on heavy superconducting strings \cite{2000csot.book.....V}.

Broadly speaking, there are two kinds of superconductivity, distinguished by whether the charge carriers are bosonic or fermionic \cite{Witten:1984eb}. However, it is more important in the context of cosmic strings to distinguish between the superconduction mechanisms supported by global strings on the one hand and local strings on the other. When a global symmetry is spontaneously broken in the early universe, global cosmic strings will form. Global strings couple to a Goldstone boson field which gives rise to long range interactions:  the pion strings of Brandenberger and Zhang are global strings.
 
Global strings can emit electromagnetic radiation in three ways \cite{Harari:1991nv}:
\begin{enumerate}
\item Global strings can emit Nambu-Goldstone bosons which can be converted into photons \cite{1988PhRvL..61..783S}.
\item Oscillating global strings give rise to a quasistationary NG boson field which can be a source of electromagnetic radiation. This is explored by Harari and Mazzitelli  \cite{Harari:1991nv}.
\item Electric current along the string can result in electromagnetic radiation. For global strings, superconduction can arise thanks to an anomalous term of the form in (\ref{wedge}) which causes charge to flow into the string, as explored by Kaplan and Manohar \cite{Kaplan:1987kh} (earlier references are \cite{Lazarides:1984zq} and \cite{Callan:1984sa}). Thus even though the Goldstone field is electrically neutral, global strings can couple to electromagnetism via an axionic $F \wedge F$ interaction, as above.
\end{enumerate}

On the other hand, for gauge strings to be superconducting, charged fermions must be created in a neutral combination when an electric field is applied along the string, with charges of opposite sign moving in opposite directions.  

The examples studied in \cite{Kaplan:1987kh} and \cite{Brandenberger:1998ew} are of global vortex strings whose anomalous coupling to electromagnetism results in charge flowing into the vortex so that current can flow. Our goal in \cite{Gwyn:2008fe} was to generalise the results of \cite{Brandenberger:1998ew} to the case of cosmic superstrings, and heterotic cosmic strings in particular. However, global strings are at best metastable, because they are axionic. The axionic instability is reviewed in Section \ref{WITTEN} below, and is the reason that heterotic cosmic strings were originally ruled out by Witten \cite{Witten:1985fp}. A loophole presented by Copeland, Myers and Polchinski \cite{Copeland:2003bj} was used by Becker, Becker and Krause \cite{Becker:2005pv} to construct stable heterotic cosmic strings. As we shall see, removal of the axionic instability also rules out superconduction via the anomalous coupling to electromagnetism above. This is to be expected, since the resulting strings are no longer global. We therefore look for gauge superconductivity, for which the existence of fermionic zero modes is a necessary condition. We shall see that although the heterotic cosmic strings constructed by Becker, Becker and Krause \cite{Becker:2005pv} are local, they are not superconducting. A more general set-up is required in order for fermionic zero modes to be permitted, which we give in Section \ref{modspace}.

\section{Heterotic cosmic strings}
\label{hetero}
We would like to reproduce the results of \cite{Brandenberger:1998ew} for cosmic strings arising in a string theoretic setting. We would then have an explanation for seed magnetic fields with the required coherence scale that was consistent with string theory as the theory of the early universe. 

We begin by considering heterotic cosmic strings. We require our cosmic strings to be stable, and that they carry charged zero modes. We consider heterotic strings because charge is evenly distributed over them instead of being localised at the end-points. The gauge group (either SO(32) or $\mathrm{E}_8 \times \mathrm{E}_8$) comes from charged modes that propagate only on the string. In addition, compactifications of the heterotic string have led to the most phenomenologically attractive vacua in the string/M--theory landscape. Vacua containing exactly the MSSM spectrum from heterotic compactifications were constructed in \cite{Braun:2005nv}, and other realistic vacua have been constructed (see \cite{Blumenhagen:2006ux, Lebedev:2006kn, Lebedev:2007hv} for instance).
However, as we shall see in Section \ref{WITTEN}, fundamental heterotic strings cannot give rise to stable cosmic strings upon dimensional reduction. Instead we have to use wrapped M-branes in a higher dimensional theory. Their stability is discussed in Section \ref{Mbranes} and the existence of charged zero modes on the suitable candidates is discussed in Section \ref{zeromodes}.

\subsection{The axionic instability}
\label{WITTEN}
\label{axion}
Fundamental heterotic strings were ruled out as candidates for cosmic strings by Witten in 1985 \cite{Witten:1985fp}. Not only is their tension (in a flat background) too large to be compatible with
the existing limits \cite{Wyman:2005tu, Fraisse:2005hu, Bevis:2007gh}, the strings are also unstable. Although simple decay is ruled out because there are no open strings in the theory,\footnote{Note that this is not necessarily the case for the $SO(32)$ heterotic string which can end on monopoles. This was pointed out by Polchinski \cite{Polchinski:2005bg}.}Witten argues that the fundamental heterotic string is actually an axionic string, and as a result is unstable. The argument runs as follows:  first, the worldsheet theory is anomalous because current is carried in one direction only. Then anomaly cancellation generically demands the presence of axions. During phase transitions as the universe cools axionic domain walls are formed, the boundaries of which must be superconducting. The heterotic strings become the boundaries of an axionic domain walls. The tension of the domain wall leads to an instability of the string towards its contraction. The instability can be seen by considering the energy of a large circular string \cite{Witten:1985fp}
\begin{eqnarray}
E & = & \frac{R}{\alpha \prime} + \pi R^2 \sigma \, ,
\end{eqnarray}
where $R$ is the radius of the string, $\sigma$ is the wall tension, and the second term thus represents the energy due to the domain wall tension (the tension $\sigma$ being the energy per unit area of the domain wall). This term dominates when $R > \frac{1}{\alpha' \sigma}$, and in this regime the string therefore tends to collapse. As the domain wall shrinks, strings intersect and chop each other off, until 
$R < \frac{1}{\alpha' \sigma}$. Then the string mass alone determines the energy of the string. Microscopic strings will decay away quickly through gravitational radiation. The string is prevented from growing to the cosmic scales at which it could survive by the domain wall \cite{Vilenkin:1982ks}.


\subsection{Loopholes via M theory and the BBK construction}
\label{Mbranes}

The possibility of obtaining stable cosmic superstrings was resurrected by Copeland, Myers and Polchinski \cite{Copeland:2003bj} (see also \cite{Leblond:2004uc} and the review in \cite{Polchinski:2004ia}). The existence of extended objects of higher dimension, namely branes of various types, 
provides a way to overcome the instability problems pointed out by Witten \cite{Witten:1985fp}, as we shall see for the heterotic string in particular. As we have already discussed, string tensions can in general be lowered by placing the strings in warped throats of the internal manifold and using the gravitational redshift to reduce the string tensions, so that this constraint no longer rules out all cosmic superstrings. 
 
Using the axionic instability loophole presented in \cite{Copeland:2003bj}, Becker, Becker and Krause \cite{Becker:2005pv} studied the possibility of cosmic strings in heterotic theory, pointing out that suitable string candidates can arise from wrapped branes in M theory. When compactified on a line segment $S^1/{\mathbb Z}_2$, M theory reduces to heterotic string theory \cite{Horava:1996ma}. 
Compactifying a suitable configuration to $3+1$ dimensions could give us heterotic cosmic strings in our world.  Note that because brane tensions are significantly lower than the fundamental string tension, the cosmic strings arising from such wrapped branes can also avoid the tension bound mentioned above.

\begin{figure}[htp]
\centering
\includegraphics[scale=0.4]{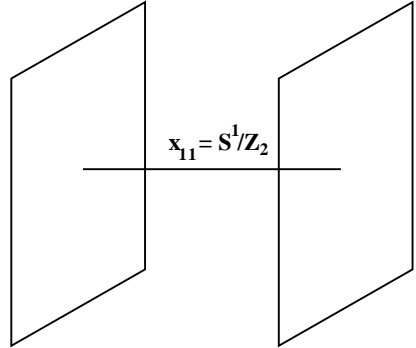}
\caption[Heterotic M-theory]{The heterotic M--theory construction: M theory is compactified on a line segment $S^1/{\mathbb Z}_2$ to give heterotic string theory in 10 dimensions.}
\label{Mtheory}
\end{figure}

There are two kinds of M--theory branes to consider as potential cosmic string candidates: M2-- and M5--branes. In descending to $3+1$ dimensions, suitable candidates must be extended along the time direction and one of the large spatial dimensions. They must therefore wrap 1- or 4-cycles 
respectively in the internal dimensions. Heterotic string theory is obtained by compactifying M theory on $S^1/{\mathbb Z}_2$, so the internal dimensions are naturally separated into $x^{11}$ along the circle, and $x^4,..,x^9 \in CY_3$ on the 10-dimensional boundaries of the space, which we can think of as M9--branes. This is shown in Figure \ref{Mtheory}.
Thus there are 4 possible wrapped-brane configurations, which can be labelled (following the notation of \cite{Becker:2005pv}) as M$2_\perp$, M$2_{\parallel}$, M$5_\perp$ and M$5_\parallel$, where the designations perpendicular and parallel refer to the brane wrapping and not wrapping the orbifold direction $x^{11}$ respectively. Their viability as cosmic string candidates is discussed below; the results are given in the following table:

\vspace{0.5cm}
\begin{table}[htbp]
\begin{center}
\TableCaption{M-brane cosmic string candidates}
\begin{tabular}{c|cccc}
& topology & tension & stability & production\\
\hline
$M2_\perp$ & \checkmark & $\times$ &\checkmark &$\times$ \\
$M2_{||}$ & $\times$ & -& - & - \\
$M5_\perp$ & \checkmark & \checkmark &$\times$& $\times$\\
$M5_{||}$ & \checkmark & \checkmark &\checkmark& \checkmark
\end{tabular}
\end{center}
\end{table}
\vspace{0.5cm}


\subsubsection{Wrapped M2--branes}
There is no 1--cycle available in a Calabi-Yau threefold, so the M2--brane candidates can only wrap $x^{11}$. We can check their viability by comparing the tension of the resulting cosmic strings with the constraint given by anisotropy measurements of the CMB (\ref{CMB}): $\mu G_N \leq 2 \times 10^{-7}$, 
where $G_N$ is Newton's gravitational constant.

The M2--brane action is given by
\begin{eqnarray} \label{M2action}
S_{M2} & = & \tau_{M2} \int dt \int dx \int_0^L dx^{11} \sqrt{-\det h_{ab}} + ... \, ,\nonumber\\
\end{eqnarray}
where $\tau_{M2}$ is the tension of the brane, and $h_{ab}$ denotes the worldsheet metric. The 11-dimensional metric $G_{IJ}$ of spacetime is found by considering the internal manifold to be compactified by the presence of G-fluxes \cite{Curio:2000dw}. The result is
\begin{eqnarray}
ds_{11}^2 & = & e^{-f(x^{11})} g_{\mu \nu} dx^{\mu}dx^{\nu} \\
& & + e^{f(x^{11})} \left (g_{mn} dy^m dy^n + dx^{11}dx^{11} \right ) \, ,
\nonumber
\end{eqnarray}
where
\begin{eqnarray}
e^{f(x^{11})} \, = \, ( 1 - x^{11} Q_v )^{2/3} \, .
\end{eqnarray}
In the above $g_{\mu \nu}$ is the metric in our four dimensional spacetime, and $g_{mn}$ is the metric on the Calabi-Yau threefold. There is warping along the orbifold direction given by the function $f(x^{11})$, and $Q_v$ is the two-brane charge. Making use of the above metric, we obtain from (\ref{M2action}) the following cosmic string action:
\begin{eqnarray}
S_{M2} & = & \mu_{M2} \int dt \int dx \sqrt{-g_{tt} g_{xx}} + ...,\\
\nonumber \mu_{M2} & = & \tau_{M2} \int_0^L dx^{11} e^{-f(x^{11})/2},\\
\nonumber  & = & \frac{3 \tau_{M2}}{2 Q_v} \left[1- (1-LQ_v)^{2/3} \right].
\end{eqnarray}
Upon evaluation, this gives a brane tension of 
\begin{eqnarray}
\mu_{M2} & \approx & 9 (2^{10} \pi^2)^{1/3} M_{GUT}^2,
\end{eqnarray}
which is too large to satisfy the bound (\ref{CMB}). Thus wrapped M2--branes are ruled out as candidates for heterotic cosmic strings. However, they \em are \em stable (see \cite{Becker:2005pv}). If produced in a cosmological context, they would therefore have disastrous consequences.
\subsubsection{Wrapped M5--branes: Tension}

For the case of the M5--brane, there are two possible types of configurations. Following \cite{Becker:2005pv} we label them M$5_{||}$ and M$5_{\perp}$. The M$5_{||}$--brane is confined to the 
10-dimensional boundary of the space, wrapping a 4-cycle $\Sigma_4$, while the M$5_{\perp}$--brane wraps $x^{11}$ and a 3-cycle $\Sigma_3$.  By similar analyses to those outlined above one obtains the
brane action for the parallel five-brane:
\begin{eqnarray}
S_{M5_{||}} \,  = \,  \tau_{M5} \int dt dx 
\int_{\Sigma_4}d^4 y \sqrt {- \det h_{ab}} + ... \, ,
\end{eqnarray}
where $\tau_{M5}$ is the brane tension. The effective string tension from the point of view of four-dimensional spacetime is given by
\begin{eqnarray}
\mu_{M5_{||}}  \, = \, 
64 \left (\frac{\pi}{2} \right ) ^{1/3} 
{\left (1 - \frac {x^{11}}{L_c} \right ) }^{2/3}M_{GUT}^2 r_{\Sigma_4}^4,
\end{eqnarray}
where $r_{\Sigma_4}$ measures the mean radius of the 4-cycle $\Sigma_4$ in units of the inverse GUT scale.  $L_c$ is a critical length of the $S^1/\mathbb{Z}_2$ interval determined by $G_N$.\footnote{See \cite{Curio:2000dw} and \cite{Curio:2003ur, Krause:2001qf} for the derivations.}

Similarly, for the orthogonal five-brane one obtains 
\begin{eqnarray}
S_{M5_\perp} & = & \tau_{M5} \int dt dx \int_0^L dx^{11} 
\int_{\Sigma_3} d^3 y \sqrt{- \det h_{ab}} 
\nonumber \\
& & + ..... \, ,
\end{eqnarray}
and the associated cosmic superstring tension is
\begin{eqnarray}
\mu_{M5_\perp} \, = \, \frac{1152}{5} {\frac{\pi}{2}}^{1/3} M_{GUT}^2 r_{\Sigma_3}^3 \, ,
\end{eqnarray}
where $r_{\Sigma_3}$ measures the mean radius of the 3-cycle $\Sigma_3$ in units of the inverse GUT scale. Although there is some dependence on the size of the wrapped space, it is not hard for the M$5_{||}$--brane to pass the CMB constraint. With a little more difficulty, the M$5_{\perp}$--brane also passes this test (although the numerical coefficient given in (3.23) of \cite{Becker:2005pv} is about an order of magnitude too small).

\subsubsection{Wrapped M5--branes: stability}

The next check is a stability analysis, which shows that only the M$5_{||}$--brane is stable. The reason is that axionic branes are unstable \cite{Witten:1985fp}. The massless axion that is responsible for this 
instability can only be avoided in the case of the M5--brane on the boundary: M$5_{||}$. The argument is presented in detail in \cite{Becker:2005pv} and is sketched below (see also \cite{Copeland:2003bj, Leblond:2004uc}).

To begin with, the presence of a massless axion is generally implied by the existence of the branes. M5--branes are charged under $C_6$ (the Hodge dual to $C_3$  in 11 dimensions). This form descends to $C_2$ in the 4-dimensional theory and, via
\begin{eqnarray}
\star d C_2 & = & d \phi,
\end{eqnarray}
this implies the presence of an axionic field. However, the presence of the M9 boundaries leads to a modification of $G = dC_3$ on the boundaries. Together with appropriate U(1) gauge fields, this leads 
to a coupling of $C_2$ to the gauge fields. This amounts to a Higgsing of the gauge field which then acquires a mass given by the axion term.

To see how this happens, recall that because of the presence of the boundaries on which a 10-dimensional theory lives, an anomaly cancellation condition must be satisfied. Writing the 10-dimensional anomaly as $I_{12} = I_4 I_8$ we require for anomaly cancellation the existence of a two-form $B_2$ such that $H = d B_2$ satisfies
\begin{eqnarray}
d H & = & I_4 \, .
\end{eqnarray}
In addition, it is required that the interaction term
\begin{eqnarray}
\Delta L \, = \, \int B_2 \wedge I_8 
\end{eqnarray}
be present \cite{Horava:1996ma}. In M theory the four-form $I_4$ is promoted to a five-form $I_5$, and although $dG = 0$ (a Bianchi identity) in the absence of boundaries, we must have
\begin{eqnarray}
dG & \sim &  \delta ( x^{11}) d x^{11} I_4 
\end{eqnarray}
in the presence of boundaries. Thus, the Bianchi identity acquires a correction term which turns out to be \cite{Horava:1996ma}
\begin{eqnarray}
dG & = &  c {\kappa}^{\frac{2}{3}} \delta\left(\frac{x^{11}}{L}\right) 
\left ( d \omega_Y - \frac{1}{2} d \omega_L \right ),
\end{eqnarray}
written in terms of the Yang-Mills three-form $\omega_Y$ and the Lorentz Chern-Simons three-form $\omega_L$ given by
\begin{eqnarray}\label{csterms}
d \omega_Y & = & \mathrm{tr}~ F \wedge F;\nonumber\\
d \omega_L & = & \mathrm{tr}~ R \wedge R.
\end{eqnarray}
Then 
\begin{eqnarray}
G \, = \, d C_3 + \frac{c}{2} \kappa^{\frac{2}{3}} 
\left ( \omega_Y - \frac{1}{2} \omega_L \right ) \epsilon (x^{11})  \wedge d x^{11}
\nonumber
\end{eqnarray}
which implies
\begin{eqnarray}
H \, = \, d B_2 - \frac{c}{2 L} k^{\frac{2}{3}} 
\left ( \omega_Y - \frac{1}{2} \omega_L \right ). 
\end{eqnarray}
It follows that $H \wedge \star H$ contains the term
\begin{eqnarray}
\left ( \omega_Y - \frac{1}{2} \omega_L \right ) \wedge d C_6,
\end{eqnarray}
which upon integration (and integrating by parts) yields
\begin{eqnarray}
\int C_6 \wedge 
\left ( {\rm tr}~ F \wedge F - \frac{1}{2} {\rm tr}~ R \wedge R \right )  .
\end{eqnarray}
Note that $C_6$ is in the M5--brane directions here.  

The gauge group is generically broken to something containing a U(1) factor, so there exists some $F_2$ on the boundary. Then the 11D action is
\begin{eqnarray}
S_{11D} \, = \, - \frac{1}{2 \times 7! \kappa_{11}^2}\int_{{\cal M}^{11}} |d C_6 |^2 &+& 
\frac{c}{2 \kappa_{11}^{\frac{4}{3}}}\int_{{\cal M} ^{10}} C_6 
\wedge {\rm tr}~ F \wedge F \nonumber \\
&-& \frac{1}{4 g_{10}^2} \int_{{\cal M} ^{10}} |F|^2
\end{eqnarray}
which dimensionally reduces to
\begin{eqnarray}\label{kore}
S_{4D}\, = \, - \frac{1}{2} \int_{{\cal M}^4} |d C_2 |^2 &+& 
m \int_{{\cal M}^4} C_2 \wedge F_2 \nonumber \\
&-& \frac{1}{2} \int_{{\cal M}^4} |F_2|^2 
\end{eqnarray}
where
\begin{eqnarray}
m & \propto & \frac{L_{\mathrm{top}}^4}{V^{\frac{1}{2}} V_h^{\frac{1}{2}}} \, ,
\end{eqnarray}
V being the CY volume averaged  over the ${S^1}/{\mathbb{Z}_2}$ interval and $V_h$ the CY volume at the boundary. $L_{\mathrm{top}}$ is a length parameter defined by
\begin{eqnarray*}
\int_{{\cal M}^{10}} C_6 \wedge tr(F \wedge F_2) & = & L_{\mathrm{top}}^4 \int_{{\cal M}^4} C_2 \wedge F_2. 
\end{eqnarray*}

The equations of motion for $A_1$ and $C_2$ are found to be
\begin{eqnarray}
d \star_4 d A_1 & = & - m d C_2;\\
\label{c2} d \star_4 d C_2 & = & - m F_2.
\end{eqnarray}
(\ref{c2}) is solved by taking $d C_2 = \star (d \phi - m A_1 )$ which gives
\begin{eqnarray}
d \star d A_1 & = &  \star ( - m d \phi + m^2 A_1).
\end{eqnarray}
For the ground state in which $\phi = 0$ or by picking a gauge which sets $d \phi = 0 $, this result shows that $A_1$ has acquired a mass $m$.
\begin{eqnarray}
\label{higgs}
A_1 \rightarrow A_1 - \frac{d \phi}{m} \, . 
\end{eqnarray}
The U(1) gauge field has swallowed the axion $\phi$ and become massive. The theory no longer contains an axion. 

In order for this anomaly cancellation mechanism (which swallows the axion and thus eliminates the instability of the strings) to work, the gauge field must be on the boundary and so the brane must be parallel to the boundary. Thus, only the M$5_{||}$--brane is stabilized, and the M$5_\perp$--brane
remains unstable. 
\section{Superconductivity}
\label{spercon}
\subsection{Fermionic zero modes}
\label{zeromodes}
The pion strings studied in \cite{Brandenberger:1998ew} are global vortex strings on which current can arise thanks to an anomalous coupling to electromagnetism. However, as commented above, global strings suffer from an axionic instability. Thus the only way to construct stable heterotic cosmic strings is by removing the instability. The axionic instability can be removed for string set-ups in which there is a $C_2 \wedge F_2$ coupling. This higgses the axion, as in (\ref{higgs}). It is easy to see that when (\ref{higgs}) holds, the term $\phi F \wedge F$ breaks gauge invariance,\footnote{We would like to thank Louis Leblond for pointing this out to us.}so that superconduction via the anomalous couping to electromagnetism is impossible.

The stable heterotic cosmic strings constructed by Becker, Becker and Krause are thus gauge strings. The only superconduction mechanism available to them requires the existence of charged fermionic zero modes on the strings. Then charged fermions are created in a neutral combination when an electric field is applied along the string, with charges of opposite sign moving in opposite directions. 


We begin by deriving the coupling to electromagnetism that can arise on the worldsheet of the heterotic cosmic string and argue using inverse bosonisation (fermionisation) that this can be recast in a more familiar form by writing it in terms of fermions. What results is an explicit kinetic term for charged fermions on the worldsheet.\footnote{We would like to thank Ori Ganor for directing our attention to the applicability of bosonisation in our case.}We shall see that although the heterotic cosmic strings constructed by Becker, Becker and Krause \cite{Becker:2005pv} are local, they are not superconducting. A more general set-up is required in order for fermionic zero modes to be permitted, which we proceed to construct. We are thus able to give local superconducting strings, with the superconductivity clearest in a fermionic basis, as in (\ref{charge}).

\subsection{Coupling to Electromagnetism}

Consider a wrapped M$5_\parallel$--brane. It can be taken to be along the following directions:
$$ \begin{array}{cccccccc} 
M5_\parallel&& 0& 1 & 4 &5&6&7
\end{array} $$
Let the $0,1$ co-ordinates be labelled by $x$ and the remaining co-ordinates wrapped on $\Sigma_4$ be labelled by $y$. The massless field content on the five-brane worldvolume is given by the tensor multiplet $(5 \phi, B_{mn}^+)$  \cite{Gibbons:1993sv, Kaplan:1995cp, Dasgupta:1995zm, Witten:1995em}, where the scalars correspond to excitations in the transverse directions and the tensor is antisymmetric and has antiselfdual field strength $H_3 = d B^+$.  Thus it has $3 = \frac{1}{2} \times { }^4C_2 $ degrees of freedom which, together with the scalars, make up the required 8 bosonic degrees 
of freedom.\footnote{A $D = 11$ Majorana spinor has 32 real components, which are reduced to 16 by the presence of the M5--brane. This means the M5--brane theory will have 16 fermionic zero modes and 8 bosonic zero modes \cite{Kaplan:1995cp}.}

The field strength $H_3$ couples to $C_3$, the bulk three-form field sourced electrically by the M2--brane and magnetically by the M5--brane, as given in \cite{Townsend:1995af}:
\begin{eqnarray}
S & = & -\,  \frac{1}{2} \int d^6 \sigma \sqrt{-h}  [ h^{ij}\partial_i X^M \partial_j X^N g_{MN}  
\\  \nonumber 
&& + \, \frac{1}{2} h^{ij} h^{jm} h^{kn} (H_{ijk} - C_ {ijk}) (H_{lmn} - C_{lmn} ) - 4],
\end{eqnarray}
which can be rewritten in terms of differential forms as
\begin{eqnarray}
\label{action}
S & = & - \, \frac{1}{2} \int d^6 \sigma \sqrt{-h}\left (  h^{ij} g_{ij} - 4 \right ) \\ \nonumber
&&  - \,  \frac{3}{2} \int (H_3 - C_3) \wedge \star (H_3 - C_3).
\end{eqnarray}
Here $i, j = 0, 1, ..., 5$ are indices on the brane worldvolume and $M, N = 0, ..., 9, 11$ are indices in the full eleven-dimensional theory. $g_{ij}$ is the pullback of the 11-dimensional metric, $C_{ijk}$ is the pullback of the 11-dimensional three-form, and $h$ is the auxiliary worldvolume metric. Explicitly,
\begin{eqnarray}
g_{ij} & =& \partial_i X^M \partial_j X^N g_{MN}^{(11)};\\
C_{ijk} & = & \partial_i X^M \partial_j X^N \partial_k X^P C_{MNP}^{(11)}.
\end{eqnarray}

$B^+$ and $C_3$ are both functions of $y$ as well as $x$. To find the massless modes on the string upon compactification on $X$, we decompose them in terms of harmonic forms. For a harmonic differential form $\beta$ on a closed compact manifold (such as $\Sigma_4$) we have $d \beta = d \star \beta = 0$. The two-form is decomposed as
\begin{eqnarray}\label{bdecom}
B^+ & = & \phi^a(x) \otimes \Omega_2^a (y) + b_2 (x) \otimes \Phi(y);\\
dB^+ & = & d \phi ^a (x) \otimes \Omega_2^a (y),
\end{eqnarray}
where $a$ runs over the two-cycles on the $\Sigma_4$ which the M5-brane wraps.\footnote{We take $\Omega_2^a$ to be antiselfdual, so that $a$ runs from 1 to $b_{-}$, where we have chosen a basis of $H^2(\Sigma_4)$ made of ($b_+$) forms which are entirely selfdual and ($b_-$) forms which are entirely antiselfdual. This imposes the property of antiselfduality mentioned earlier for the two-form living on the five-brane. (Clearly then, \( \mathrm{Dim}\,H^2(\Sigma_4) = b_- + b_+\).)}We have taken $H^1(\Sigma_4) = 0$ for simplicity. $\Omega_2^a$ are the harmonic two-forms on $\Sigma_4$ and $b_2$ is a two-form in the $0, 1$ directions. Similarly we want $C_3$ to be decomposable as
\begin{eqnarray}\label{c3decom}
C_3 & = & A^a(x) \otimes \tilde \Omega_2^a(y) + \varphi^p(x) \otimes \tilde \Omega_3^p(y),
\end{eqnarray}
where the $\tilde \Omega_2^a$ are now harmonic two-forms on the CY base, as this decomposition could give rise to the required $U(1)$ gauge fields $A^a$ in $x$-space. This time $a$ runs over the $h^{(1,1)}$ possible two-cycles on the internal space, while $p$ runs over the $2 h^{(2,1)}$ possible three-cycles. We have also denoted the harmonic three-forms by $\tilde \Omega_3^q$. 

\subsection{Generalised M--theory compactifications}
\label{modspace}
The M-theory description of the $E_8 \times E_8$ string that we have been using so far now leads to the following puzzle. To allow 
a decomposition of the three-form field of the kind that we want
means that the background $C_3$ flux would have to be switched on parallel to the 
M5$_{||}$--brane. 
This is impossible for M theory compactified on $S^1/\mathbb{Z}_2$
because the $\mathbb{Z}_2$ projection demands
\begin{eqnarray}\label{z2proj}
C_3 ~ \to ~ -C_3,
\end{eqnarray}
and therefore all components of the background G-flux with no legs along the 
$S^1/\mathbb{Z}_2$ direction are projected out! Our naive compactification of 
M theory on CY $\times S^1/\mathbb{Z}_2$ 
thus cannot give rise to charged modes propagating on the string.

However, in a cosmological setting an $E_8 \times E_8$ heterotic string in the limit of 
strong coupling cannot 
simply be described by a time-independent M--theory background. Instead the 
description should be in terms of 
a much bigger moduli space of M--theory compactifications, with the moduli themselves 
evolving with time.  Specifically, we require a large moduli space of M--theory 
compactifications that would include the heterotic compactification above, at least 
for $t=0$. Such a picture can be motivated  
from the  F-theory/heterotic duality which relates F theory compactified on a 
K3 manifold to  heterotic string theory compactified on a two-torus $T^2$ 
\cite{Vafa:1996xn, Morrison:1996na, Morrison:1996pp}. From here it follows immediately that M theory 
compactified on K3 will be dual to heterotic string theory compactified on a 
three-torus $T^3$. Fibering both 
sides of the duality by another $T^3$ gives us 
\begin{eqnarray}\label{duality}
&& {\rm M~theory~on~a}~ G_2 ~{\rm holonomy~manifold} ~ \equiv \nonumber\\
&& {\rm Heterotic ~string~theory~on} ~{\cal M}_6,
\end{eqnarray}
where the $G_2$ holonomy manifold is a seven-dimensional manifold given by a 
non-trivial $T^3$ fibration over a K3 base, and ${\cal M}_6$ is a six-dimensional 
manifold given by a non-trivial $T^3$ fibration over a $T^3$ 
base. Note that ${\cal M}_6$ is not in general a CY space. This duality has been 
discussed in the literature \cite{Atiyah:2001qf, Acharya:2001gy, Witten:2001uq}.

To confirm that there exists a point in the M--theory moduli space that describes the 
$E_8 \times E_8$ heterotic string, one needs to study the 
degeneration limits of the elliptically fibred base K3 (which can be written as a 
$T^2$ fibration over a $P^1$ base).  
Elliptically fibred K3 surfaces can be described by the family of elliptic curves 
(called Weierstrass equations)
\begin{eqnarray}
y^2 & = & x^3 + f(z) x + g(z),
\end{eqnarray}
where $(x,y)$ are the co-ordinates of the $T^2$ fibre of K3, $z$ is a co-ordinate on 
$P^1$, and $f$ and $g$ are polynomials of degree 8 and 12 respectively. Different 
moduli branches exist for which the modulus $\tau$ of the elliptic fibre is constant 
\cite{Dasgupta:1996ij}. Gauge symmetries arise from the singularity types of the fibration on 
these branches.  $E_8\times E_8$ \em can \em be realised: 
The specific degeneration limit of K3 that produces an $E_8\times E_8$ heterotic string 
 corresponds to the Weierstrass equation \cite{Dasgupta:1996ij, Morrison:1996na}
\begin{equation}
 y^2 = x^3 + (z-z_1)^5(z - z_2)^5 (z-z_3) (z-z_4).
\end{equation}
The two zeroes of order 5 each give rise to an $E_8$ factor, while the simple zeroes 
give no singularity.\footnote{This point in the moduli space of the M--theory compactification 
could as well be 
locally an $S^1/\mathbb{Z}_2$ fibration over a six-dimensional base $\widetilde{\cal M}_6$ 
(we haven't verified this here). 
Then the theory is dual to the $E_8 \times E_8$ heterotic
string compactified on $\widetilde{\cal M}_6$, and there is a clear distinction between 
M5$_{||}$ and M5$_\perp$. Our earlier stability analysis could then be used to 
eliminate M5$_\perp$.}

Given the existence of such a point in the moduli space of M--theory compactification, 
the future evolution of the system 
will in general take us to a different point in the moduli space. The picture that emerges 
from here is rather interesting. We start with heterotic $E_8 \times E_8$ theory. The 
strong coupling effects take us to the M--theory picture. From here cosmological evolution 
will drive us to a general point in the moduli space of $G_2$ manifolds. In fact, no matter 
where we start off, we will 
eventually be driven to some point in the vast moduli space of $G_2$ manifolds.

With M--theory compactified on a $G_2$ manifold, turning on fluxes becomes easy. However 
there are still a 
few subtleties that we need to address. Firstly, in the presence of fluxes we only expect the 
manifolds to have a  
$G_2$ {\it structure} and not 
necessarily $G_2$ holonomy.\footnote{For details on $G_2$ structure, see for example 
\cite{Chiossi:2002tw}.}Thus the moduli space becomes the moduli space of $G_2$ 
structure manifolds.\footnote{As should be clear,
we are no longer restricted to K3 fibered cases only. This situation is a bit like that of 
conifold transitions where 
we go from one CY moduli space to another in a cosmological setting governed by rolling 
moduli \cite{Candelas:1989ug, Strominger:1995cz}. Furthermore,  the constraint of $G_2$ 
structure comes from demanding low-energy supersymmetry. Otherwise we could 
consider any seven-manifold.}Secondly, due to the Gauss' law constraint  
we will have to consider a non-compact 
seven manifold, much like the one considered in 
\cite{Becker:2000rz}.\footnote{Note that although the seven manifold is 
non-compact, the six-dimensional base is 
always compact here. Thus our earlier arguments depending on the existence of closed 
compact cycles on a $CY_3$ still hold, for an undetermined number of such cycles on 
some compact six-dimensional base. This is a construction we are free to choose.}
Finally, since our M5--brane wraps a four-cycle inside the 
seven-manifold and we are switching on $G$ fluxes parallel to the directions of the 
wrapped M5--brane, we need to address
the concern of \cite{Duff:1996rs} that this is not permitted.

 In the presence of a G-flux on the four-cycle a wrapped 
M5--brane has the following equation of motion:\footnote{This can be seen from 
(\ref{action}): one has to find the equation of motion for $B^+$ and then impose 
the antiselfduality of $H_3$.}
\begin{equation}\label{h3g}
dH_3 = G.
\end{equation}
For a four-cycle with no boundary this implies $G = 0$, as in \cite{Duff:1996rs}. However, 
our case is slightly different. We have a 
wrapped M5--brane on a four-cycle, but the G-flux has two legs along the wrapped 
cycle (the $x^{4, 5}$ directions, say) and two legs in the 
$x^{0, 1}$ directions. Therefore the G-flux is defined on a {\it non-compact} four-cycle 
and we can turn it on if we  modify the above equation \eqref{h3g} by inserting $n$ 
M2--branes ending on the wrapped M5--brane. The M2--branes end on the M5 in small 
loops of string in the $x^{4, 5}$ directions, with their other ends at some point along the 
non-compact direction inside the seven-manifold, which the M2-branes are extended along.
These strings will change \eqref{h3g} to
\begin{equation}\label{h3gchange}
dH_3 = G - n \sum_{i = 1}^n \delta^4_{{\bf W}^i},
\end{equation}
where the $\delta^4_{{\bf W}^i}$ denote the localised actions of 
$n$ worldsheets on the M5--brane.\footnote{From the Type IIB point of 
view, this is analogous to the baryon vertex with spikes coming out from the wrapped 
D3--brane on a ${\bf S}^3$ 
with $H_{RR}$ fluxes in the 
geometric transition set up \cite{Cachazo:2001jy, Becker:2004qh, Alexander:2004eq, Becker:2005ef, Gwyn:2007qf}.} Then $G$ need no longer be vanishing. In fact,  
\begin{equation}
\int_{\widetilde\Sigma_4} G = n,
\end{equation}
where $\widetilde\Sigma_4$ is the non-compact 4-cycle. This way we see that (a) we 
can avoid the $\mathbb{Z}_2$ projection \eqref{z2proj} by going to a generic point 
in the moduli space of $G_2$ --structure manifolds, and (b) we can switch on a non-trivial 
$G$-flux along an M5--brane wrapped on a non-compact 4-cycle. 
Using the decompositions \eqref{bdecom} and \eqref{c3decom} we can now factorise the 
interaction term:
\begin{eqnarray}\label{sint}
\nonumber S_{int} & = & - \,  \frac{3}{2} \int (H_3 - C_3) \wedge \star (H_3 - C_3) + ...\\
& = & - \, \frac{3}{2} \int (dB^+ - C_3) \wedge \star (dB^+ - C_3) + ...\\
\nonumber & = & - \, \frac{3}{2} \int \left (d \phi^a - A^a \right ) \wedge \star \left (d \phi^b - A^b \right ) \otimes \Omega_2^a \wedge \star \tilde \Omega_2^b
\\ \nonumber && -\,  \frac{3}{2} \int d^2 x \sqrt{- h_x} \varphi^p \varphi^q \Omega_3^p \wedge 
\star \tilde \Omega_3^q + ...
\end{eqnarray}
where the dotted terms above involve the $n$ tadpoles coming from the worldvolume 
strings. These tadpoles are proportional to $\phi^a$. The variables
$h_x$ and $h_y$ denote the determinants of the worldvolume metrics along the 
$x$ and $y$ directions respectively.  
We are interested in the coupling to electromagnetism, so we focus on the first term 
of \eqref{sint} and take the number of 2-cycles on $\Sigma_4$ to be 1.\footnote{In the presence of multiple 2-cycles we will have more
abelian fields. This doesn't change the physics of our discussion here.}Then we have
\begin{eqnarray}
\label{boson}
S_{int} & = & - \frac{3}{2} \kappa \int d^2 \sigma |d \phi - A|^2 \sqrt{- h_x} + \ldots,
\end{eqnarray}
where
\begin{eqnarray}
 \kappa &=& \int_y \Omega_2 \wedge \star \Omega_2 
\end{eqnarray}
is a constant factor.\footnote{Note that there are also non-abelian gauge fields 
coming from $G$ fluxes 
{\it localised} at the singularities of the $G_2$ structure 
manifolds in the limit where some of the singularities are merging. 
The $G$ flux that we have switched on is non-localised. 
This picture is somewhat similar to the story developed in \cite{Dasgupta:1999ss, Becker:2003yv, Becker:2003sh} where 
heterotic gauge fields were generated 
from localised $G$ fluxes on an eight manifold. In a time-dependent background all these fluxes would also evolve with 
time, but for our present case it will suffice to assume 
a slow evolution so that the gauge fields (abelian and non-abelian) 
do not fluctuate very fast.}

\subsection{Fermionisation}

The coupling in (\ref{boson}) implies that the action can be expressed more conveniently as 
one generating fermionic superconductivity along the string.  We can see this by rewriting 
the term in terms of fermions, using a process known as fermionisation.

 In $1 + 1$ dimensions, the degrees of freedom of free fermions and free bosons match, and the corresponding conformal field theories can be shown to be equivalent. This is not the case in higher dimensions, where spin degrees of freedom distinguish between them. This observation is at the heart of bosonisation, the process of going from a fermionic basis to a bosonic basis. In evaluating the superconductors on the string resulting from the wrapped M5--brane, we find that the correct basis is a charged fermionic one, implying fermionic superconductivity.

Fermionisation\footnote{Canonical references are \cite{Coleman:1974bu},  
\cite{Mandelstam:1975hb} and \cite{Witten:1983ar}. [17] of \cite{Siegel:1985tw} gives a 
comprehensive list of the early references. A useful textbook treatment is given in 
\cite{Polchinski:1998rr}.}is possible because of the equivalence in $1+1$ dimensions of the 
conformal field theories of $2n$ Majorana fermions and $n$ bosons.\footnote{This has been 
shown to hold in the infinite volume limit as well as in the finite volume case, where care must 
be taken to match the boundary conditions correctly \cite{Green:1987sp, Green:1987mn}. Our long cosmic strings 
correspond to the infinite volume case.} 

The correlator for the bosonic field can be found from the action,\footnote{We use the 
conventions of Polchinski \cite{Polchinski:1998rr}, working in units where $\alpha' = 2.$}
\begin{eqnarray}
\label{scalaraction}
S_B & = & \frac{1}{4 \pi} \int d^2 z \, \eta_{\mu \nu}  \, \partial X^ \mu (z, \bar z)\bar \partial X^ \nu( z, \bar z),
\end{eqnarray}
to be
\begin{eqnarray}
\label{one} \langle X^\mu (z) X^\nu(w)\rangle & = & - \eta^{\mu \nu} \ln (z - w);\\
\langle X^\mu (z) \partial X^\nu (w) \rangle & = & \eta^{\mu \nu} \frac{1}{(z - w)};\\
\langle \partial X^\mu ( z) \partial X^\nu (w) \rangle & = & - \eta^{\mu \nu} \frac{1}{ (z - w)^2},
\end{eqnarray}
where $z$ and $w$ are local complex co-ordinates on the worldsheet and the correlators 
are all for the holomorphic (left-moving) parts of the bosonic fields only.  The kinetic term 
for Majorana fermions on the worldsheet is 
\begin{eqnarray}
\label{fermionicaction}
S_F & = & \frac{1}{4 \pi} \int d^2 z \left ( \psi^\mu \bar \partial \psi_\mu + \tilde \psi^\mu \partial \tilde \psi_\mu \right).
\end{eqnarray}
The fields $\psi$ and $\tilde \psi$ are holomorphic and antiholomorphic respectively, with the holomorphic correlator given by
\begin{eqnarray}
\langle \psi^\mu (z) \psi ^\nu (w) \rangle & = & \eta ^{\mu \nu} \frac{1}{(z - w)}.
\end{eqnarray}
Equivalently we could write the action and correlators in terms of 
\begin{eqnarray} \label{bar}
\psi & = & \frac{1}{\sqrt{2}} ( \psi^1 + \imath \psi^2 ),\\
\nonumber \bar \psi &= & \frac{1}{\sqrt{2}} (\psi^1 - \imath \psi^2 ),
\end{eqnarray}
as
\begin{eqnarray}
S_F & = & \frac{1}{4 \pi} \int d^2 z ( \bar \psi \bar \partial \psi + \psi \bar \partial \bar \psi)
\end{eqnarray}
(writing the holomorphic terms only). 
Then \[ \langle \psi (z) \bar \psi(w) \rangle~ =~ \frac{1}{ (z - w)}.\]
These correlators lead one to make the identification
\begin{eqnarray}
\label{id} \psi(z) & \equiv & e ^ {\imath \phi (z)};\\
\nonumber \bar \psi (z) & \equiv& e^ {- \imath \phi (z)},
\end{eqnarray}
where $\phi$ is the holomorphic part of one bosonic field. Now we consider the OPEs \cite{Polchinski:1998rr},
\begin{eqnarray}
\label{OPE} e^{\imath \phi(z)} e^{- \imath \phi(-z)} & = & \frac{1}{2 z} + \imath \partial \phi (0) + 2 z T_B^\phi(0) + ...;\\
\nonumber \psi (z) \bar \psi(-z) & = & \frac{1}{2z} + \psi \bar \psi (0) + 2 z T_B^\psi (0) + ...,
\end{eqnarray}
where $T_B^\phi$ and $T_B^\psi$ are the energy-momentum tensors arising from the actions 
(\ref{scalaraction}) and (\ref{fermionicaction}):
\begin{eqnarray}
T_B & = & - \frac{1}{2} \partial X^\mu \partial X_\mu - \frac{1}{2} \psi^\mu \partial \psi_\mu.
\end{eqnarray}
The identification (\ref{id}) implies that the OPEs (\ref{OPE}) should be equivalent, since 
all local operators in the two theories can be built from operator products of the fields 
being identified. This implies that the energy-momentum tensors of the two theories 
must be the same, allowing us to identify the theories as CFTs. This allows us to rewrite 
the kinetic term for $n$ scalars as the kinetic term of a theory containing $2n$ fermions. 
Furthermore, we have the identification
 \begin{eqnarray}
 \psi \bar \psi & \equiv & \imath \partial \phi.
\end{eqnarray}
We can now rewrite our wrapped M-brane term
\begin{eqnarray*}
|d \phi - A|^2 & = & (\partial_\mu \phi - A_\mu) (\partial^\mu \phi - A^\mu) \\
& = & \partial_\mu \phi \partial^\mu \phi - A_\mu \partial^\mu \phi - A^\mu \partial_\mu \phi + A^2
\end{eqnarray*}
in a fermionic basis:\footnote{We make use of the fact that $\phi$ is holomorphic, as discussed below.}
\begin{eqnarray}\label{charge}
 |d \phi - A|^2 & = & 2 ( \bar \psi \bar \partial \psi + \psi \bar \partial \bar \psi ) + 2 \imath  A \psi \bar \psi + 2 A \bar A\\
\nonumber & = & 2 \psi_1 (\bar \partial + \frac{\imath}{2}A) \psi_1 + 2 \psi_2 ( \bar \partial + \frac{\imath}{2} A) \psi_2 \\  \nonumber && + 2 A \psi_1 \psi_2 + 2 A \bar A,
\end{eqnarray}
which makes it clear that the worldsheet supports charged fermionic modes. 
Here $A$ and $\bar A$ are defined in terms of components as in (\ref{bar}). 
Each boson is replaced by one $\psi$ fermion and one $\bar \psi$ fermion at the 
same point, moving left at the speed of light, and carrying charge as shown 
explicitly by (\ref{charge}). This proves the existence of charged fermionic zero 
modes on the string obtained by suitably wrapping an M5-brane. Note that \cite{Witten:1984eb} gives a similar discussion, relating a theory describing charged fermionic zero modes trapped on a string to a bosonised dual with an interaction of the form $|d \phi - A|^2$.

One might worry that the above analysis should hold equivalently for the antiholomorphic 
part of the bosonic fields, leading to an equal number of right-moving fermionic modes. 
This is not the case, since $\phi$ is in fact holomorphic. From the antiselfduality of $dB^+$ 
it follows that $d\phi = - \star d \phi$ in $1+1$-dimensions.\footnote{This conclusion also 
depends on the fact that we have chosen a Calabi-Yau (or 6-d base of our 7-manifold) 
with only one 2-cycle on the 4-cycle $\Sigma_4$.}Writing $d\phi$ as 
$(\partial + \bar \partial) \phi$ one can show that $\bar \partial \phi = 0$ is implied by 
the antiselfduality condition. This is just the condition that $\phi$ does not depend on 
$\bar z$, i.e. it is holomorphic or, in worldsheet terms, left-moving. 

\section{Stability and  Production}
\label{stab}
\subsection{Axionic Stability}
Finally, we should argue that the axionic instability is also removed for 
our case. This can easily be seen either directly from M--theory or from 
its type IIA limit. From a type IIA point of view the wrapped M5-brane can appear as
a D4-brane or an NS5-brane in ten dimensions depending on which direction we 
compactify in  M theory. First, assume that the 
four-cycle $\Sigma_4$ on which we have the wrapped M5-brane is locally of the form 
$\Sigma_3 \times S^1$. Then M theory 
can be compactified along the $S^1$ direction to give a wrapped D4-brane on 
$\Sigma_3$ in ten dimensions.\footnote{One might worry at this stage that this is 
not the standard M5$_{||}$ that we want. Recall however that 
at a generic point of the moduli space M5$_{||}$ and 
M5$_\perp$ cannot be distinguished.} We can now eliminate the axion following Becker, Becker and Krause \cite{Becker:2005pv}. The 
axion here appears from the D4-brane source i.e. the five-form RR-charge $C_5$. 
This form descends to an axion in 
four dimensions exactly as we discussed before ($dC_5$ descends to $dC_2$ in four 
dimensions, which in turn is 
Hodge dual to $d\phi$, the axion). What are the gauge fields that will eat the axion?
In the BBK case \cite{Becker:2005pv}, the gauge fields arose on the 
ten-dimensional boundary. Here, instead of the boundary, we can insert coincident
D8 branes\footnote{Such D8 branes are allowed in massive type IIA theory. They correspond to M9-branes when lifted to M theory \cite{Bergshoeff:1996ui, Bergshoeff:1998bs}. One can reduce an M9 either as a nine-brane in type IIA theory or as a D8-brane. The nine-brane
configuration is exactly dual to the $E_8 \times E_8$ theory that we discussed before, where the 
required O9-plane comes from 
Gauss' law constraint. To avoid the orientifold of the nine-brane configuration in type IIA, 
we consider only D8-branes in type IIA.}that allow gauge fields to propagate on their worldvolume $\Sigma_8$. 
Therefore the relevant parts of the action are
\begin{eqnarray}
&&-{1\over \kappa_{10}^2}\int \vert dC_5\vert^2 + \mu_8\int_{\Sigma_8} C_5 \wedge {\rm tr}~F \wedge F \nonumber\\ 
&&~~~~~~~~~~~~~~~~~ - {1\over g_{\rm YM}^2}\int_{\Sigma_8} \vert F\vert^2,
\end{eqnarray}
which dimensionally reduce to an action similar to \eqref{kore}. This implies that the 
D8-brane gauge fields can 
eat up the axion to become heavy, and in turn eliminate the axionic instability. One 
subtlety with this process is
the global D8-brane charge cancellation once we compactify. In fact a similar charge 
cancellation condition should also arise 
for the M2-branes that we introduced earlier to allow non-trivial fluxes on the M5 branes. 
We need to keep one of 
the internal directions non-compact to satisfy Gauss' law.\footnote{A fully compactified version would 
require a much more elaborate framework that we do not address here.}   

If instead we dimensionally reduce in a direction orthogonal to the wrapped M5 brane, 
then one can show that it is 
impossible to eliminate the axionic instability by the above process. 

\subsection{Production of M$5_\parallel$--branes}

Whether strings or branes of a particular type will be present at late cosmological times relevant to the generation of seed galactic magnetic fields will depend on the history of the early universe.
We can distinguish between cosmological models which underwent a phase of cosmic inflation and those which did not, for instance standard Big Bang cosmology, Pre-Big-Bang cosmology \cite{Gasperini:1992em}, Ekpyrotic cosmology \cite{Khoury:2001wf} and string gas cosmology \cite{Nayeri:2005ck, Brandenberger:1988aj}.

In models without inflation in which there was a very hot thermal stage in the very early universe, all types of stable particles, strings and branes will be present. In such models one expects all stable branes to be present. Since the wrapped M$2_\perp$--branes are stable but have too large a tension for the values of the parameters considered here, we conclude that there is a potential problem for our proposed magnetogenesis scenario without a period of inflation which would eliminate the M$2_\perp$--branes present in the hot early universe. However, if the temperature was never hot enough to thermally produce the M$2_\perp$--branes,  which is possible in string gas cosmology or in bouncing
cosmologies, there would be no cosmological M$2_\perp$--brane problem.\footnote{Another way to get rid of the potential M$2_\perp$--brane problem might be to change the parameters of the model in order to reduce the M$2_\perp$--brane tension to an acceptable level.}

In inflationary universe scenarios, the pre-inflationary number densities of all particles, strings and branes are diluted by inflation. To have any strings or branes present within our Hubble patch after inflation, these objects must be generated at the end of inflation. Which objects are produced will depend critically on the details of the inflationary model. We consider first the model of M--theory inflation suggested by Becker, Becker and Krause \cite{Becker:2005sg}. In this model, several M5--branes are distributed along the ${S^1}/{\mathbb{Z}_2}$ interval. During the inflationary phase they are sent towards the boundaries by repulsive interactions. Slow-roll conditions are satisfied as long as the distance $d$ between the M5 branes is much less than $L$ the orbifold length. Once $d \sim L$
non-perturbative contributions (which stabilise the orbifold length and Calabi-Yau volume at values consistent with a realistic value for $G_N$ and a SUSY-breaking scale close to a TeV) come into effect. This stabilisation was used in the argument above and also leads to a small M$5_\parallel$ tension, so that while wrapped M5--branes will be produced at the end of inflation there is insufficient
energy density to produce the M$2_\perp$--branes. 

In our model, where cosmological evolution takes us to a generic point
in the moduli space of $G_2$ structure manifolds (by rolling moduli), there may not be 
a problem with M$2_\perp$--branes $-$ at least in the limit of compact $G_2$ structure 
manifolds with $G_2$ holonomy. 
This is because compact manifolds with $G_2$ holonomy have finite fundamental group. 
This implies vanishing of the first Betti number \cite{joyce}, which in turn means that M2--branes 
have no 1-cycles to wrap on. Once we make the $G_2$ manifolds non-compact 
(keeping the six-dimensional base compact 
with vanishing first Chern class\footnote{The base doesn't have to be a Calabi-Yau 
manifold to have vanishing first Chern class. See for example constructions in \cite{Dasgupta:1999ss, Becker:2003yv, Becker:2003sh}.}), we can still argue the non existence of finite 1-cycles, and therefore we don't expect a 
cosmological M2--brane problem.  The question of generalising M--theory inflation to this picture is still open.

\section{Magnetogenesis}
\label{magnetogenesis}
To estimate the amplitude of the resulting magnetic fields, we recall that a current-carrying conductor is surrounded by an azimuthal magnetic field whose strength at a distance $r$ from the string is
\begin{eqnarray}
\label{magn}
\nonumber B(r) & = & \frac{J}{2 \pi r}\\
B(r) & \sim & \frac{e n} {2 \pi r},
\end{eqnarray}
where $J$ is the current on the conductor (or string), given by the electric charge multiplied by $n$, the number density per unit length of charge carriers on the string. This is what the expression found by \cite{Brandenberger:1998ew} and \cite{Kaplan:1987kh} reduces to for the case $\alpha = 0$, i.e. when there is no anomalous $F \wedge F$ term present. We expect that the coefficient in (\ref{magn}) will depend on the parameters of the compactification. It has not been our aim here to fix these in any sense, but merely to make a plausibility argument for the use of heterotic cosmic strings to generate primordial fields. 

We want to calculate the magnetic field at a time $t$ after decoupling in the matter-dominated
epoch (specifically, at the beginning of the period of galaxy formation)
at a distance $r$ from the string. We will take this distance to be a
typical galactic scale. During the formation of the string network at time $t_c$, the number 
density of charge carriers per unit length is of the order of $T_c = T(t_c)$ \cite{Brandenberger:1998ew}:
\begin{eqnarray} \label{initial}
n(t_c) \, \sim \, T_c \, .
\end{eqnarray}
As the correlation length $\xi(t)$ of the
string network expands, the number density drops proportionally to
the inverse correlation length. However, mergers of string loops onto
the long strings leads to a buildup of charge on the long strings which
can be modelled as a random walk \cite{Brandenberger:1998ew} and partially cancels the
dilution due to the expansion of the universe.\footnote{Note that without
string interactions, the correlation length $\xi(t)$ would not scale
as $t$.}These effects lead to the expression
\begin{eqnarray} 
\nonumber n(t) & \sim & \frac{\xi(t_c)}{\xi(t)} \left (\frac{\xi(t)}{\xi(t_c)} \right ) ^{\frac{1}{2}} n(t_c)\\
n(t) & \sim & \left(\frac{\xi(t_c)}{\xi(t)}\right)^{1/2} n(t_c) \, .
\end{eqnarray}
In the first line the first factor arises from the stretching of the strings, and the second from the random walk superposition \cite{Brandenberger:1998ew}. 
Assuming that the universe is dominated by radiation between $t_c$
and $t_{eq}$ and by matter from $t_{eq}$ until $t$, we can express
the ratio of correlation lengths in terms of ratios of temperatures:
\begin{eqnarray}\label{ndens}
n(t) \, \sim \, \left[\frac{T(t)}{T_{eq}} \right]^{3/4} 
\frac{T_{eq}}{T_c} n(t_c) \, .
\end{eqnarray}
To see this, note that $\xi(t) \sim t$ at late times, that $a(t) \sim t^{\frac{1}{2}}$ during the radiation-dominated era, $a(t) \sim t^{\frac{2}{3}}$ during the matter-dominated era, and $T^{-1} \sim a(t)$. In fact, for early times $\xi(t) \sim t^p$ where $p$ is slightly greater than 1 \cite{Brandenberger:1998ew}. For the case of the pion strings, it was assumed that $p=1$ for all times because these strings are not stable and their production should be suppressed. In our case this is not a good assumption, and one should keep in mind that a power of $p$ should probably appear on the $\frac{T_{eq}}{T_c}$ factor. However, we have no good estimate for $T_c$ for our case, and are therefore unable to estimate the strength of the magnetic fields produced by heterotic cosmic strings unless we take $p=1$ so that the factors of $T_c$ will cancel. We note that $p$ is only slightly greater than 1 ($p = \frac{5}{4}$ according to \cite{Kibble:1981gv}.) and that $T_c < T_{eq}$ so that the neglected factor of $\left ( \frac{T_{eq}}{T_c}\right )^{\frac{1}{4}}$ should give a small amplification.\footnote{Detailed simulations of cosmic superstring networks may change this estimate.} To proceed, we follow \cite{Brandenberger:1998ew} and take $\xi(t) \sim t$ for all times.  

Inserting  (\ref{ndens}) and (\ref{initial}) into (\ref{magn}) gives
\begin{eqnarray}
B(t) \, \sim \, \frac{e}{2\pi} 
\frac{T_{eq}}{r} \left[\frac{T(t)}{T_{eq}} \right]^{3/4}.
\end{eqnarray}
By expressing the temperature in units of GeV and the radius in units
of $1$ m, and converting from natural units to physical units making
use of the relation
\begin{eqnarray}
\frac{e}{2\pi} \frac{GeV}{m} \, = \, 10^5 ~{\rm Gauss},
\end{eqnarray}
we obtain 
\begin{eqnarray}
B(t) \, \sim \, 10^5 ~{\rm Gauss} 
\frac{T_{eq}}{\rm{GeV}} r_M^{-1} \left[\frac{T(t)}{T_{eq}} \right]^{3/4},
\end{eqnarray}
where $r_M$ is the radius in units of metres.

Evaluated at the time of recombination $t_{rec}$
(shortly after the time $t_{eq}$)
and at a radius of $1$ pc, the physical length which turns into
the current galaxy radius after expansion from $t_{rec}$ to the
current time, we obtain
\begin{eqnarray}
B(t) \, \sim \, 10^{-20}~{\rm Gauss} \left(\frac{r}{r_c}\right)^{\alpha \pi}.
\end{eqnarray}
%
Thus this value is of the right order of magnitude as is required to
yield the seed magnetic field for an efficient galactic dynamo. In the case of a global string with anomalous coupling to electromagnetism, the amplitude would be enhanced by a factor of $ \left ( \frac{r}{r_0} \right ) ^ {\alpha \pi}$, where $r_0$ is the width of the string.  This factor could be large, since
$r_c$ is a microscopic scale whereas $r$ is cosmological, but it cannot arise for heterotic strings whose axionic instability is removed as in BBK \cite{Becker:2005pv}.

\section{Discussion and Conclusions}
\label{magconc}
We have attempted to construct heterotic cosmic strings which could produce seed magnetic fields coherent on galactic scales at the time of galaxy formation. Instead of employing the Harrison-Rees mechanism to generate magnetic fields in the plasma around the strings, we have explored the question of whether strings can produce magnetic fields directly, as in the Kaplan-Manohar mechanism. This requires that the strings be superconducting. In the case of global strings such as the pion strings considered by Brandenberger and Zhang, superconduction is achieved by anomalous coupling to electromagnetism. This coupling is not allowed for heterotic cosmic strings which are stable thanks to removal of the axionic instability. For these strings to be superconducting, fermionic zero modes must be supported on the strings. As we have seen, the heterotic cosmic strings constructed by Becker, Becker and Krause \cite{Becker:2005pv} are not superconducting. A more general set-up is required in order for fermionic zero modes to be permitted, which is what we have constructed. Thus ours are local superconducting strings, where the superconductivity is clearest in a fermionic basis, as in (\ref{charge}).



The existence of charged fermionic zero modes is sufficient to give current on the strings only if magnetic fields are present at the time of string formation. This is generically true because phase transitions will produce small randomly oriented magnetic fields by the Kibble mechanism \cite{1991PhLB..265..258V}. These are exactly the primordial fields that have been suggested as candidates for seeding the magnetic fields in galaxies, but they lack the required coherence property. They can however induce currents on superconducting cosmic strings which give rise to magnetic fields. The cosmic string network continues to generate magnetic fields for all times, with the network of magnetic field lines stretching as the string network stretches. The correlation length of the magnetic fields scales as $\xi(t)$.

This scaling property of the magnetic fields generated by cosmic strings is true both for the pion strings and for our superconducting heterotic cosmic strings. However, pion strings are unstable \cite{Nagasawa:1999iv, Nagasawa:2002at} and will decay at some time $t_d$, after galaxy formation.  The final correlation length of the magnetic fields set up by these strings will be given by the comoving distance corresponding to $\xi(t_d)$, which is of the order of the Hubble radius at that time. Provided that pion strings decay later than the time corresponding to a temperature of 1 ${\rm MeV}$, this final correlation length will be of the size of a galaxy, giving a cutoff on the scale of coherent magnetic fields. In this model, magnetic fields on supergalactic scales can arise only as a random superposition of galactic scale fields, and hence the power spectrum of magnetic fields will be Poisson-suppressed on these scales.

By contrast, our superconducting heterotic strings are stable, so there is no upper bound on the coherence length of the fields produced. Coherent magnetic fields on the scale of galaxy clusters and above will result. This means our mechanism is in principle distinguishable from that of \cite{Brandenberger:1998ew}. However, it is only seed fields on scales which undergo gravitational collapse which can be amplified by the galactic dynamo mechanism. The fields which we predict on larger scales will not have been amplified and thus will have a very small amplitude. These weak coherent fields are therefore a prediction of our set-up, but their amplitude is presumably beyond our current detection abilities.  New data from the SKA (Square Kilometre Array) may shed some light on this issue, and the question of the origin of seed magnetic fields in galaxies in general, by improving our knowledge of extra-galactic systems in early stages of formation and allowing us to map the evolution of magnetic fields in this system \cite{Gaensler:2004gk, Gaensler:2006tj}. It is an interesting question for future study to ask whether these observations could rule out or constrain our model. 

To summarise, our proposal uses wrapped M5--branes at some point in the moduli space of $G_2$ structure manifolds, which act as superconducting cosmic strings from the point of view of our four-dimensional universe. These branes are stable, and carry charged zero modes which are excited via the Kibble mechanism in the early universe. Because of the scaling properties of cosmic string networks,  the currents on the strings resulting from the charged zero modes generate magnetic fields which are coherent on the scale of the cosmic string network. This scale is proportional to the Hubble distance at late times, which means that the scale increases much faster in time than the physical length associated with a fixed comoving scale. It is this scaling which enables our mechanism to generate magnetic fields that are coherent on galactic scales at the time of  galaxy formation.

\newchapter{Conclusion}
In the course of this thesis we have explored the intersection of string theory with particle physics and cosmology. To begin with, in Chapter \ref{chapter:axion}, we considered the problem of constructing string theory realisations of axions compatible with the observational bounds. Axions come about naturally in string theory as some of the scalar moduli that arise upon compactification. However, their decay constant is generically far too large, at $10^{16}$ GeV in most string models. Values this high would imply an overabundance of axionic dark matter, which is ruled out observationally. Assuming string theory is correct, finding a realistic axion model in string theory is a string theoretic problem with stringent experimental constraints, which is uncommon and very interesting. We found that in warped heterotic compactifications, the axion decay constant is essentially given by the mass scale in the throat. This means that in a suitably engineered compactification geometry, the axion decay constant can be lowered to experimentally acceptable values. A small amount of fine tuning is required, but can be removed by allowing for a number of throats with different warp factors in the compactification. This construction is thus a strong sign that our `real-world' physics takes place in a warped region of the compactification geometry. It would be misleading to say that we are constrained to a specific warped heterotic construction since, as discussed in Chapter \ref{introduction}, heterotic string theory is related to the other 10-dimensional string theories by duality transformations. It is however an encouraging result in the sense that we have shown it is possible to construct a string realisation of the axion which is consistent with the data. 

In the second part of the thesis, we have focussed on signatures of cosmic superstrings. Networks of cosmic superstrings are generically produced at the end of inflation in string models. They have some properties in common with the GUT cosmic string networks originally studied in the eighties and nineties, but can also be qualitatively distinguished from them, opening a potential experimental window onto string theory in the early universe. This is because cosmic superstring networks are actually $(p,q)$-string networks in which the intercommutation probability can be much less than one, and junctions can arise when two strings meet and form a bound state. In Chapter \ref{lumps} we were able to study these junctions in a warped geometry which is the natural habitat of brane inflation (and of realistic string theory, as we learnt from Chapter \ref{chapter:axion}). Our result reduces to the known formula in the flat space limit and is consistent with the description of $(p,q)$-strings in warped geometries. We expect that this should be consistent with SUSY considerations, a calculation we leave for future work. In deriving our result we were also able to provide a formalism in which $(p,q)$-strings, junctions, semilocal strings and cosmic necklaces can all be constructed by wrapping D3-branes with suitable fluxes on 0,1,2 or 3-cycles. We have not explicitly calculated the observable consequences of these results, which will most likely require detailed numerical simulations adapted to take these junction tension properties into account.

Finally, in Chapter \ref{magneto}, we asked a difficult and ambitious question concerning the possibility of using cosmic superstrings to generate primordial magnetic fields. This has been attempted before, but only for the case of magnetic field generation by the Harrison-Rees mechanism, i.e. by vortices in the wakes of superconducting strings. By contrast, we have tried to construct a model in which magnetic fields are generated directly by the superstrings, as by a current-carrying conductor. We have neglected the interaction of these strings with the plasma, instead focussing on the M-theory construction of suitable strings. Stable local heterotic cosmic strings can be constructed by wrapping M5-branes appropriately. However, these strings do not support charged zero modes, which are needed for the gauge strings to be superconducting. In order to construct strings with charged zero modes, it is necessary to go to a more general M-theory compactification picture in which the compactification moduli are time-dependent and evolve cosmologically. This is very suggestive. A consistent M-theory picture of cosmology would go a long way towards a complete string description of our universe, particularly since it is not ambiguous up to a duality transformation as other string models are. 

We have not attempted to combine the results of these three investigations into one consistent picture. However, it is interesting to note that the warped compactification demanded by axion data is generic in brane inflation scenarios, including the M5-brane inflation picture of Becker, Becker and Krause. Many questions still remain: What are the properties of a network of superconducting heterotic cosmic strings? Is such a network differentiable from a $(p,q)$-string network, or a GUT cosmic string network? We expect that a string network in heterotic theory should not be distinguishable from a string network in IIA or IIB since they are related by duality transformations, but this remains to be checked. Lastly, although inflation has only been mentioned tangentially in this thesis, it is one of the most important problems in string cosmology. It would be extremely interesting to see how the existing models of inflation in M-theory cold be generalised to the case of a cosmologically varying M-theory compactification, as is demanded for our superconducting heterotic cosmic strings. 

As discussed in the introduction, the universe is essentially a huge once-off particle accelerator experiment which has probed higher energy scales than we can ever reach with particle accelerators here on earth. We do not have direct access to the very early universe where string theory, if correct, should apply. However, we do have indirect access via what we know of the universe's subsequent evolution.  Cosmological observations constrain string models, since the string vacuum describing our world must be consistent with known cosmology and astrophysics, as well as known terrestrial physics. Furthermore, there may remain signatures of an early stringy regime in the universe today. With sufficient data and a better theory of string cosmology, we may hope to eventually be able to make general statements about which string models are consistent with the universe's evolution and possibly even be able to amass evidence in favour of string theory. The fact that new particle physics, astrophysics and cosmological data is expected very soon makes this a fascinating time to study string theory in the early universe.

\bibHeading{\bf{References}}
\bibliography{thesisbib}
\bibliographystyle{hieeetr}


\end{document}